%% file: main_arviv_new.tex
\documentclass[ reprint,amsmath,amssymb, aps,pra,]{revtex4-2}
\usepackage{subcaption} %
\usepackage{graphicx} % for pdf, bitmapped graphics files
\usepackage{amsmath} % assumes amsmath package installed
\usepackage{amssymb,amsthm} % assumes amsmath package installed
\usepackage[colorlinks=true]{hyperref}
\usepackage{relsize} % use \mathlarger{}
\usepackage{bbold}
\usepackage[percent]{overpic}
\usepackage[export]{adjustbox}
\usepackage{lmodern}
\usepackage{blochsphere}
\usepackage{xcolor}
\usepackage{ifpdf}
\usepackage{qcircuit}

\definecolor{sphere_color}{HTML}{F5F5F5}
\definecolor{bleu}{HTML}{005377}
\definecolor{vert}{HTML}{18A089}
\definecolor{violet}{HTML}{721D68}
\definecolor{indigo}{HTML}{8D86C9}

\newcommand{\bg}{\widehat{g}}
\newcommand{\ba}{\widehat{a}}
\newcommand{\bb}{\widehat{b}}
\newcommand{\bc}{\widehat{c}}

\newcommand{\bH}{\widehat{H}}
\newcommand{\bI}{\widehat{I}}
\newcommand{\bJ}{\widehat{J}}

\newcommand{\bL}{\widehat{L}}
\newcommand{\bM}{\widehat{M}}

\newcommand{\bR}{\widehat{R}}
\newcommand{\bS}{\widehat{S}}
\newcommand{\bU}{\widehat{U}}

\newcommand{\bW}{\widehat{W}}
\newcommand{\bX}{\widehat{X}}
\newcommand{\bY}{\widehat{Y}}
\newcommand{\bZ}{\widehat{Z}}

\newcommand{\cD}{\mathcal{D}}

\newcommand{\cL}{\mathcal{L}}
\newcommand{\cK}{\mathcal{K}}
\newcommand{\ocK}{\overline{\mathcal{K}}_0}

\newcommand{\ocR}{\overline{\mathcal{R}}_0}

\newcommand{\cH}{\mathcal{H}}

\newcommand{\dt}{\delta t}

\newcommand{\ket}[1]{\left|#1\right\rangle}
\newcommand{\bra}[1]{\left\langle #1\right|}
\newcommand{\bket}[1]{\left\langle #1 \right\rangle}

\newcommand{\ketbra}[2]{\left|#1 \right\rangle\!\left\langle #2 \right|}
\newcommand{\dotex}{\frac{d}{dt}}

\newcommand{\Tr}[1]{\rm{Tr}\left(#1\right)}

\newcommand{\nbar}{|\alpha|^2}

\newcommand{\Cp}{\ket{\mathcal{C}_\alpha^+}}
\newcommand{\Cm}{\ket{\mathcal{C}_\alpha^-}}
\newcommand{\Cpm}{\ket{\mathcal{C}_\alpha^\pm}}
\newcommand{\Cmp}{\ket{\mathcal{C}_\alpha^\mp}}
\newcommand{\Cpd}{\bra{\mathcal{C}_\alpha^+}}
\newcommand{\Cmd}{\bra{\mathcal{C}_\alpha^-}}

\newcommand{\CpCp}{\ketbra{\mathcal{C}_\alpha^+}{\mathcal{C}_\alpha^+}}

\usetikzlibrary{positioning,calc,decorations.pathmorphing}

\definecolor{bleuZ}{HTML}{005377}
\definecolor{vertY}{HTML}{18a089}
\definecolor{rougeX}{HTML}{D1103A}
\definecolor{roseZZ}{HTML}{18a089}

\definecolor{bleu}{HTML}{0652ff}
\definecolor{vert}{RGB}{0,158,0}
\definecolor{rouge}{RGB}{253,0,0}
\definecolor{jaune}{HTML}{fedf08} % xkcd pissenlit
\definecolor{gris}{HTML}{929591} % xkcd grey

\definecolor{bleu1}{HTML}{0c3953}
\definecolor{bleu2}{HTML}{055a8c}
\definecolor{bleu3}{HTML}{0079bf}
\definecolor{bleu4}{HTML}{5Ba4cf}
\definecolor{bleu5}{HTML}{8bbdd9}

\definecolor{stab0}{HTML}{ffffff} % white
\definecolor{stab1}{HTML}{929591} % gris
\definecolor{stab-1}{HTML}{929591} % white
\definecolor{back}{HTML}{f1f1f1} % white
\usetikzlibrary{calc}
\definecolor{xlogical}{HTML}{ffffff} % white
\definecolor{zlogical}{HTML}{929591} % white

\tikzset{data/.style={fill=black,circle,minimum size=0.1pt, inner sep=1pt}}
\tikzset{ancilla/.style={draw,circle,minimum size=0.1pt, inner sep=1pt}}
\tikzset{link/.style={draw,decorate,decoration=snake}}

\graphicspath{ {./fig/} }

\begin{document}

\newcommand{\mytitle}{Adiabatic elimination for composite open quantum systems: \\ reduced model formulation and numerical simulations}

\title{\mytitle}

\author{Francois-Marie Le R\'egent}
\email{francois-marie.le-regent@alice-bob.com}
% \orcid{0000-0002-5229-7155}
\affiliation{Alice\&Bob, 53 boulevard du G\'{e}n\'{e}ral Martial Valin, 75015 Paris}
\affiliation{Laboratoire de Physique de l'Ecole normale sup\'{e}rieure, ENS-PSL, CNRS, Inria, Mines-Paris - PSL, Universit\'{e} PSL, Paris, France.  }

\author{Pierre Rouchon}
\email{pierre.rouchon@minesparis.psl.eu}
\affiliation{Laboratoire de Physique de l'Ecole normale sup\'{e}rieure, ENS-PSL, CNRS, Inria, Mines-Paris - PSL, Universit\'{e} PSL, Paris, France.  }

%%%%%%%%%%%%%%%%%%%%%%%%%%%%%%%%%%%%%%%%%%%%%%%%%%%%%%%%%%%%%%%%%%%%%%%%%%%%%%%

\begin{abstract}

A numerical  method is proposed for simulation of  composite open quantum systems. It  is  based on Lindblad  master equations and adiabatic elimination. Each subsystem is assumed to converge exponentially  towards a stationary subspace, slightly impacted by some decoherence channels and weakly coupled to the other subsystems. This numerical method is based on a perturbation analysis with an asymptotic expansion. It  exploits the formulation of the slow dynamics with reduced dimension. It relies on the invariant operators of the local and nominal dissipative dynamics attached to each subsystem. Second-order expansion can be computed only with local numerical calculations. It  avoids  computations on the tensor-product  Hilbert space  attached to the full  system. This numerical method is particularly well suited for  autonomous quantum error correction schemes. Simulations of such reduced models  agree with complete full model simulations for typical  gates acting on one and two  cat-qubits (Z, ZZ and  CNOT) when the mean photon number of each cat-qubit is less than 8. For larger mean photon numbers and gates with  three cat-qubits (ZZZ and CCNOT), full model  simulations  are almost impossible  whereas reduced model simulations remain accessible. In particular, they capture both the dominant phase-flip error-rate  and the  very small bit-flip error-rate   with  its  exponential suppression versus the mean photon number.

\end{abstract}

\maketitle
\tableofcontents
%%%%%%%%%%%%%%%%%%%%%%%%%%%%%%%%%%%%%%%%%%%%%%%%%%%%%%%%%%%%%%%%%%%%%%%%%%%%%%%%

\section{Introduction}

Quantum processors rely on controllable quantum systems~\cite{PhysRevLett.95.060501, BlaisRMP2021}, which are prone to 
 errors, mainly due to the environment, and therefore require quantum error correction with a very large number of physical resources to operate~\cite{Shor1995, PhysRevLett.81.2152, PhysRevA.63.042307, PhysRevA.73.012340, McEwen_2021, chen_exponential_2021, Krinner2022}.
To reduce errors hence resource overheads, bosonic encodings have emerged, taking advantage of the infinitely large Hilbert space of harmonic oscillators for intrinsic autonomous error correction~\cite{Joshi2021, CAI202150, GKP-PRA2001, hu_quantum_2019, gertler_protecting_2021, OfekPetrenkoHeeresEtAl2016}.

However, with such infinite systems, capturing the physics of gates and error processes becomes challenging. Classical numerical simulations require taking into account many  states of the Hilbert space to model their dynamics~\cite{Sivak2023, sellem2023gkp}.
In addition, simulations of composite systems with more than two modes are often intractable, as the dimension of the total Hilbert space is exponential in the number of modes, each mode description  requiring an Hilbert-space of large dimension~\cite{GuillaudMirrahimiPRX2019, AmazonPRXQ2022}.
The computational requirements even quickly surpass the capabilities of classical computers when considering only two bosonic qubits, and simulating gates involving three bosonic qubits with high precision becomes unfeasible.
Model reduction techniques have thus been developed and can use a more suitable basis of the Hilbert space to describe the physical systems via a subsystem decomposition~\cite{AmazonPRXQ2022, pantaleoni2023zak, schlegel2023coherentstate}.

Other methods, such as adiabatic elimination, are used to analyze the dynamics of open and dissipative quantum systems under a deterministic Lindblad master equation. Adiabatic elimination  corresponds  to a perturbation technique known in dynamical and control system theory as singular perturbations for slow/fast systems.
It is related to the Tikhonov approximation theorem (see, e.g., ~\cite{VerhulstBook2005,kokotovic-book-1}) and its coordinate-free formulation due to Fenichel~\cite{Fenichel79} with the
notion of invariant slow manifold of a dynamical system having  two time-scales
dynamics: the fast and exponentially converging ones and the slow ones of
reduced dimension.
Adiabatic elimination produces low
dimensional dynamical models via the derivation of the slow differential equation governing the
evolution on the invariant slow manifold~\cite{BrionJPA07,ZanarC2014PRL,AzouitCDC15,AzouitQST2017,BurgarthQ2019,TokiedaIFAC23}.

In this context, we propose here  an original  numerical method based on adiabatic elimination to simulate on a classical computer, quantum master equations modeling  composite systems  having  fast  and  local dissipation  with   weak coupling between the sub-systems and slow  decoherence. These calculations  are  simplified  by exploiting   the invariant operators  attached  the fast dynamics.
The resulting reduced model of the
slow evolution yields an
efficient numerical method for classical simulations of composite slow/fast
systems having a too large Hilbert space for brute-force numerical integration
of the original slow/fast master equations.
In particular, we show how to perform classical simulations involving three bosonic qubits with high precision.

Such low-dimensional reduced models are particularly well suited for numerical simulation of
autonomous quantum error correction schemes developed for  bosonic codes.  
In particular, for cat-qubit systems, around 50 to 100 photons per cat-qubit are required for simulating experimental setups, corresponding to a mean photon number of 10 to 15. Two-qubit quantum process tomography \cite{Smithey1993, Chuang1997a} is manageable via standard simulation methods for a small mean photon number but becomes infeasible when it exceeds 10. In the case of a three-qubit gate with a truncation of 100, standard simulations are impossible as they require storing density matrices of dimension $100^6$ and quantum process tomography would present even greater challenges. 
For such cat-qubit systems, several numerical simulations based on formal adiabatic calculations   and their numerical implementations are presented. They succeed in capturing both the macroscopic phase-flip errors associated with finite gate time   and  photon losses (the dominant error process for  harmonic oscillators),  and also the exponentially small bit-flip errors known to be much harder to estimate~\cite{GuillaudMirrahimiPRX2019, AmazonPRXQ2022}.
This method enables reduced computations with low-dimensional density operator for the global system state ($2^6$ for three-qubit gate).

In Section~\ref{sec:order2}, we recall for quantum master differential equations  the formalism of stationary states and invariant operators, and detail the formal adiabatic calculations up to the second-order of the continuous-time slow dynamics. These formal calculations are then exploited numerically to simulate the resulting    second-order slow model  for  a Z-gate on a  single cat-qubit. Comparison  with  numerical simulations of the full slow/fast model  are given.
In Section~\ref{sec:composite}, we then extend these second-order  calculations  to a composite system of locally  stabilized subsystems. We show how their  numerical  implementations can be done  with only local computations on the  Hilbert space  of each subsystem. This    avoids  computations on the  full Hilbert space of the complete  system.
For the composite system made of  two (resp. three) cat-qubits, numerical simulations  of  a ZZ  (resp. ZZZ) gate are presented with an emphasis on the different error rates.
In  Section~\ref{sec:hybrid}, we adapt  this simulation method to   composite systems for which one of the subsystem  is not  stabilized.
For  two (resp. three) cat-qubits, numerical simulations provide the error probabilities of a CNOT (resp. CCNOT)  gate   where the target qubit is not stabilized during the gate.
Sections in appendix are  mainly devoted to high-order  adiabatic calculations, additional simulation results,   discrete-time formulations with Kraus maps and the derived  time-discretization schemes underlying the  numerical simulations.

\section{Second-order expansion and Z-gate simulations} \label{sec:order2}

\subsection{Invariant manifold and slow dynamics approximation} \label{ssec:order2continuous}
The calculations of this sub-section  are very similar to section 2 and 3 of~\cite{FMRPR-CDC23}.

Consider the time-varying density operator $\rho_t$ on underlying Hilbert space
$\cH$ obeying to the following dynamics

\begin{equation}\label{eq:dynL0L1}
  \centering
 \dotex \rho_t = \cL_0(\rho_t) + \epsilon \cL_1(\rho_t)
\end{equation}
with two Gorini–Kossakowski–Sudarshan–Lindblad (GKSL) linear superoperators $\cL_0$ and $\cL_1$ where $\epsilon $ is a small positive parameter. For $\sigma=0,1$ one has
\begin{equation}
  \begin{split}
 & \cL_{\sigma}(\rho) = \\
 & - i [\bH_\sigma,\rho] + \sum_{\nu} \bL_{\sigma,\nu}\rho \bL_{\sigma,\nu}^\dag - \tfrac{1}{2} \Big( \bL_{\sigma,\nu}^\dag \bL_{\sigma,\nu}\rho+
 \rho \bL_{\sigma,\nu}^\dag \bL_{\sigma,\nu} \Big)
  \end{split}
\end{equation}
with $\bH_\sigma$ Hermitian operator and $\bL_{\sigma,\nu}$ any operator not necessarily Hermitian.

Assume that for $\epsilon=0$ and any initial condition $\rho_0$, the solution of~\eqref{eq:dynL0L1}
 converges exponentially towards a steady-state depending a priori on $\rho_0$. This means that we have a quantum channel $\ocK$ defined by
\begin{equation}\label{eq:Kchanel}
 \lim_{t\mapsto +\infty} e^{t\cL_0}(\rho_0) \triangleq \ocK(\rho_0)
.
\end{equation}
The range of $\ocK$ is denoted by $\cD_{0}$, the set of steady-states corresponding to the kernel of $\cL_0$, a vector subspace of Hermitian operators. Denote by $\bar d$ the dimension of $\cD_{0}$ and consider an orthonormal basis of $\cD_{0}$ made of $\bar d$ Hermitian operators $\bS_1$, \ldots, $\bS_{\bar d}$ such that $\Tr{\bS_d\bS_{d'}}=\delta_{d,d'}$.
To each $\bS_d$ is associated an invariant operator $$\bJ_d=\lim_{t\mapsto +\infty} e^{t \cL_0^*}(\bS_d)$$ being a steady-state of the adjoint dynamics (according to the Frobenius Hermitian product) $\dotex \bJ = \cL_0^{*}(\bJ)$ where $\cL_0^*$ is the adjoint of $\cL_0$ (see, e.g.,~\cite{AlberJ2014PRA}). For any solution $\rho_t$ of~\eqref{eq:dynL0L1} with $\epsilon=0$, $\Tr{\bJ_d\rho_t}$ is constant. This gives the following expression for $\ocK$:
\begin{equation}\label{eq:cK0}
 \lim_{t\mapsto +\infty}\rho_t = \sum_{d=1}^{\bar d} \Tr{\bJ_d \rho_0} \bS_d \triangleq\ocK(\rho_0).
\end{equation}
Moreover, $\Tr{\bJ_d \bS_{d'}} = \delta_{d,d'}$ since for any $t >0$
\begin{equation} \begin{split}
 & \Tr{e^{t \cL_0^*}(\bS_d)~ \bS_{d'}}=\Tr{\bS_d~ e^{t \cL_0}(\bS_{d'}) } \\
 & =\Tr{\bS_d \bS_{d'} } =\delta_{d,d'}
\end{split}\end{equation}
using the fact that $e^{t \cL_0}(\bS_{d'})= \bS_{d'}$.

For $\epsilon >0$ and small, Eq.~\eqref{eq:dynL0L1} also admits a $\bar d$
dimensional linear subspace denoted by $\cD_{\epsilon}$ invariant and close to
$\cD_{0}$ (see~\cite{kato-book-66} for a mathematical justification in finite dimension). Thus,
the set of $\bar d$ real variables
$$x_1=\Tr{\bJ_1 \rho}, \ldots, x_{\bar
  d}=\Tr{\bJ_{\bar d}\rho}
  $$
 can be chosen to be local coordinates on $\cD_{\epsilon}$:
any density operators $\rho\in \cD_{\epsilon}$ reads
$\rho= \sum_{d=1}^{\bar d} x_d \bS_{d}(\epsilon)$ with  the perturbed basis $\bS_{1}(\epsilon)$, \ldots
$\bS_{\bar d}(\epsilon)$ and $\bar d$ real numbers $x_d$.

 Invariance of $\cD_{\epsilon}$ with respect to~\eqref{eq:dynL0L1} means that, if at some time $t$, the solution $\rho_t$~of the
perturbed system~\eqref{eq:dynL0L1} belongs to $\cD_{\epsilon}$, it remains on
$\cD_{\epsilon}$ at any time: $\dotex \rho_t =(\cL_0+\epsilon\cL_1)(\rho_t) $
with $\rho_t= \sum_{d=1}^{\bar d} x_d(t) \bS_{d}(\epsilon)$. For any
$(x_1(t), \ldots, x_{\bar d}(t)) \in\mathbb{R}^{\bar d}$, this invariance
property reads
\begin{equation}\label{eq:InvCond}
 \sum_{d=1}^{\bar d} \frac{dx_d}{dt} ~ \bS_{d}(\epsilon)
 = \left(\cL_0+ \epsilon \cL_1\right)\left(\sum_{d=1}^{\bar d} x_d \bS_{d}(\epsilon)
 \right)
.
\end{equation}
Thus, for any $d\in\{1,\ldots,\bar d\}$, $\frac{dx_d}{dt}$ depends linearly on $x=(x_1,\ldots,x_{\bar d})$, i.e.
\begin{equation}\label{eq:dynXF}
 \dotex x_d = \sum_{d'}F_{d,d'}(\epsilon) x_{d'}.
\end{equation}
The invariance condition reads now,
\begin{equation}
  \begin{split}
\forall (x_1,\ldots,x_{\bar d})\in\mathbb{R}^{\bar d}, ~
\sum_{d,d'} x_{d'} F_{d,d'}(\epsilon) \bS_{d}(\epsilon) \\
 \equiv \sum_d x_d (\cL_0+\epsilon\cL_1)(\bS_d(\epsilon))
\end{split}
\end{equation}
which is equivalent to
\begin{equation} \begin{split}
  \forall d \in\{1,\ldots,\bar d\}, \quad
 \sum_{d'=1}^{\bar d} F_{d',d}(\epsilon) \bS_{d'}(\epsilon) = (\cL_0+\epsilon\cL_1)(\bS_d(\epsilon))
.
\end{split}\end{equation}
With the asymptotic expansion
\begin{equation} \begin{split}
  F_{d,d'}(\epsilon) =\sum_{n\geq 0} \epsilon^n F_{d,d'}^{(n)}, \quad \bS_{d}(\epsilon) = \sum_{n\geq 0} \epsilon^n \bS_{d}^{(n)}
\end{split}\end{equation}
one can compute recursively $F_{d,d'}^{(n)}$ and $\bS_{d}^{(n)}$ from $F_{d,d'}^{(m)}$ and $\bS_{d}^{(m)}$ with $m < n$. The recurrence relationship is based on the identification of terms with same orders versus $\epsilon$ in the following equations
\begin{equation} \begin{split}
  & \forall d \in\{1,\ldots,\bar d\},\quad \sum_{d'=1}^{\bar d} \left(\sum_{n\geq 0} \epsilon^n F_{d',d}^{(n)}\right) \left(\sum_{n'\geq 0}\epsilon^{n'} \bS_{d'}^{(n')} \right) \\
 & = \left(\cL_0 +\epsilon\cL_1\right)
 \left(\sum_{n\geq 0} \epsilon^n\bS_{d}^{(n)} \right).
\end{split}\end{equation}

The zero-order condition is satisfied with $F^{(0)}_{d,d'}=0$ and $\bS_{d}^{(0)} = \bS_d$.
First-order condition reads
\begin{equation} \begin{split}
  \forall d \in\{1,\ldots,\bar d\},\quad \sum_{d''=1}^{\bar d} F_{d'',d}^{(1)} \bS_{d''}^{(0)} = \cL_0(\bS_{d}^{(1)} ) + \cL_1( \bS_{d}^{(0)})
.
\end{split}\end{equation}
Left multiplication by operator $\bJ_{d'}$ and taking the trace yields
\begin{equation}\label{eq:F1}
 F_{d',d}^{(1)} = \Tr{\bJ_{d'} \cL_1( \bS_{d}{^{(0)}})}
\end{equation}
since $\Tr{\bJ_{d'} \bS_{d''}^{(0)}}= \delta_{d',d''}$ and $\Tr{\bJ_{d'}\cL_0(\bW)}=0$ for any operator $\bW$ because $\cL_0^*(\bJ_{d'})=0$.
Thus, $\bS_{d}^{(1)}$ is a solution $\bX$ of the following equation:
\begin{equation} \begin{split}
  \cL_0(\bX) & = \sum_{d'} \Tr{\bJ_{d'} \cL_1(\bS^{(0)}_d )} \bS_{d'} - \cL_1(\bS_d^{(0)}) \\
  & = \ocK\big(\cL_1(\bS_d^{(0)}) \big) - \cL_1(\bS_d^{(0)})
\end{split}\end{equation}
where the quantum channel $\ocK$ is defined in~\eqref{eq:Kchanel}. Following~\cite{AzouitQST2017}, the general solution $\bX$ is given by the absolutely converging integral,
\begin{equation} \begin{split}
  \bX = \int_{0}^{+\infty} e^{s\cL_0}\left(\cL_1(\bS_d^{(0)}) - \ocK\big(\cL_1(\bS_d^{(0)}) \big) \right) ~ds +\bW
\end{split}\end{equation}
where $\bW$ belongs to $\cD_0$ the kernel of $\cL_0$. We consider the solution with $\bW=0$ and thus
\begin{equation}
 \bS_{d}^{(1)}= \int_{0}^{+\infty} e^{s\cL_0}\left(\cL_1(\bS_d^{(0)}) - \ocK\big(\cL_1(\bS_d^{(0)}) \big) \right) ~ds
 \label{eq:S1}
\end{equation}
where for all $d'$, $\Tr{\bJ_{d'} \bS_{d}^{(1)}}=0$.
The superoperator $\ocR$ defined for any operator $\bW$ by
\begin{equation}\label{eq:R}
 \ocR(\bW) = \int_{0}^{+\infty} e^{s\cL_0}\left(\bW - \ocK\big(\bW \big) \right) ~ds
\end{equation}
provides thus the unique solution $\bX=\ocR(\bW)$ of
$\cL_0(\bX)= \ocK(\bW) - \bW $
such that for all $d$, $\Tr{\bJ_d \bX}=0$. To summarize, the first-order terms in $\epsilon$ are
\begin{equation}\label{eq:F1S1}
 F_{d',d}^{(1)} = \Tr{\bJ_{d'} \cL_1( \bS_{d}{^{(0)}})} \text{ and } \bS_{d}^{(1)}= \ocR\big(\cL_1(\bS_d) \big)
.
\end{equation}

Second-order conditions are
\begin{equation} \begin{split}
  &\forall d \in\{1,\ldots,\bar d\}, \\
  & \sum_{d''=1}^{\bar d} F_{d'',d}^{(1)} \bS_{d''}^{(1)}+ F_{d'',d}^{(2)} \bS_{d''}^{(0)} = \cL_0(\bS_{d}^{(2)} ) + \cL_1( \bS_{d}^{(1)}).
\end{split}\end{equation}
Left multiplication by operator $\bJ_{d'}$ and taking the trace yields:
\begin{equation}\label{eq:F2}
 F_{d',d}^{(2)} = \Tr{\bJ_{d'} \cL_1( \bS_{d}^{(1)})}= \Tr{\cL_1^{*}(\bJ_{d'}) ~\bS_{d}^{(1)}}.
\end{equation}
Computations similar to the ones performed for the first-order conditions yield
\begin{equation}
\bS_{d}^{(2)}= \ocR\left(\cL_1(\bS_d^{(1)})-\sum_{d''=1}^{\bar d} F_{d'',d}^{(1)} \bS_{d''}^{(1)} \right).
\label{eq:S2}
\end{equation}

Higher order formulae are given in appendix~\ref{ssec:order_n}.
The equivalent of Eqs.~\eqref{eq:F1S1} and~\eqref{eq:F2} for a slow time dependency are detailed in appendix \ref{sec:TimeVarying} and for discrete-time setting in appendix \ref{sec:discrete_single}.

\subsection{Z-gate simulations for a single cat-qubit}\label{ssec:Zgate}

For a cat-qubit system~\cite{MirrahimiCatComp2014,LeghtTPKVPSNSHRFSMD2015S,GuillaudMirrahimiPRX2019}, the quantum state $\rho$ is attached to a harmonic oscillator. It is confined through an engineered
two-photon driven dissipation process to have its range close to a two-dimensional subspace spanned by two coherent wave functions $\ket{\pm\alpha}$ of opposite complex  amplitudes $\pm \alpha$. This means that the support of $\rho$ remains close to the sub-Hilbert space of dimension $2$ spanned by the orthonormal wave functions (the Schr\"{o}dinger cat states)
\begin{equation} \ket{\mathcal{C}_\alpha^\pm} := \mathcal{N}_\pm (\ket{\alpha}
 \pm \ket{-\alpha}), \label{eq:encoding} \end{equation}
where $\mathcal{N}_\pm = (2(1\pm \exp(-2 |\alpha|^2)))^{-1/2} $ are normalizing constants. The computational wave-function  are given by the following equations:
\begin{align}
 \ket{0}_C & = (\ket{\mathcal{C}_\alpha^+} +
 \ket{\mathcal{C}_\alpha^-})/\sqrt{2} = \ket{\alpha} +
 \mathcal{O}(e^{-2|\alpha|^2})        \\
 \ket{1}_C & = (\ket{\mathcal{C}_\alpha^+} -
 \ket{\mathcal{C}_\alpha^-})/\sqrt{2} = \ket{-\alpha} +
 \mathcal{O}(e^{-2|\alpha|^2}).
\end{align}
The engineered two-photon driven dissipation process can be effectively modeled by as single Lindblad term of the form
\begin{equation}\label{eq:catL0}
 \cL_{0}(\rho)= \cD_{\bL_0}(\rho) \triangleq \left( \bL_0 \rho \bL_0^\dag - \tfrac{1}{2} (\bL_0^\dag\bL_0\rho+ \rho \bL_0^\dag\bL_0 ) \right)
\end{equation}
with $\bL_0=\sqrt{\kappa_2}(\ba^2-\alpha^2)$, $\kappa_2 >0$ and $\ba$ being the photon annihilator operator.
Such a process can be engineered in a superconducting platform~\cite{MirrahimiCatComp2014}.
It stabilizes exponentially the cat-qubit subspace corresponding then to $\cD_0$ (called the code subspace in the context of bosonic codes)~\cite{AzouitSarletteRouchon2016}.
Its real dimension is $\bar d=4$ with the following orthonormal operator basis
\begin{equation}
 \begin{aligned}
  \bS_1=(\Cp \Cpd+\Cm \Cmd )/ \sqrt{2}, \\
  \bS_2=(\Cp\Cpd-\Cm\Cmd) / \sqrt{2},  \\
  \bS_3=(i\Cp\Cmd-i\Cm\Cpd) / \sqrt{2}, \\
  \bS_4=(\Cp\Cmd+\Cm\Cpd ) /\sqrt{2}.
 \end{aligned}
\end{equation}

Among the errors and decoherence processes, the dominant one is the undesired single-photon loss, modelled by
\begin{equation} \begin{split}
  \cD_{\sqrt{\kappa_1}\ba} (\rho) \triangleq \kappa_1 \left( \ba \rho \ba^\dag - \tfrac{1}{2} (\ba^\dag\ba\rho+ \rho \ba^\dag\ba ) \right)
\end{split}\end{equation}
where $\kappa_1 >0$.
Usually the ratio $\kappa_1/\kappa_2$ is small:  $\kappa_1$ is the  single-photon loss rate, much smaller than $\kappa_2$ the rate of mechanism stabilizing the code-space $\cD_0$.

A Z-gate  corresponds to a unitary  transformation exchanging $\Cp$ and $\Cm$. Following~\cite{MirrahimiCatComp2014, Touzard-et-al-PRX2018}, it can be approximately engineered via the propagator of time duration $T>0$ associated to  the  Hamiltonian $\bH_1=\epsilon_Z\left(\ba+\ba^{\dagger}\right) $  where $\epsilon_Z=\frac{\pi}{4\alpha T}$ has to be much smaller than $\kappa_2$.
The superoperators $\cL_0$ and $\cL_1$ corresponding here to Eq.~\eqref{eq:dynL0L1} are thus
\begin{equation}\begin{split}
  \label{eq:L0L1Zgate}
 & \cL_0(\rho)=\kappa_2 D_{\ba^2-\alpha^2}(\rho), \\
 & \epsilon \cL_1(\rho)= \kappa_1 D_{\ba}(\rho)-i\tfrac{\pi}{4\alpha T}\left[\ba+\ba^{\dagger}, \rho\right]
\end{split}
\end{equation}
where $\kappa_1/\kappa_2$ and $T \kappa_2$ are much smaller than $1$, ensuring the scaling based on the small parameter $\epsilon$.
Moreover, replacing $\cL_1$ in formulae~\eqref{eq:F1S1} and~\eqref{eq:F2} by the superoperator $\kappa_1 D_{\ba}(\bullet)-i\tfrac{\pi}{4\alpha T}\left[\ba+\ba^{\dagger}, \bullet\right]$ corresponding to $\epsilon \cL_1$, provides directly $\epsilon F^{(1)}_{d',d}$, $\epsilon \bS_d^{(1)}$ and $\epsilon^2 F^{(2)}_{d',d}$ without defining precisely $\epsilon$.

\begin{figure}[!h]
 \centering
 \includegraphics[width=0.4\textwidth]{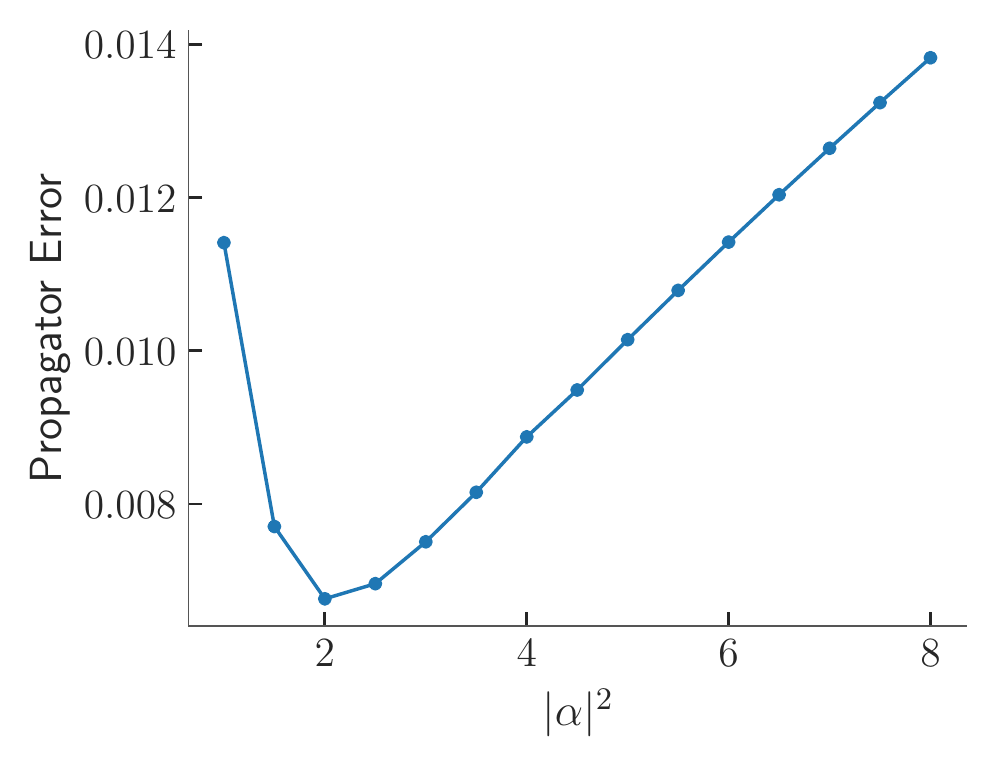}
 \caption{
  Propagator error between the full-model~\eqref{eq:L0L1Zgate} and reduced model~\eqref{eq:RedZgate} for the Z gate with the mean photon number $\alpha^2$ between 1 and 16.
  The error is computed as $\sqrt{\Tr{(G_{\text{\tiny red}} G_{\text{\tiny full}}^{-1} - I_4)(G_{\text{\tiny red}} G_{\text{\tiny full}}^{-1} - I_4)^\dag}}$.
 }
 \label{fig:comp_propagator_Z}
\end{figure}

\begin{figure}[!h]
 \centering
 \begin{subcaptiongroup}
  \subcaptionlistentry{}
  \label{fig:idealZchi}
  \begin{overpic}[
    width=0.49\textwidth,
    % grid
    % ]{fig/comp_CNOT_bit-flip_nbsteps20000.pdf}
   ]{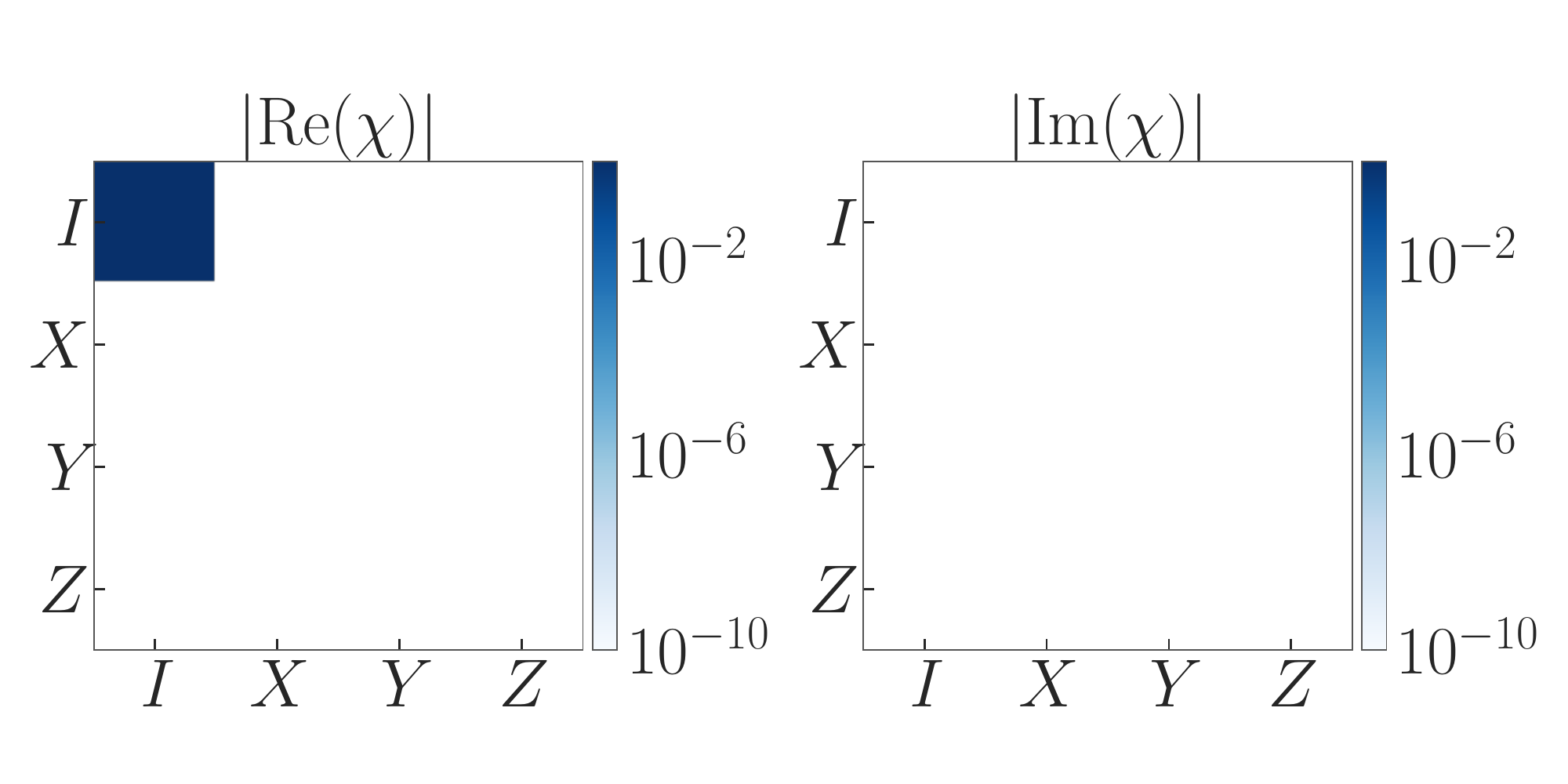}
   % ]{example-image-a}
   \put(0, 40){\captiontext*{}}
  \end{overpic}
  \subcaptionlistentry{}
  \label{fig:ScrodZchi}
  \begin{overpic}[
    width=0.49\textwidth,
    % grid
    % ]{fig/comp_CNOT_bit-flip_nbsteps20000.pdf}
   ]{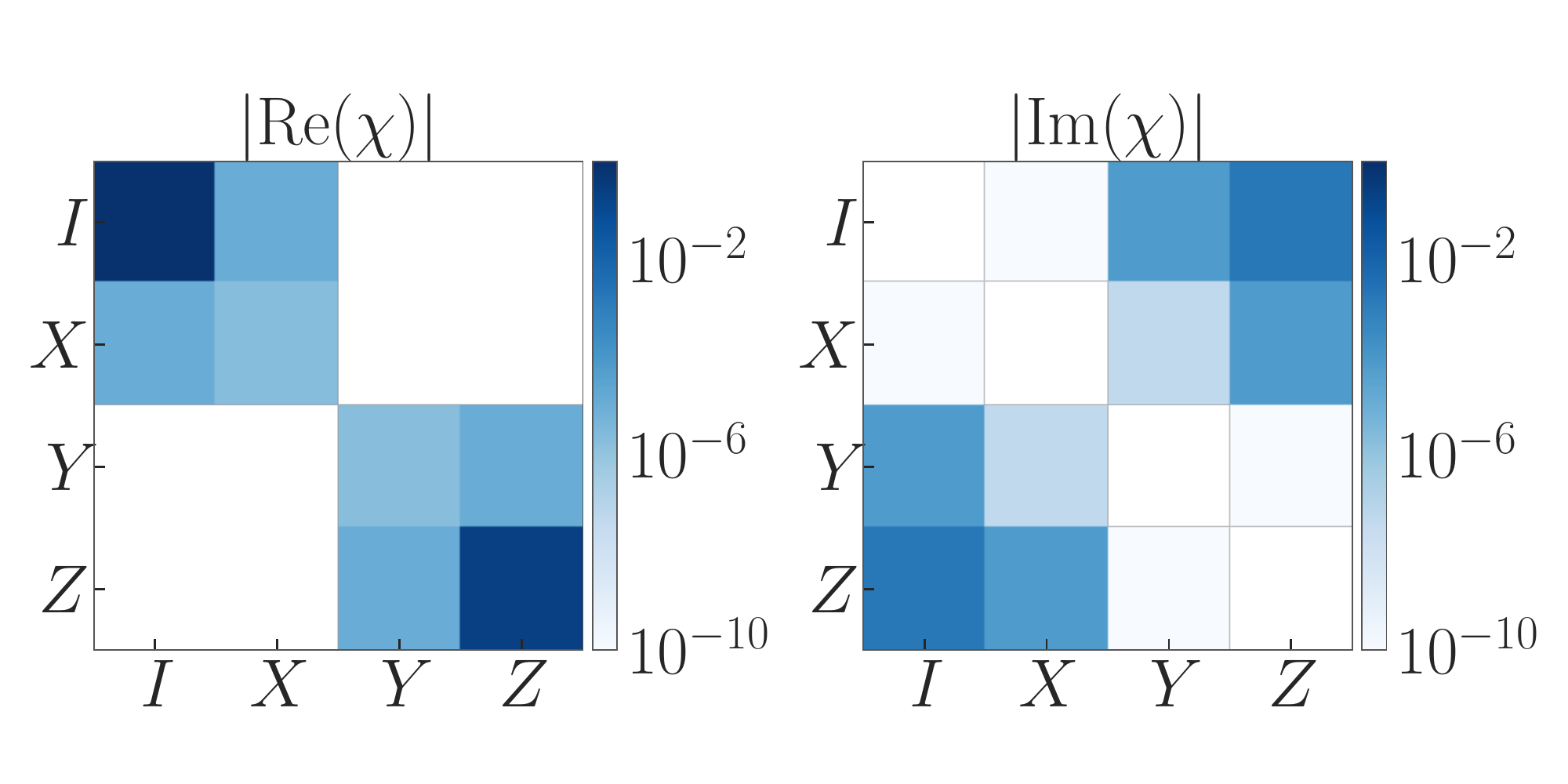}
   % ]{example-image-a}
   \put(0, 40){\captiontext*{}}
  \end{overpic}  \subcaptionlistentry{}
  \label{fig:HeisZchi}
  \begin{overpic}[
    width=0.49\textwidth,
    % grid
    % ]{fig/comp_CNOT_bit-flip_nbsteps20000.pdf}
   ]{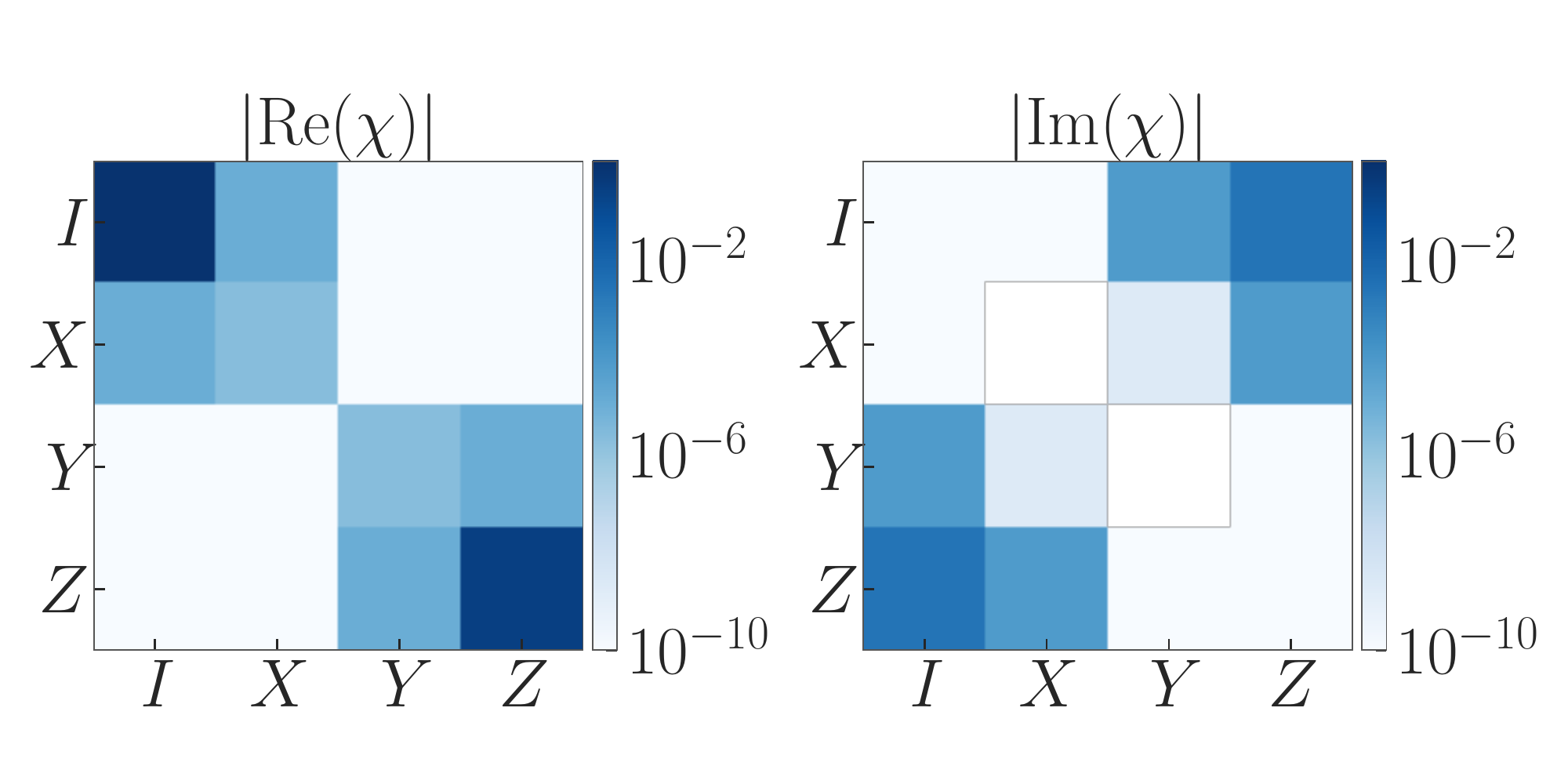}
   % ]{example-image-a}
   \put(0, 40){\captiontext*{}}
  \end{overpic}
 \end{subcaptiongroup}
 \captionsetup{subrefformat=parens}
 \caption{
 Real (left) and imaginary (right) part of the quantum-error matrix $\chi^E$. \subref{fig:idealZchi} corresponds to no-error with $E=I_4$; \subref{fig:ScrodZchi} corresponds to full model simulations~\eqref{eq:L0L1Zgate} with $\alpha^2=4$ and  $E=E_{\text{\tiny full}}$; \subref{fig:HeisZchi} corresponds to reduced model simulations~\eqref{eq:RedZgate} with $\alpha^2=4$ and  $E=E_{\text{\tiny red}}$.}
 \label{fig:Zchi}
\end{figure}

\begin{figure}
 \centering
 \begin{subcaptiongroup}
  \subcaptionlistentry{}
  \label{fig:Zx}
  \begin{overpic}[
    width=0.35\textwidth,
    % width=100pt,
    % grid
   ]{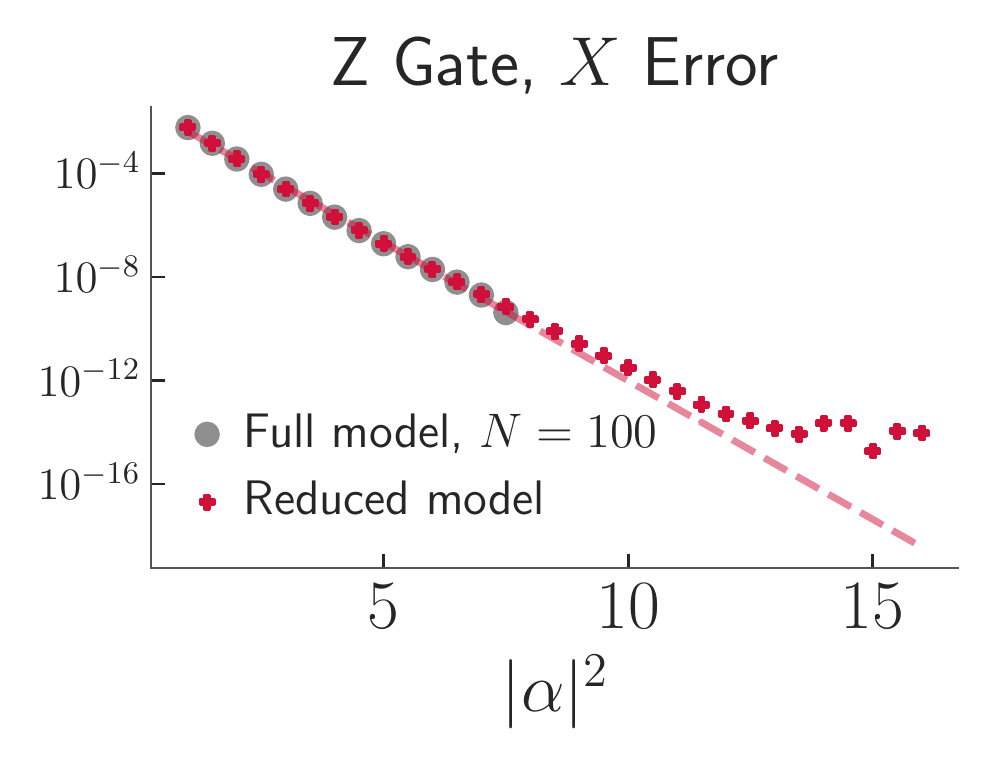}
   % ]{example-image-a}
   \put(5,70){\captiontext*{}}
  \end{overpic}
  \subcaptionlistentry{}
  \label{fig:Zz}
  \begin{overpic}[
    width=0.35\textwidth,
    % grid
   ]{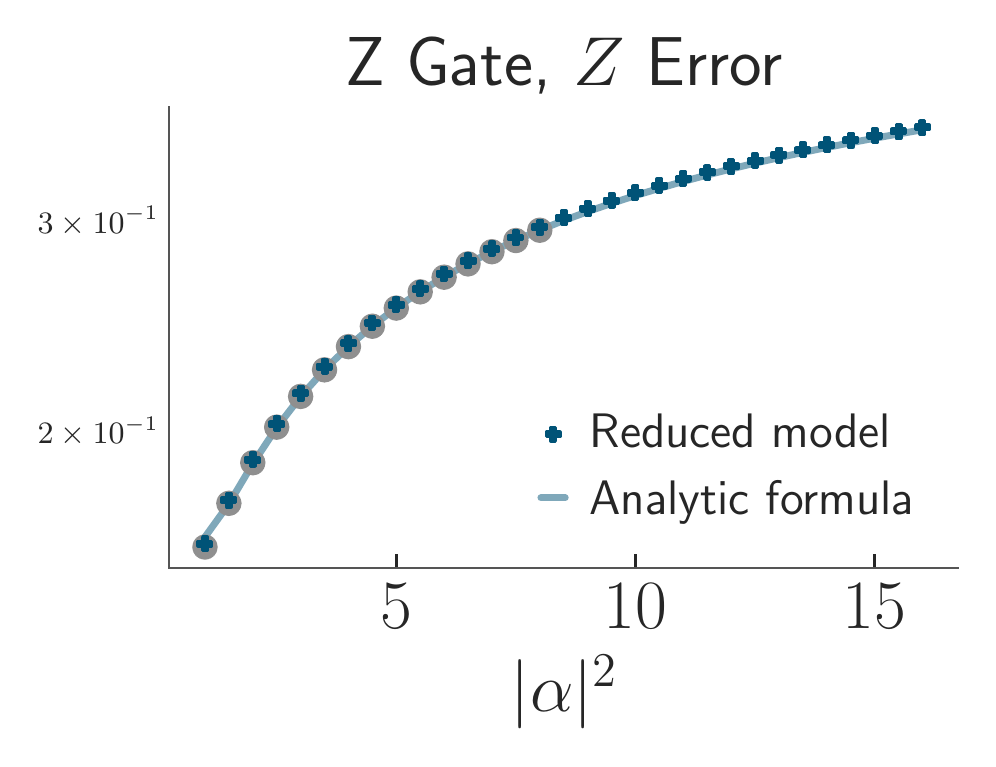}
   % ]{example-image-a}
   \put(5, 70){\captiontext*{}}
  \end{overpic}
 \end{subcaptiongroup}
 \captionsetup{subrefformat=parens}
 \caption{Comparison between \subref{fig:Zx} $X$ and \subref{fig:Zz} $Z$ error probabilities of a Z gate obtained via full model simulations for $|\alpha|^2\leq 8$~\eqref{eq:L0L1Zgate} (shown as gray circles) and the reduced model simulations~\eqref{eq:RedZgate} (colored plus) for different mean photon numbers $\alpha^2$. A simple fit yields an exponential suppression of bit-flips with an exponential coefficient of $2.46 \pm 0.03$ (dashed line).}
 \label{fig:Zerror}
\end{figure}

Numerical simulations of figures~\ref{fig:comp_propagator_Z},~\ref{fig:Zchi} and~\ref{fig:Zerror} are based on a Galerkin approximation of the Hilbert space relying on the photon-number state $\ket{n}$ with $n$ between $0$ to $N$. The integer $N$ is   chosen large enough to ensure that $|\bket{\alpha|N}|^2= e^{-|\alpha|^2}|\alpha|^{2 N}/ N!$ remains  negligible. The time discretization of the resulting  finite-dimensional system of ordinary differential equations is based on the  numerical scheme described in appendix~\ref{sec:kraus_map}. It provides a discrete-time setting
$\rho(t+\delta t) = \cK_0(\rho(t)) + \epsilon \cK_1(\rho(t)) $ where $\cK_0$ is an exact quantum channel close to identity with
\begin{equation} \begin{split}
  & \kappa_2 \dt= \frac{1}{1000}, \quad
 \kappa_1=\frac{\kappa_2}{100}, \\
 & \epsilon_Z= \frac{\pi}{4\alpha T}=\frac{\kappa_2}{20}, \quad  1\leq \alpha^2 \leq 16  \text{ and } N=100.
\end{split}\end{equation}
The operators $\bS_d^{(0)}=\bS_d$ with $d=1,\ldots,4=\bar d$ are obtained from truncated approximations of coherent states
$\ket{\pm \alpha} \approx e^{-\alpha^2/2} \sum_{n=0}^{N} \frac{(\pm\alpha)^n}{\sqrt{n!}} \ket{n}$.
The associated invariant operators $\bJ_d$ are obtained numerically via the discrete-time formulation given in appendix~\ref{sec:discrete}.
Similarly, the entries of $\epsilon F^{(1)}_{d',d}$ and $\epsilon^2F^{(2)}_{d',d}$ are given by discrete-time formulae~\eqref{eq:F12discrete} divided by $\dt$ and  where $\cK_1$ stands for $\epsilon \dt \cL_1$. These matrices provide, up
to third-order terms, the generator of the  continuous-time reduced dynamics:
\begin{equation}\label{eq:RedZgate}
 \dotex x = (\epsilon F^{(1)}+\epsilon^2 F^{(2)}) x = F(\epsilon) x +
 O(\epsilon^3)
\end{equation}
where $x_d=\Tr{\bJ_d \rho}$ for $d=1,\ldots,4$.

On figure~\ref{fig:comp_propagator_Z}, the reduced model propagator $G_{\text{\tiny red}}=e^{T (\epsilon F^{(1)}+\epsilon^2 F^{(2)})}$, a $4\times 4$ real matrix, is then compared to the full model propagator
$G_{\text{\tiny full}}$, another $4\times 4$ real matrix with entries given by $\Tr{\bJ_{d'} \bW_d(T)}$ where $\bW_d(t)$ is the numerical solution of the full model~\eqref{eq:dynL0L1} truncated to $N=100$ photons and starting from initial condition $\bW_d(0)=\bS_d$. We observe an error $\sqrt{\Tr{(G_{\text{\tiny red}} G_{\text{\tiny full}}^{-1} - I_4)(G_{\text{\tiny red}} G_{\text{\tiny full}}^{-1} - I_4)^\dag}}$ of less than $0.014$ for mean-photon number $\alpha^2$ between $1$ and $16$ ($I_4$ is the $4\times 4$ identity matrix).

Both $G_{\text{\tiny red }}$ and $G_{\text{\tiny full}}$ are close to the ideal Z-gate matrix
$$ G_{\text{\tiny ideal }} = \left(
 \begin{array}{cccc}
   1 & 0 & 0 & 0 \\
   0 & -1 & 0 & 0 \\
   0 & 0 & -1 & 0 \\
   0 & 0 & 0 & 1 \\
  \end{array}
 \right)
.
$$
Thus, the reduced model error propagator $E_{\text{\tiny red }}=G_{\text{\tiny ideal }}^{-1}G_{\text{\tiny red }}$ and the full model error propagator $E_{\text{\tiny full }}=G_{\text{\tiny ideal }}^{-1} G_{\text{\tiny full }}$ are close to identity matrix $I_4$: they correspond in fact to quantum channels usually close to identity and characterizing the errors. These channels can be decomposed according to the basis $(\bS_1, \ldots,\bS_4)$.
This means that for $E=E_{\text{\tiny red }}, E_{\text{\tiny full}}$, the identity
 \begin{equation}
 \begin{split}
 & \forall x\in\mathbb{R}^4, \\
 & \sum_{d,d'=1}^{4} E_{d,d'} x_{d'} \bS_d = \sum_{m,n=1}^4 \chi^E_{m,n} \bS_m \left(\sum_{d=1}^4 x_d \bS_d\right) \bS_n
 \label{eq:chi1}
 \end{split}
 \end{equation}
uniquely defines the  $\chi^E$ matrix, a  $4\times 4$-matrix,  characterizing the errors and close to $\chi^{I_4}$ having a single non-zero entry $\chi^{I_4}_{1,1}=1$. This is illustrated on figure~\ref{fig:Zchi}.

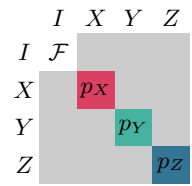
\begin{figure}
  \centering
  \input{fig/chi_matrix_1_qubit.tikz}
 \caption{
  One-qubit $\chi$ matrix representing the noise channel of an imperfect gate reduced to a two-level system.
  The off-diagonal elements shown in gray are ignored since they do not cause symmetric Pauli errors.
 }
  \label{fig:chi_1}
 \end{figure}

Since $ \bX=\sqrt{2}\bS_2 $, $ \bY=\sqrt{2}\bS_3$ and $ \bZ=\sqrt{2}\bS_4$ correspond to three Pauli operators on the code-space, $\chi^E_{2,2}$ (resp. $\chi^E_{3,3}$, $\chi^E_{4,4}$) gives roughly-speaking the X-error (resp. Y-error, Z-error) probability, see figure~\ref{fig:chi_1}.
These error probabilities have to be less than some thresholds in order to be cancelled by high-level error correction code. For cat-qubit, Z-error probability is usually much larger than the two other ones, X-error and Y-error probabilities, called bit-flip errors. Simulations of figure~\ref{fig:Zx}
show that the reduced model  captures the very small error probabilities associated to bit-flip errors known to be exponentially
suppressed for large $|\alpha|^2$ as shown experimentally in~\cite{LescanneZaki2019}.
We found an exponential suppression of bit-flips proportional to $ \exp^{-a |\alpha|^2}$ with $a=2.46 \pm 0.03$.
The reduced model also captures the phase-flips (Z-error) in figure~\ref{fig:Zz}. It matches well with full model simulations and also with an analytical formula obtained via a perturbation expansion given in~\cite{AmazonPRXQ2022}: $p_Z=|\alpha|^2 \kappa_1 T + \frac{\epsilon_Z^2
  T}{|\alpha|^2 \kappa_2}$.

Z-gate simulations up to order 5 are discussed in appendix~\ref{ssec:Zgate_order5}, showing the convergence of the $X$, $Y$ and $Z$ error probabilities by increasing the order of the pertubative analysis in figure \ref{fig:order_expansion}.
The equation~\eqref{eq:S1} allows performing leakage computation, defined as the population outside the code space, see appendix~\ref{sec:leakage_one_mode} and figure~\ref{fig:Zleakage}.

\section{Composite systems and ZZ/ZZZ-gate simulations} \label{sec:composite}

\subsection{Second-order approximation with only local computations}

Take a bipartite system made of sub-systems $A$ and $B$ with Hilbert spaces
$\cH_A$ and $\cH_B$. The system Hilbert space is $\cH= \cH_A\otimes\cH_B$.
Assume that the unperturbed  dynamics $ \cL_0$ in~\eqref{eq:dynL0L1} admit the following structure:
\begin{equation}\label{eq:LA0LB0}
 \cL_0=\cL_{A,0} + \cL_{B,0}
\end{equation}
where $\cD_{A,0}$ and $\cD_{B,0}$ are the
steady-state subspaces of operators on $\cH_A$ and $\cH_B$ of local Lindblad
superoperators $\cL_{A,0}$ and $\cL_{B,0}$. These local nominal dynamics
stabilize the subspaces of dimensions $\bar d_A$ and $\bar d_B$, having
$(\bS_{A,d_A})_{1\leq d_A\leq \bar d_A}$ and $(\bS_{B,d_B})_{1\leq d_B\leq \bar
  d_B}$ as orthonormal basis of Hermitian operators. Thus, all Hermitian operators in $\cD_0$, the kernel of $\cL_A+\cL_B$, read
  \begin{equation}\label{eq:rhoABparam}
   \sum_{d_A,d_B} x_{d_A,d_B} \bS_{A,d_A}\otimes \bS_{B,d_B}
  \end{equation}
  where $x_{d_A,d_B}$ are arbitrary real numbers.

  We assume that $\cL_{A,0}$
and $\cL_{B,0}$ ensure exponential convergence towards $\cD_{A,0}$ and
$\cD_{B,0}$: for any operators $\rho$ on $\cH$,
\begin{equation}
\begin{split}
  \label{eq:cKAB0}
 & \lim_{t\mapsto +\infty} e^{t(\cL_{A,0}+\cL_{B,0})} (\rho) = \ocK(\rho) \\
 & = \sum_{d_A,d_B} \Tr{\bJ_{A,d_A}\otimes \bJ_{B,d_B}~\rho } \bS_{A,d_A}\otimes \bS_{B,d_B}
\end{split}
\end{equation}
where  $\bJ_{A,d_A}$ and $\bJ_{B,d_B}$ are  local invariant operators
\begin{equation}
\begin{split}
 & \bJ_{A,d_A} = \lim_{t\mapsto +\infty} e^{t \cL_{A,0}^*}(\bS_{A,d_A}), \\
 & \bJ_{B,d_B} = \lim_{t\mapsto +\infty} e^{t \cL_{B,0}^*}(\bS_{B,d_B}).
\end{split}
\end{equation}

Assume that $\bH_1$ and $\cL_{1,\nu}$ defining the super operator
$\cL_1$ in~\eqref{eq:dynL0L1} only involve finite sums of tensor products of
operators on $\cH_A$ and $\cH_B$. This means that for any $\bX_A$ and $\bX_B$
local operators on $\cH_A$ and $\cH_B$,
\begin{equation}\label{eq:L1AB}
 \cL_{1} (\bX_A\otimes \bX_B) = \sum_{\nu=1}^{\bar \nu} \bL_{A,\nu}\bX_A \bR_{A,\nu} \otimes \bL_{B,\nu}\bX_B \bR_{B,\nu}
\end{equation}
where $\bar \nu$ is a positive integer, where $\bL_{A,\nu}$, $\bR_{A,\nu}$ are operators on $\cH_A$ and where $\bL_{B,\nu}$, $\bR_{B,\nu}$ are operators on $\cH_B $.

The operator $\bJ_{d'}$ appearing in~\eqref{eq:F1} corresponds here to
$\bJ_{A,d'_A} \otimes \bJ_{B,d'_B}$ with $d'=(d'_A,d'_B)$. Similarly, $\bS_{d}$
reads here $\bS_{A,d_A} \otimes \bS_{B,d_B}$ with $d=(d_A,d_B)$.
With~\eqref{eq:L1AB} one obtains

\begin{multline}
  \label{eq:F1AB}
F^{(1)}_{(d'_A,d'_B),(d_A,d_B)} =  \sum_{\nu=1}^{\bar\nu} \Tr{\bJ_{A,d'_A} \bL_{A,\nu}\bS_{A,d_A} \bR_{A,\nu}} \ldots
\\
 \Tr{\bJ_{B,d'_B} \bL_{B,\nu}\bS_{B,d_B} \bR_{B,\nu}}.
\end{multline}

For $X=A,B$, consider the local operators
\begin{equation} \label{eq:JXdXnu}
   \bJ_{X,d'_X,\nu}=
 \bR_{X,\nu}\bJ_{X,d'_X} \bL_{X,\nu}, \quad \bS_{X,d_X,\nu}=
 \bL_{X,\nu}\bS_{X,d_X} \bR_{X,\nu}.
\end{equation}
 Then
\begin{multline}\label{eq:F1AB_reformule}
 F^{(1)}_{(d'_A,d'_B),(d_A,d_B)} \\
 = \sum_{\nu=1}^{\bar\nu} \Tr{\bJ_{A,d'_A} \bS_{A,d_A,\nu}} \Tr{\bJ_{B,d'_B} \bS_{B,d_B,\nu}} \\
  = \sum_{\nu=1}^{\bar\nu} \Tr{\bS_{A,d_A} \bJ_{A,d'_A,\nu}} \Tr{\bS_{B,d_B} \bJ_{B,d'_B,\nu}}.
\end{multline}

 This gives the first-order approximation of the  reduced dynamics  for which the coordinate vector $x = (x_{d'_A,d'_B})_{d'_A,d'_B}$ evolves according to:
 \begin{equation} \label{eq:x_order1}
 \dotex x_{d'_A,d'_B} = \sum_{d_A,d_B}\epsilon  F^{(1)}_{(d'_A,d'_B),(d_A,d_B)} x_{d_A,d_B}
 .
 \end{equation}

Take the second-order correction $F^{(2)}$ given by the general formula~\eqref{eq:F2}.
We have
\begin{equation} \begin{split}
   \cL_1^*(\bJ_{d'}) & = \sum_{\nu'=1}^{\bar\nu} \bJ_{A,d'_A,\nu'} \otimes \bJ_{B,d'_B,\nu'}, \\
 \cL_1(\bS_{d}) & = \sum_{\nu=1}^{\bar\nu} \bS_{A,d_A,\nu}\otimes \bS_{B,d_B,\nu}
\end{split}\end{equation}
where $d'=(d'_A,d'_B)$ and $d=(d_A,d_B)$.
By linearity of $\ocR$
\begin{equation} \begin{split}
  & \cL_1^*(\bJ_{d'}) \ocR\big(\cL_1(\bS_{d})\big) \\
 & =\sum_{\nu,\nu'=1}^{\bar\nu}
 \Big(\bJ_{A,d'_A,\nu'} \otimes \bJ_{B,d'_B,\nu'}\Big) \ocR\Big(\bS_{A,d_A,\nu}\otimes\bS_{B,d_B,\nu}\Big)
.
\end{split}\end{equation}
Combining $e^{s\cL_{A,0}+s\cL_{B,0}}= e^{s\cL_{A,0}}\otimes e^{s\cL_{B,0}}$ with~\eqref{eq:R} and~\eqref{eq:cKAB0} gives

\begin{widetext}

\begin{equation} \begin{split}
 \ocR\Big(\bS_{A,d_A,\nu}\otimes\bS_{B,d_B,\nu}\Big)
   = \int_{0}^{+\infty} \Bigg(&e^{s\cL_{A,0}}\big(\bS_{A,d_A,\nu} \big) \otimes e^{s\cL_{B,0}}\big( \bS_{B,d_B,\nu} \big) \\
  & - \sum_{d''_A,d''_B} \Tr{\bJ_{A,d''_A} \bS_{A,d_A,\nu}} \Tr{\bJ_{B,d''_B} \bS_{B,d_B,\nu}} \bS_{A,d''_A}\otimes \bS_{B,d''_B} \Bigg) ~ds \\
\end{split}\end{equation}

Here, we are only interested in the trace of the product with $\bJ_{A,d'_A,\nu'} \otimes \bJ_{B,d'_B,\nu'}$:

\begin{multline*}
 \Tr{\bJ_{A,d'_A,\nu'} \otimes \bJ_{B,d'_B,\nu'} ~\ocR\Big(\bS_{A,d_A,\nu}\otimes\bS_{B,d_B,\nu}\Big)}
 \\
 = \mathlarger{\mathlarger{\int}}_{0}^{+\infty}
 \Bigg(
 \Tr{\bJ_{A,d'_A,\nu'}~e^{s\cL_{A,0}}\big(\bS_{A,d_A,\nu} \big)} \Tr{ \bJ_{B,d'_B,\nu'} ~ e^{s\cL_{B,0}}\big( \bS_{B,d_B,\nu} \big)}
 \ldots \\ \ldots
 - \sum_{d''_A,d''_B} \Tr{\bJ_{A,d''_A} \bS_{A,d_A,\nu}} \Tr{\bJ_{B,d''_B} \bS_{B,d_B,\nu}}
 \Tr{\bJ_{A,d'_A,\nu'}~ \bS_{A,d''_A}} \Tr{ \bJ_{B,d'_B,\nu'} ~ \bS_{B,d''_B}} \Bigg) ~ds
 \\
 =
 \mathlarger{\mathlarger{\int}}_{0}^{+\infty}
 \Bigg(
 \Tr{\bJ_{A,d'_A,\nu'}~e^{s\cL_{A,0}}\big(\bS_{A,d_A,\nu} \big)} \Tr{ \bJ_{B,d'_B,\nu'} ~ e^{s\cL_{B,0}}\big( \bS_{B,d_B,\nu} \big)}
 \ldots \\ \ldots
 - G_{A,d'_A,d_A,\nu,\nu'}G_{B,d'_B,d_B,\nu,\nu'} \Bigg) ~ds,
\end{multline*}
\end{widetext}
where for $X=A,B$

\begin{equation*}
\begin{split}
 & G_{X,d'_X,d_X,\nu,\nu'}  \\
& = \sum_{d''_X} \Tr{\bJ_{X,d'_X}~ \bS_{X,d''_X,\nu'}} \Tr{\bJ_{X,d''_X} \bS_{X,d_X,\nu}}
\end{split}
\end{equation*}
and using identities like $ \Tr{\bJ_{X,d'_X}~ \bS_{X,d''_X,\nu'}}\equiv \Tr{\bS_{X,d''_X}~ \bJ_{X,d'_X,\nu'}}$.

To conclude, one gets each entry of $F^{(2)}$ with only local numerical
computations on $\cH_A$ and $\cH_B$:

\begin{widetext}
\begin{multline} \label{eq:F2AB}
 F^{(2)}_{(d'_A,d'_B),(d_A,d_B)} =
 \mathlarger{\mathlarger{\sum}}_{\nu,\nu'=1}^{\bar\nu}
 \mathlarger{\int}_{0}^{+\infty}
 \Bigg(
 \Tr{\bJ_{A,d'_A,\nu'}~e^{s\cL_{A,0}}\big(\bS_{A,d_A,\nu} \big)} \Tr{ \bJ_{B,d'_B,\nu'} ~ e^{s\cL_{B,0}}\big( \bS_{B,d_B,\nu} \big)}
 \ldots
 \\ \ldots
 - G_{A,d'_A,d_A,\nu,\nu'}G_{B,d'_B,d_B,\nu,\nu'} \Bigg) ~ds
.
\end{multline}
\end{widetext}
The equivalent of Eqs.~\eqref{eq:F1AB_reformule} and~\eqref{eq:F2AB} for a discrete time setting are given in appendix \ref{sec:discrete_composite}.

\subsection{ZZ gate}\label{ssec:ZZgate}

A ZZ-gate  corresponds to a unitary transformation changing  $\Cpm \Cpm$ to $\Cmp \Cmp$ (parity change).
As for the Z-gate implementation, it can be approximately engineered via the propagator of time duration $T>0$ associated to the Hamiltonian $\bH_1=\epsilon_{ZZ}\left(\ba \bb^{\dagger}+\ba^{\dagger} \bb\right)$ where
$\ba$ (resp. $\bb$) is the photon annihilation operator on sub-system $A$ (resp. $B$) and where $\epsilon_{ZZ}=\frac{\pi}{4\alpha^2 T}$ has to be much smaller than $\kappa_2$.
The superoperators $\cL_0$ and $\cL_1$ corresponding here to Eq.~\eqref{eq:dynL0L1} are thus
\begin{equation}
\begin{aligned}\label{eq:L0L1ZZgate}
 \cL_0(\rho)&=\kappa_2 \left[ D_{\ba^2-\alpha^2} +  D_{\bb^2-\alpha^2}\right](\rho), \\
 \epsilon \cL_1(\rho) &= \kappa_1 \left[ D_{\ba} +  D_{\bb}\right](\rho)-i\tfrac{\pi}{4\alpha^2 T}\left[\left(\ba \bb^{\dagger}+\ba^{\dagger} \bb\right) , \rho\right] \\
\end{aligned}
\end{equation}
where $\kappa_1/\kappa_2$ and $T \kappa_2$ are much smaller than $1$.

\begin{figure*}
 \centering
 \begin{subcaptiongroup}
 \subcaptionlistentry{ZZ x}
 \label{fig:ZZx}
 \begin{overpic}[
  width=0.35\textwidth,
  % grid,
  ]{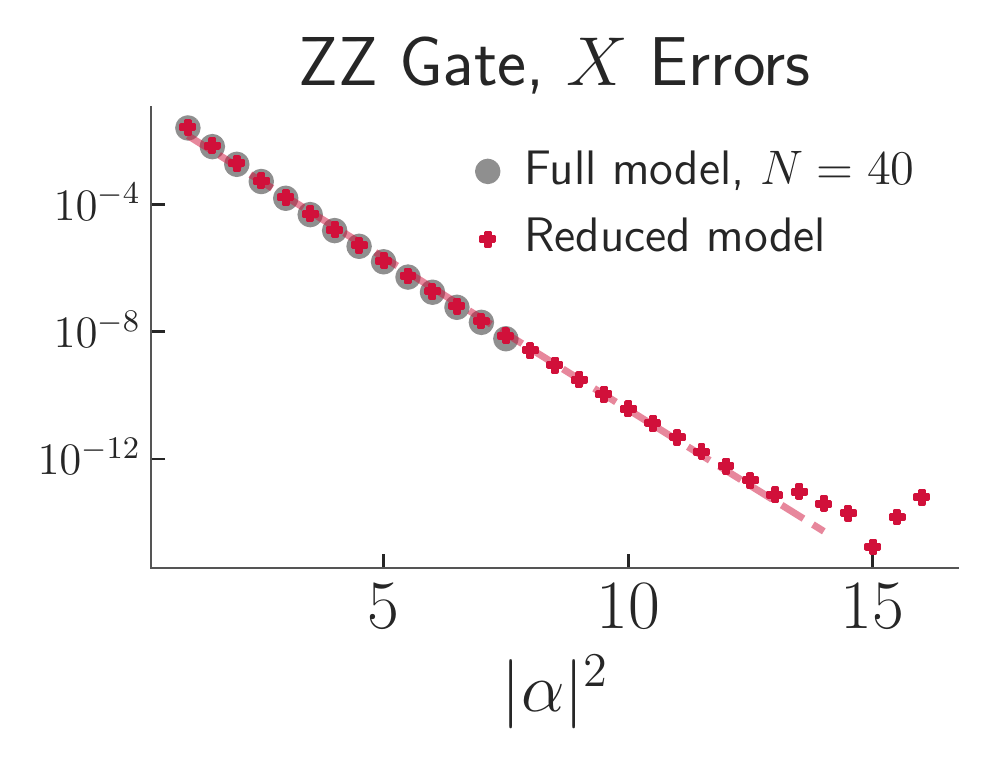}
  % ]{example-image-a}
 \put(5, 70){\captiontext*{}}
 \end{overpic}
 \subcaptionlistentry{S01}
 \label{fig:ZZz}
 \begin{overpic}[
  width=0.35\textwidth,
  % grid
  ]{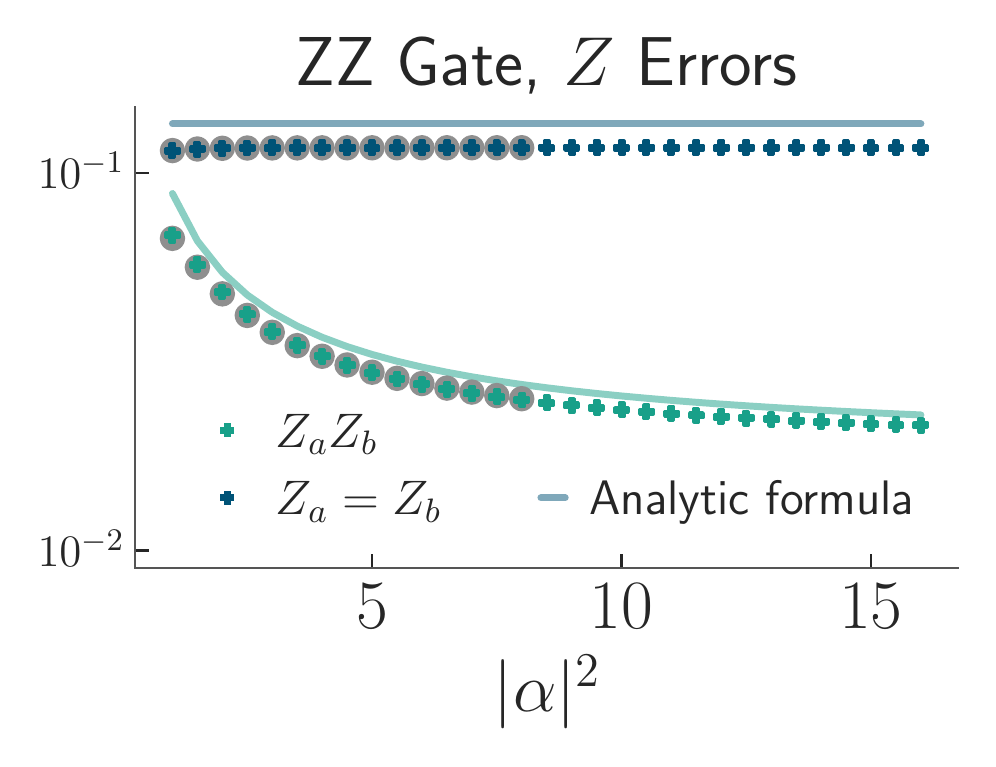}
  % ]{example-image-a}
 \put(5, 70){\captiontext*{}}
 \end{overpic}
 \end{subcaptiongroup}
 \captionsetup{subrefformat=parens}
 \caption{
  Comparison between \subref{fig:ZZx} $X$ and \subref{fig:ZZz} $Z$ error probabilities of a ZZ gate, $p_{ZZ}$ in green and $p_Z$ in blue, obtained via full model simulations for $|\alpha|^2\leq 8$~\eqref{eq:L0L1ZZgate} (shown as gray circles) and the reduced model simulations~\eqref{eq:RedZgate} (colored plus) for different mean photon numbers $|\alpha|^2$. A simple fit yields an exponential suppression of bit-flips with an exponential coefficient of $2.20 \pm 0.01$ (dashed line).
 }
 \label{fig:ZZerror}
\end{figure*}

 $\epsilon F^{(1)}_{(d_A', d_B'),(d_A, d_B)}$ and $\epsilon^2F^{(2)}_{(d_A', d_B'),(d_A, d_B)}$ defined in~\eqref{eq:F1AB_reformule} and~\eqref{eq:F2AB}  are computed using discrete-time formulae and provide, up
to third-order terms, the generator of the  continuous-time reduced dynamics of Eq.~\eqref{eq:RedZgate} with the coordinate-vector $x$
$$
x=\left(x_{(d_A, d_B)}=\Tr{\bJ_{d_A} \otimes \bJ_{d_B} \rho}\right)_{d_A, d_B=1,\ldots,4}.
$$
The parameters of the numerical simulations of figures~\ref{fig:ZZerror} are

\begin{equation*}
  \begin{split}
&\kappa_2 \dt= \frac{1}{1000}, \quad
 \kappa_1=\frac{\kappa_2}{100}, \\
 & \epsilon_{ZZ}= \frac{\pi}{4\alpha^2 T}=\frac{\kappa_2}{20} \text{ with } 1\leq \alpha^2 \leq 16
\end{split}
\end{equation*}
where photon-number truncation $N$ is equal to $100$ for the reduced-model  and to $40$ for the full-model.

As for the Z gate, the reduced model error propagator $E_{\text{\tiny red }}=G_{\text{\tiny ideal }}^{-1}G_{\text{\tiny red }}$ and the full model error propagator $E_{\text{\tiny full }}=G_{\text{\tiny ideal }}^{-1} G_{\text{\tiny full }}$ are close to identity matrix $I_{16}$ and characterize the errors. These channels can also be decomposed according to the basis $(\bS_{A, 1}, \ldots,\bS_{A, 4})\otimes(\bS_{B, 1}, \ldots,\bS_{B, 4})$.
This means that for $E=E_{\text{\tiny red }}, E_{\text{\tiny full}}$, the following identity

\begin{widetext}
\begin{multline}  \label{eq:chiAB}
 \forall x\in\mathbb{R}^{16}, \quad
 \sum_{d_A, d_B,d_A', d_B'=1}^{4} E_{(d_A, d_B),(d_A', d_B')} x_{(d_A', d_B')} \bS_{A, d_A}\bS_{B, d_B}  \\
 = \sum_{m_A, m_B,n_A, n_B=1}^4 \chi^E_{(m_A, m_B),(n_A, n_B)} \bS_{A, m_A} \bS_{B, m_B} \left(\sum_{d_A, d_B=1}^4 x_{(d_A, d_B)} \bS_{A, d_A}\bS_{B, d_B}\right) \bS_{A, n_A}\bS_{B, n_B}
\end{multline}
\end{widetext}
uniquely defines the $16\times 16$, $\chi^E$ matrix characterizing the errors (close to $\chi^{I_{16}}$ having a single non-zero entry $\chi^{I_{16}}_{1,1}=1$).
An illustration of  $\chi^{E_{\text{\tiny red }}}$ and  $\chi^{E_{\text{\tiny full }}}$ is given in appendix \ref{sec:error_models}, figure~\ref{fig:ZZchi} for $\alpha=2$.

\begin{figure*}
  \centering
  \input{fig/chi_matrix_2_qubits.tikz}
 \caption{
  Two-qubit $\chi$ matrix representing the noise channel of an imperfect gate reduced to two two-level systems.
  The off-diagonal elements shown in gray are  not considered in such a rough analysis based  on  symmetric Pauli errors.
 }
  \label{fig:chi_2}
 \end{figure*}
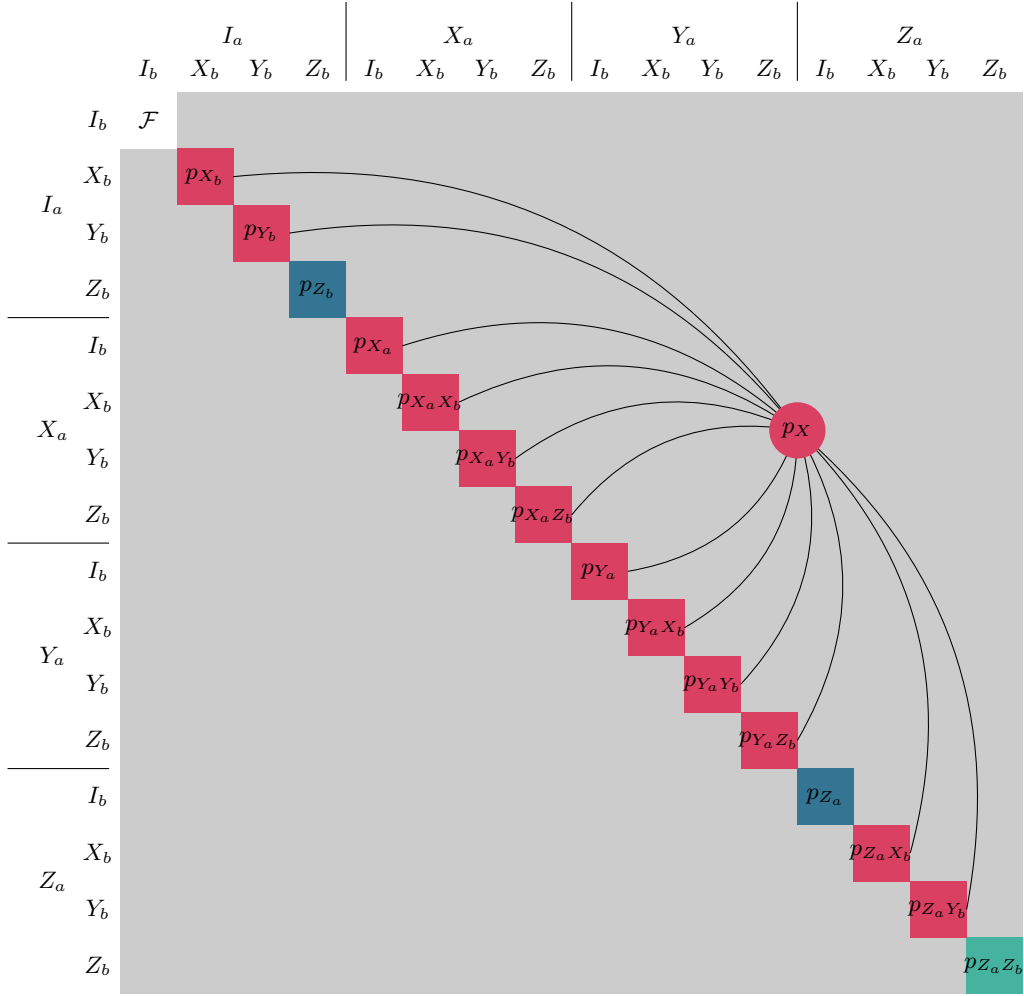

Coefficients of the diagonal of $\chi^E$ give the Pauli errors of the gate. In figure~\ref{fig:chi_2}, the total bit-flip error probability is displayed as the sum of all the 12 Pauli errors involving a bit flip ($X$- or $Y$-error) and simulated in figure~\ref{fig:ZZx}.
We found an exponential suppression of bit flips proportional to $ \exp^{-a |\alpha|^2}$ with $a=2.20 \pm 0.01$.
The reduced model also captures the phase flips (Z-error) in figure~\ref{fig:ZZz}.
It matches well with full-model simulations and also with an analytical formula obtained via a perturbation expansion  derived from the formula of the Z-gate errors: $p_{Z_a}=p_{Z_b}= \nbar \kappa_1 T= \frac{\pi \kappa_1}{4
\epsilon_{ZZ}}, p_{Z_aZ_b} =
\frac{\pi\epsilon_{ZZ}}{2|\alpha|^4\kappa_2} + p_{Z_a} p_{Z_b}$ (the last term coming from second-order effects of the single photon losses), see appendix~\ref{sec:SFB}.

The equation~\eqref{eq:S1} allows to perform a first-order computation of the leakage, see appendix~\ref{sec:leakage_composite} and figure~\ref{fig:ZZleakage}.

\subsection{ZZZ gate} \label{ssec:ZZZgate}

A ZZZ-gate unitary corresponds to a transformation changing  $\Cpm \Cpm\Cpm$ to $\Cmp \Cmp\Cmp$.
As for the Z- and ZZ-gate, it can be approximately engineered via the propagator of time duration $T>0$ associated to the Hamiltonian $\bH_1=\epsilon_{Z Z
Z}\left(\ba \bb \bc^{\dagger}+\ba^{\dagger} \bb^{\dagger} \bc\right)$ where
$\ba$ (resp. $\bb$, $\bc$) is the photon annihilation operator on sub-system $A$ (resp. $B$, $C$) and where $\epsilon_{ZZZ}=\frac{\pi}{4\alpha^3 T}$ has to be much smaller than $\kappa_2$.
The superoperators $\cL_0$ and $\cL_1$ corresponding here to Eq.~\eqref{eq:dynL0L1} are thus
\begin{equation}
\begin{aligned}\label{eq:L0L1ZZZgate}
 \cL_0(\rho)&=\kappa_2 \left[ D_{\ba^2-\alpha^2} +  D_{\bb^2-\alpha^2}+  D_{\bc^2-\alpha^2}\right](\rho), \\
 \epsilon \cL_1(\rho) &= \kappa_1 \left[ D_{\ba} +  D_{\bb}+  D_{\bc}\right](\rho)\\
 & -i\tfrac{\pi}{4\alpha^3 T}\left[\left(\ba \bb \bc^{\dagger}+\ba^{\dagger} \bb^{\dagger} \bc\right) , \rho\right] \\
\end{aligned}
\end{equation}
where $\kappa_1/\kappa_2$ and $T \kappa_2$ are much smaller than $1$.

\begin{figure*}
    \centering
    \begin{subcaptiongroup}
    \subcaptionlistentry{ZZZ x}
    \label{fig:ZZZx}
    \begin{overpic}[
     width=0.35\textwidth,
     % grid,
     ]{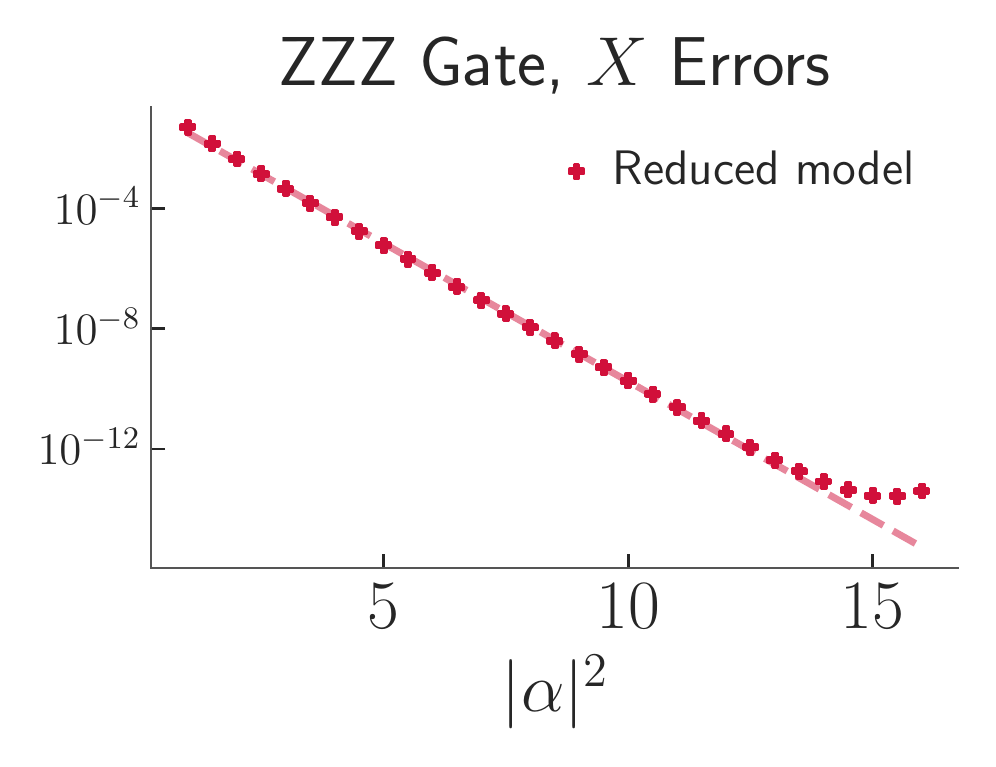}
     % ]{example-image-a}
    \put(5, 70){\captiontext*{}}
    \end{overpic}
    \subcaptionlistentry{ZZZ z}
    \label{fig:ZZZz}
    \begin{overpic}[
     width=0.35\textwidth,
     % grid
     ]{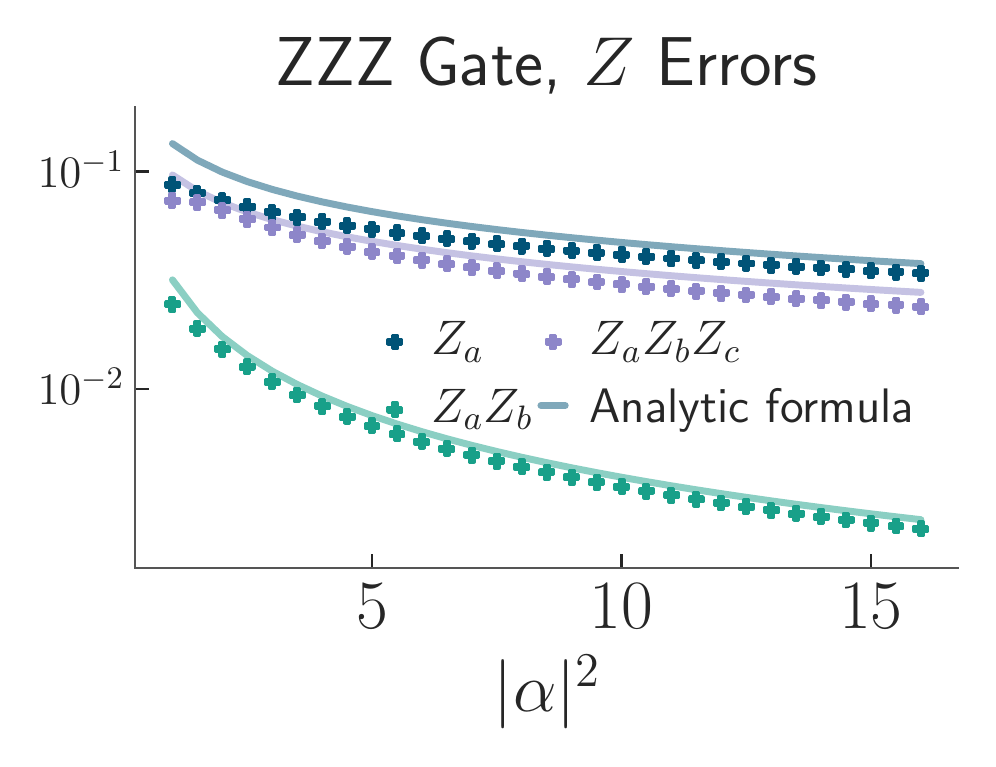}
     % ]{example-image-a}
    \put(5, 70){\captiontext*{}}
    \end{overpic}
    \end{subcaptiongroup}
    \captionsetup{subrefformat=parens}
    \caption{
  Comparison between \subref{fig:ZZZx} $X$ and \subref{fig:ZZZz} $Z$ error probabilities of a ZZZ gate, $p_Z$ ($p_{Z_a}=p_{Z_b}=p_{Z_c}$) in blue, $p_{ZZ}$ ($p_{Z_aZ_b}=p_{Z_aZ_c}=p_{Z_bZ_c}$) in green and $p_{Z_aZ_bZ_c}$ in lavender, obtained via reduced model simulations~\eqref{eq:RedZgate} (colored plus) for different mean photon numbers $|\alpha|^2$. A simple fit yields an exponential suppression of bit-flips with an exponential coefficient of $2.12 \pm 0.01$ (dashed line).
 }
    \label{fig:ZZZerror}
   \end{figure*}

 Numerical simulations of the full-model were not performed for computational limitations.  We only report  reduced-model simulations based on the direct generalization of~\eqref{eq:F1AB_reformule} and~\eqref{eq:F2AB} to a tripartite system.
The parameters of the numerical simulations of figures~\ref{fig:ZZZerror} are

\begin{equation*}
  \begin{split}
&\kappa_2 \dt= \frac{1}{1000}, \quad
 \kappa_1=\frac{\kappa_2}{100}, \\
  &\epsilon_{ZZZ}= \frac{\pi}{4\alpha^2 T}=\frac{\kappa_2}{20} \text{ with } 1\leq \alpha^2 \leq 16, \quad N=100.
  \end{split}
\end{equation*}

The total bit-flip error probability is the sum of all the 56 Pauli errors involving a bit flip ($X$- or $Y$-error) and simulated in figure~\ref{fig:ZZZx}.
We found an exponential suppression of bit flips proportional to $ \exp^{-a |\alpha|^2}$ with $a=2.12 \pm 0.01$.
The reduced model also captures the phase flips (Z-error) in figure~\ref{fig:ZZZz}.
It matches well with an analytical formula obtained via a perturbation expansion detailed in appendix~\ref{sec:SFB}: $
p_{Z_a}= p_{Z_b}= p_{Z_c}= \nbar \kappa_1 T= \frac{\pi\kappa_1}{4 |\alpha| \epsilon_{ZZZ}} ,
p_{Z_aZ_bZ_c}= \frac{3\pi \epsilon_{ZZZ}}{4 |\alpha| \kappa_2} + p_{Z_a} p_{Z_b} p_{Z_c},
p_{Z_aZ_b}=
p_{Z_aZ_c}=p_{Z_bZ_c}=p_{Z}p_{ZZZ} + p_{Z}^2 $.
An illustration of $64 \times 64$ $\chi^{E}$ matrix for the reduced propagator error $E_{\text{\tiny red }}$ is given in appendix \ref{sec:error_models}, figure~\ref{fig:ZZZchi}.
A first-order computation of the leakage is shown in figure~\ref{fig:ZZZleakage}.

\section{Composite systems with an unstabilized component} \label{sec:hybrid}

\subsection{Second-order approximation}

Assume that $\cL_{B,0} =0$ for  the  bipartite system of section~\ref{sec:composite}. Then  $(\bS_{B,d_B})_{1\leq d_B\leq \bar d_B}$ span all Hermitian operators on $\cH_B$ and $ \bJ_{B,d_B}= \bS_{B,d_B}$. Following~\eqref{eq:rhoABparam}, all operators belonging to $\cD_0$ read
$ \sum_{d_A}\bS_{A,d_A}\otimes \rho_{B, d_A}$ with  hermitian operators on $\cH_B$
\begin{equation}\label{eq:rhoBdA}
  \rho_{B, d_A}=\sum_{d_B} x_{d_A d_B} \bS_{B,d_B}
\end{equation}
 and $x_{d_A,d_B}= \Tr{\bS_{B,d_B}~\rho_{B, d_A}} $  real numbers. The mapping between $x=(x_{d_A,d_B})$ and  the set $(\rho_{B, d_A})$ of $\bar d_A$ operators on $\cH_B$ is linear and bijective. We just  translate here the formulae  of section~\ref{sec:composite} with $x_{d_A,d_B}$ variables in  $\rho_{B, d_A}$ variables.

Combining~\eqref{eq:F1AB_reformule} with \eqref{eq:x_order1}, the first-order time evolution of the coordinate vector $(x_{d'_A,d'_B})_{d'_A,d'_B}$ reads
\begin{equation*}
  \begin{split}
\dotex
 &x_{d'_A,d'_B} \\
 & = \sum_{\nu,d_A,d_B} \Tr{\bJ_{A,d'_A} \bL_{A,\nu}\bS_{A,d_A}
 \bR_{A,\nu}} \ldots \\
 & \ldots \Tr{\bS_{B,d'_B} \bL_{B,\nu}\bS_{B,d_B} \bR_{B,\nu}} x_{d_A,d_B}.
  \end{split}
\end{equation*}
Using~\eqref{eq:rhoBdA},  we get
$$
\dotex
 \rho_{B, d'_A} = \sum_{\nu, d_A} \Tr{\bJ_{A,d'_A} \bL_{A,\nu}\bS_{A,d_A}
 \bR_{A,\nu}} \bL_{B,\nu} \rho_{B, d_A} \bR_{B,\nu}.
 $$
 With $ \bar
 F^{(1)}_{d'_A,d_A,\nu}= \Tr{\bJ_{A,d'_A} \bL_{A,\nu}\bS_{A,d_A}\bR_{A,\nu}} $,
the first-order approximation of the slow dynamics reads
$$
\dotex \rho_{B, d'_A} = \sum_{\nu, d_A}
 \bar F^{(1)}_{d'_A,d_A,\nu} \bL_{B,\nu} \rho_{B, d_A} \bR_{B,\nu}
 $$

 Using in~\eqref{eq:JXdXnu},  the superoperator $\ocR$  reads
 \begin{widetext}
\begin{equation*}
  \begin{split}
 & \ocR\Big(\bS_{A,d_A,\nu}\otimes\bS_{B,d_B,\nu}\Big) \\
 & = \int_{0}^{+\infty}
 \Bigg(
 e^{s\cL_{A,0}}\big(\bS_{A,d_A,\nu} \big) \otimes \bS_{B,d_B,\nu}
 - \sum_{d''_A,d''_B} \Tr{\bJ_{A,d''_A} \bS_{A,d_A,\nu}} \Tr{\bS_{B,d''_B} \bS_{B,d_B,\nu}}
 \bS_{A,d''_A}\otimes \bS_{B,d''_B} \Bigg) ~ds \\
 & = \int_{0}^{+\infty}
 \Bigg(
 e^{s\cL_{A,0}}\big(\bS_{A,d_A,\nu} \big)
 - \sum_{d''_A} \Tr{\bJ_{A,d''_A} \bS_{A,d_A,\nu}}
 \bS_{A,d''_A} \Bigg) ~ds \otimes \bS_{B,d_B,\nu} \\
 & = \ocR\Big(\bS_{A,d_A,\nu}\Big)\otimes\bS_{B,d_B,\nu}
 .
  \end{split}
\end{equation*}
$\ocR$  is thus local on subsystem A.
 $F^{(2)}$ given by the formula~\eqref{eq:F2AB} becomes then
\begin{align*}
 F^{(2)}_{(d'_A d'_B),(d_A d_B)}
 &= \Tr{ \cL_1^{*} (\bJ_{A,d'_A} \otimes \bS_{B,d'_B}) ~ \ocR\Big( \cL_1(\bS_{A, d_A} \otimes \bS_{B, d_B} )\Big) } \\
 & = \sum_{\nu \nu'} \bar F^{(2)}_{d'_A, d_A, \nu, \nu'} \Tr{ \bS_{B, d_B'} \bL_{B,\nu'}\bL_{B,\nu}\bS_{B,d_B} \bR_{B,\nu}\bR_{B,\nu'} }
\end{align*}
where $\bar F^{(2)}_{d'_A, d_A, \nu, \nu'} = \Tr{ \bJ_{d'_A, \nu'} ~ \ocR\Big( \bS_{d_A, \nu}\Big) }$.
With
$$
\sum_{d_A,d_B} F^{(2)}_{(d'_A,d'_B),(d_A,d_B)} x_{d_A,d_B}
= \sum_{d_A,d_B,\nu \nu'} \bar F^{(2)}_{d'_A, d_A, \nu, \nu'}  x_{d_A,d_B} \Tr{ \bS_{B, d_B'} \bL_{B,\nu'}\bL_{B,\nu}\bS_{B,d_B} \bR_{B,\nu}\bR_{B,\nu'}}
$$
and
$$
\sum_{d_B} x_{d_A,d_B}  \Tr{ \bS_{B, d_B'} \bL_{B,\nu'}\bL_{B,\nu}\bS_{B,d_B} \bR_{B,\nu}\bR_{B,\nu'}}=
\Tr{ \bS_{B, d_B'} \bL_{B,\nu'}\bL_{B,\nu} \rho_{B,d_A} \bR_{B,\nu}\bR_{B,\nu'}}
$$
we get  the following expression for the second-order approximation using the parametrization based on the $\bar d_A$ set of Hermitian operators $(\rho_{B, d'_A})$  on $\cH_B$:
\begin{equation}
 \dotex \rho_{B, d'_A} = \sum_{\nu, d_A} \bar F^{(1)}_{d'_A,d_A,\nu} \bL_{B,\nu} \rho_{B, d_A} \bR_{B,\nu} + \sum_{\nu,\nu',d_A} \bar F^{(2)}_{d'_A,d_A,\nu,\nu'}
 \bL_{B,\nu'} \bL_{B,\nu} \rho_{B, d_A} \bR_{B,\nu} \bR_{B,\nu'}
.
 \label{eq:LB=0}
\end{equation}
 \end{widetext}
with
$
\bar
 F^{(1)}_{d'_A,d_A,\nu}= \Tr{\bJ_{A,d'_A} \bL_{A,\nu}\bS_{A,d_A}\bR_{A,\nu}}$ and $
 \bar F^{(2)}_{d'_A, d_A, \nu, \nu'} = \Tr{ \bJ_{d'_A, \nu'} ~ \ocR\Big( \bS_{d_A, \nu}\Big) }
$.
The discrete-time formulations of  $\bar F^{(1)}_{d'_A,d_A,\nu}$ and $\bar F^{(2)}_{d'_A,d_A,\nu,\nu'}$  can be obtained directly from appendix \ref{sec:discrete_composite}.

\subsection{CNOT gate}  \label{ssec:CNOTgate}

A CNOT-gate  corresponds to a $\pi$-rotation in the phase space of a qubit called the target qubit conditioned on the state of another qubit called the control qubit, being on the $\ket{1}_C \simeq  \ket{-\alpha}$ state.
Using cat-qubits of  complex amplitude  $\alpha$ with $|\alpha|^2 \gg 1$, it can be approximately engineered by stabilizing the control cat-qubit  via two-photon dissipation and adding the Hamiltonian $\bH_1=\frac{\pi}{4 \alpha T} \left(\ba + \ba^\dag - 2 |\alpha| \right) \left(\bb^\dag \bb - \nbar\right) $ where
$\ba$ (resp. $\bb$) is the photon annihilation operator on the control cat-qubit  $A$ (resp. the target cat-qubit  $B$) and where $T$ is the gate time.

The original  implementation of the CNOT gate~\cite{GuillaudMirrahimiPRX2019, AmazonPRXQ2022} includes the target-qubit    stabilization via a non local time-varying  two-photon dissipation.  This implementation is experimentally difficult. Thus, we  consider here an easier  one with only $\bH_1$. This corresponds to  a "stroboscopic stabilization" where the target-qubit is stabilized before and after the gate. The simulations below indicate that  the exponential suppression of bit flips   remains satisfied.

The superoperators $\cL_0$ and $\cL_1$ corresponding here to~\eqref{eq:dynL0L1} are thus
\begin{equation}
\begin{aligned}\label{eq:L0L1CNOTgate}
  \cL_0(\rho)&=\kappa_2 D_{\ba^2-\alpha^2} (\rho), \\
 \epsilon \cL_1(\rho) &= \kappa_1 \left[ D_{\ba} +  D_{\bb}\right](\rho) \\
 &-i\tfrac{\pi}{4 \alpha T}\left[\ \left(\ba + \ba^\dag - 2 |\alpha| \right) \left(\bb^\dag \bb - \nbar\right), \rho\right] \\
\end{aligned}
\end{equation}
where $\kappa_1/\kappa_2$ and $\tfrac{\pi}{4 \alpha\kappa_2 T} $ are much smaller than $1$.

\begin{figure}
 \centering
 \begin{subcaptiongroup}
 \subcaptionlistentry{CNOT x}
 \label{fig:CNOTx}
 \begin{overpic}[
  width=0.35\textwidth,
  % grid
  % ]{fig/comp_CNOT_bit-flip_nbsteps20000.pdf}
  ]{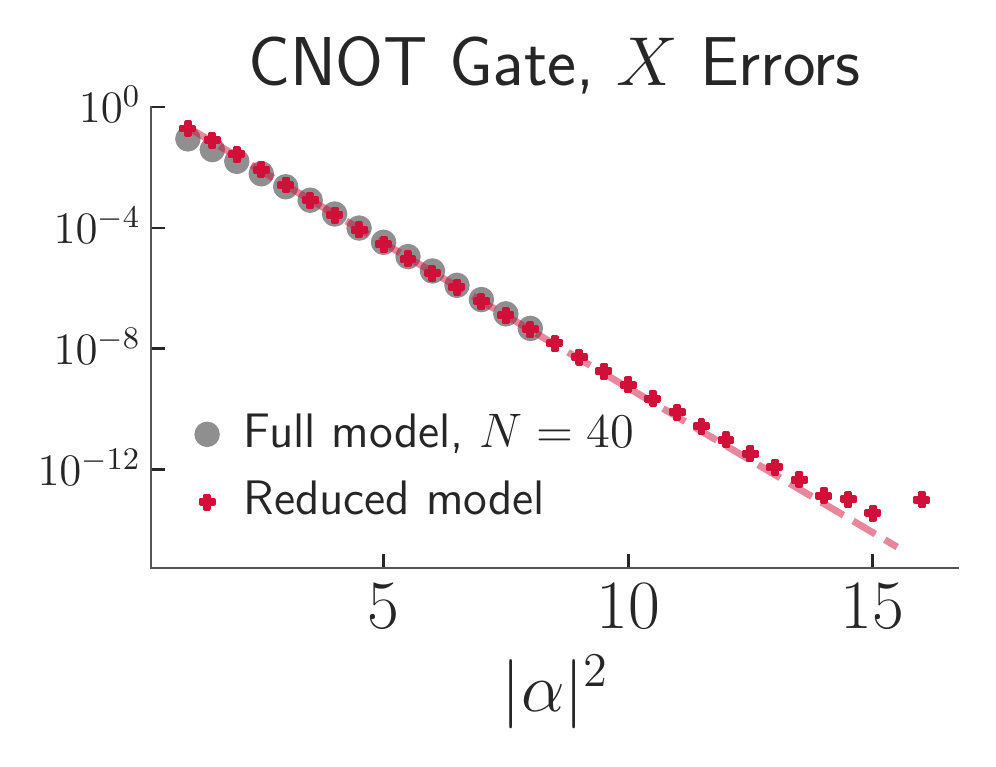}
  % ]{example-image-a}
 \put(5, 70){\captiontext*{}}
 \end{overpic}
 \subcaptionlistentry{CNOT z}
 \label{fig:CNOTz}
 \begin{overpic}[
  width=0.35\textwidth,
  % grid,
  % ]{fig/comp_CNOT_phase-flip_nbsteps20000.pdf}
  ]{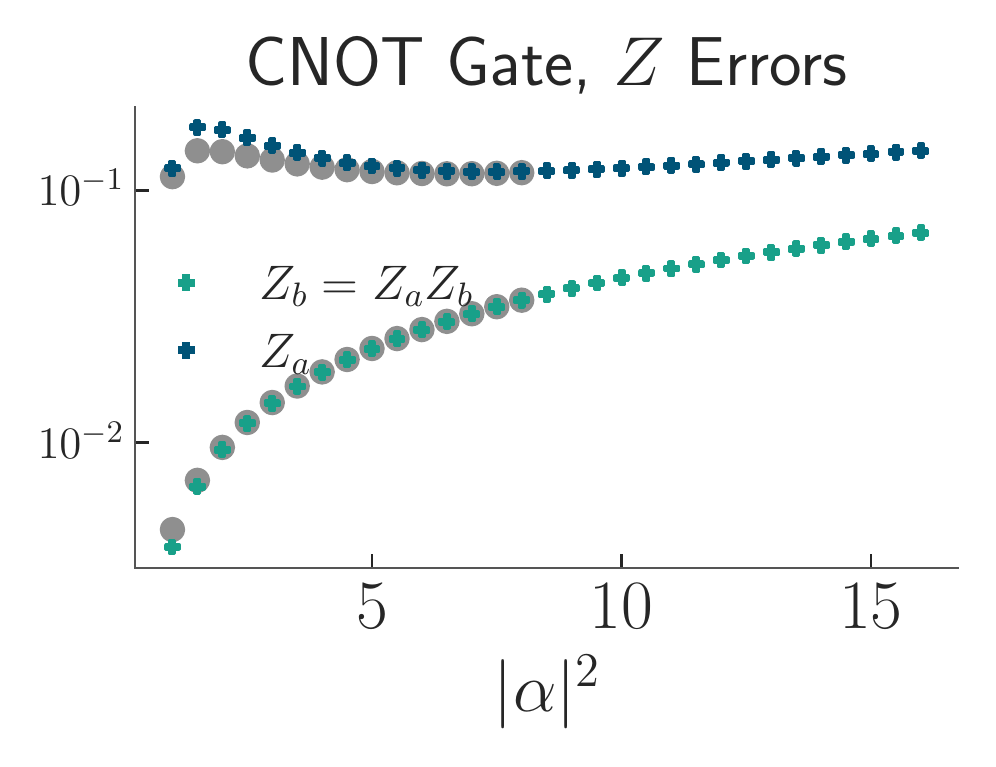}
  % ]{example-image-a}
 \put(5, 70){\captiontext*{}}
 \end{overpic}
 \end{subcaptiongroup}
 \captionsetup{subrefformat=parens}
 \caption{
  Comparison between \subref{fig:CNOTx} $X$ and \subref{fig:CNOTz} $Z$ error probabilities of a CNOT gate, $p_{Z_aZ_b} = p_{Z_b}$ in green and $p_{Z_a}$ in blue, obtained via full model simulations for $|\alpha|^2\leq 8$~\eqref{eq:L0L1CNOTgate} (shown as gray circles) and the reduced model simulations~\eqref{eq:LB=0} (colored plus) for different mean photon numbers $\alpha^2$. A simple fit yields an exponential suppression of bit-flips with an exponential coefficient of $2.204 \pm 0.009$ (dashed line).
 }
 \label{fig:CNOTerror}
\end{figure}

The complex coefficients $ F^{(1)}_{d'_A,d_A,\nu}$ and $
\bar F^{(2)}_{d'_A, d_A, \nu, \nu'}$ of~\eqref{eq:LB=0} are computed using the discrete-time formulation of appendix~\ref{sec:discrete_composite} with the following parameters
\begin{equation*} \begin{split}
  &\kappa_2 \alpha^2 \dt= \frac{1}{1000}, \quad
 \kappa_1=\frac{\kappa_2}{100}, \\
 & T= \tfrac{1}{\kappa_2} \text{ with } 1\leq \alpha^2 \leq 16, \\
 &  N(\alpha) = \max(20, \lfloor \alpha^2 + 20 \alpha \rfloor )
\end{split}\end{equation*}
where $\lfloor \cdot \rfloor$ denotes the integer part.
 Figures~\ref{fig:CNOTerror} are based on the numerical integration  via  an explicit Euler scheme  of~\eqref{eq:LB=0}, a linear  system coupling   $\bar d_A=4$ Hermitian operators on $\cH_B$:  $(\rho_{B,1},\ldots,\rho_{B,4})$.

As for the ZZ-gate simulation, the total bit-flip error probability is the sum of all the 12 Pauli errors involving a bit flip ($X$- or $Y$-error) and corresponds  in figure~\ref{fig:CNOTx}.
We found an exponential suppression of bit flips proportional to $ \exp^{-a |\alpha|^2}$ with $a=2.204 \pm 0.009$.
The reduced model also captures the phase flips (Z-error) as illustrated in  figure~\ref{fig:CNOTz}.
It matches well with full-model simulations that have been performed for  $|\alpha|^2 \leq 8$.
An illustration of $16 \times 16$ $\chi^{E}$ matrix for the reduced propagator error $E_{\text{\tiny red }}$ and the full propagator error $E_{\text{\tiny full }}$ is given in appendix \ref{sec:error_models}, figure~\ref{fig:CNOTchi}.
A first-order computation of the leakage is shown in appendix~\ref{sec:hybrid_leakage_composite} and figure~\ref{fig:CNOTleakage}.

\subsection{CCNOT gate} \label{ssec:Toffoli}

A CCNOT-gate (Toffoli gate)  corresponds to a $\pi$-rotation in the phase space of a target qubit conditioned on the state of two control qubits being on the $\ket{1}_C\ket{1}_C \simeq  \ket{-\alpha} \ket{-\alpha}$ state.
When $|\alpha|^2 \gg 1$, it  can be approximately engineered by stabilizing the two control cat-qubits via two-photon dissipation and adding the Hamiltonian $\bH_1=- \frac{\pi}{8 \alpha^2 T}\left( \left(\ba - |\alpha| \right) \left(\bb - |\alpha| \right)+\text {h.c.}\right) \left(\bc^\dag \bc - \nbar\right) $ where
$\ba$ (resp. $\bb$, $\bc$) is the photon annihilation operator on sub-system $A$ (resp. $B$, $C$) and where $T$ is the gate time.
The superoperators $\cL_0$ and $\cL_1$ corresponding here to Eq.~\eqref{eq:dynL0L1} are thus
\begin{equation}
\begin{aligned}\label{eq:L0L1CCNOTgate}
 \cL_0(\rho)&=\kappa_2 \left[ D_{\ba^2-\alpha^2} +  D_{\bb^2-\alpha^2}\right](\rho), \\
 \epsilon \cL_1(\rho) &= \kappa_1 \left[ D_{\ba} +  D_{\bb}+  D_{\bc}\right](\rho)\\
 +i \frac{\pi}{8 \alpha^2 T} & \left[\left( \left(\ba - |\alpha| \right) \left(\bb - |\alpha| \right)+\text {h.c.}\right) \left(\bc^\dag \bc - \nbar\right), \rho\right] \\
\end{aligned}
\end{equation}
where $\kappa_1/\kappa_2$ and $\tfrac{\pi}{8 \alpha^2 \kappa_2 T} $ are much smaller than $1$.

\begin{figure}
 \centering
 \begin{subcaptiongroup}
 \subcaptionlistentry{CCNOT x}
 \label{fig:CCNOTx}
 \begin{overpic}[
  width=0.35\textwidth,
  % grid,
  ]{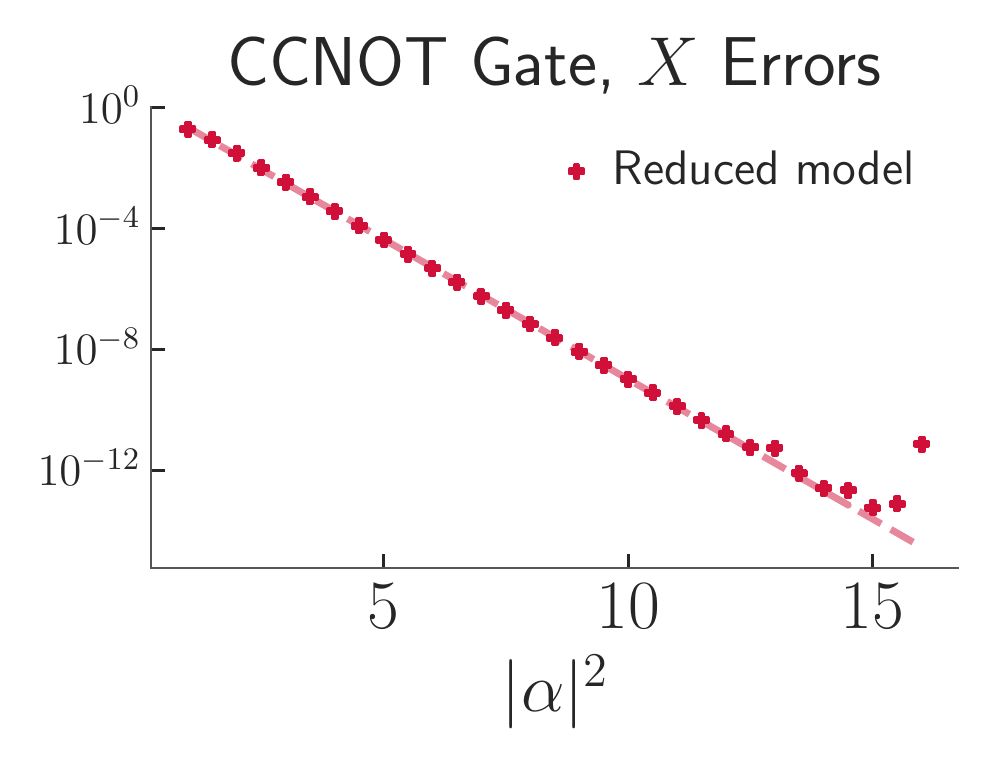}
  % ]{example-image-a}
 \put(5, 70){\captiontext*{}}
 \end{overpic}
 \subcaptionlistentry{CCNOT z}
 \label{fig:CCNOTz}
 \begin{overpic}[
  width=0.35\textwidth,
  % grid
  % ]{fig/comp_CNOT_bit-flip_nbsteps20000.pdf}
  ]{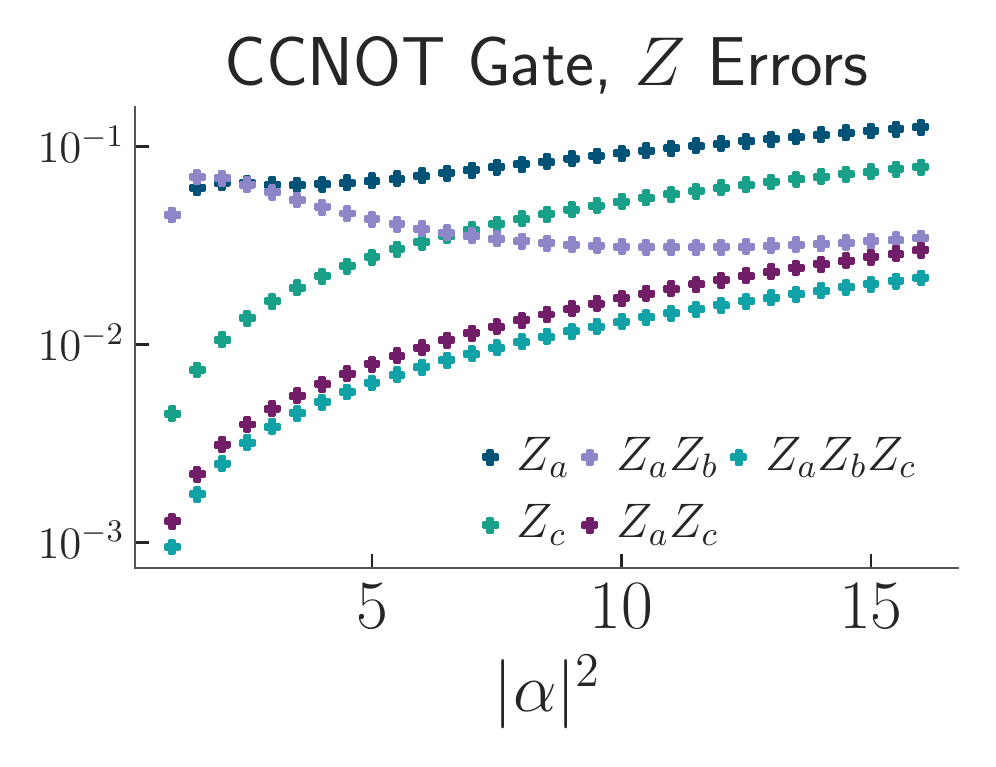}
 \put(5, 70){\captiontext*{}}
 \end{overpic}
 \end{subcaptiongroup}
 \captionsetup{subrefformat=parens}
 \caption{
  Comparison between \subref{fig:CCNOTx} $X$ and \subref{fig:CCNOTz} $Z$ error probabilities of a CCNOT gate, $p_{Z_a}=p_{Z_b}$ in dark blue, $p_{Z_c}$ in green, $p_{Z_aZ_b}$ in lavander $p_{Z_aZ_c}=p_{Z_bZ_c}$ in violet and $p_{Z_aZ_bZ_c}$ in light blue, obtained via reduced model simulations~\eqref{eq:LB=0} (colored plus) for different mean photon numbers $|\alpha|^2$. A simple fit yields an exponential suppression of bit-flips with an exponential coefficient of $2.131 \pm 0.005$ (dashed line).
 }
 \label{fig:CCNOTerror}
\end{figure}

Numerical simulations of the full-model  have not been done because of  computational  limitation. We only report here simulations based  on the direct generalisation of~\eqref{eq:LB=0} to a tripartite system where components one and two  are stabilized whereas the  third one is not.
The parameters of the numerical simulations of figures~\ref{fig:CCNOTerror} are
\begin{equation*} \begin{split}
  &\kappa_2 \alpha^2 \dt= \frac{1}{1000}, \quad
 \kappa_1=\frac{\kappa_2}{100}, \\
 & T= \tfrac{1}{\kappa_2} \text{ with } 1\leq \alpha^2 \leq 16, \\
 & N(\alpha) = \max(20, \lfloor \alpha^2 + 20 \alpha \rfloor )
\end{split}\end{equation*}
where $\lfloor \cdot \rfloor$ denotes the integer part.

As for the $ZZZ$-gate simulations, the total bit-flip error probability is the sum of all the 56 Pauli errors involving a bit flip ($X$- or $Y$-error) and simulated in figure~\ref{fig:CCNOTx}.
We found an exponential suppression of bit flips proportional to $ \exp^{-a |\alpha|^2}$ with $a=2.131 \pm 0.005$.
The reduced model provides also   the phase flips (Z-error) in figure~\ref{fig:CCNOTz}.
An illustration of $64 \times 64$ $\chi^{E}$ matrix for the reduced propagator error $E_{\text{\tiny red }}$ is given in appendix \ref{sec:error_models}, figure~\ref{fig:CCNOTchi}.
A first-order computation of the leakage is shown in figure~\ref{fig:CCNOTleakage}.

%%%%%%%%%%%%%%%%%%%%%%%%%%%%%%%%%%%%%%%%%%%%%%%%%%%%%%%%%%%%%%%%%%

\section{Concluding remarks}\label{sec:conclusion}

We have introduced a new numerical method for simulating open quantum systems
composed of several subsystems, exponentially stabilized towards
stationnary subspaces.
This numerical method is based on a perturbation
analysis with an original asymptotic expansion exploiting the reduced model
formulation of the dynamics,
relying on the invariant operators of the local and nominal dissipative
dynamics of the subsystems.
The derivation was shown up to a second-order
expansion which can be computed with only local calculations for each subsystem.
We have applied this method on several cat-qubit gates (Z, ZZ, ZZZ, CNOT and CCNOT)  and shown that   the dominant  phase-flip error rates  and the exponentially small bit-flip error rates are well described by such reduced-order models and simulations up to 16 photons in each cat-qubit.
Furthermore, this approach, which has provided significant space savings, can be used to an even larger number of bosonic qubits.

The two-photon dissipation of the cat-qubit encoding comes from a more complex master equation involving a buffer mode coupled to the memory cavity via a two-photon exchange Hamiltonian, \cite{MirrahimiCatComp2014}. Similar analysis can thus be built with such  composite cavity-buffer  description for  each cat-qubit.

The derivations shown here can be further applied to other similar  composite systems with dominant local stabilization used in autonomous quantum error correction schemes, such as   squeezed cat-qubits \cite{Schlegel2022, Xu2022} or grid-states \cite{GirvinPRL20,Sellem2022}.

This numerical method exploiting strong local dissipation with  weak coupling and decoherence  in   many-body systems 
has been presented in the context of continuous-time processes and
could also  be useful for time-discrete processes, see App.\ref{sec:discrete}
such as those appearing in  quantum error correction schemes, as the repetition code~\cite{Guillaud2021}
or the surface code~\cite{FowlerMariantoniMartinisEtAl2012}.

%%%%%%%%%%%%%%%%%%%%%%%%%%%%%%%%%%%%%%%%%%%%%%%%%%%%%%%%%%%%%%%%%%%%%%%%%%%%%%%
\section{Acknowledgments}

We thank Philippe Campagne-Ibarcq, J\'{e}r\'{e}mie Guillaud, Mazyar Mirrahimi,
Claude Le Bris, Alain Sarlette, Lev-Arcady Sellem and Antoine Tilloy for
numerous discussions and scientific exchanges on model reduction, numerical
simulations, cat-qubits and bosonic codes.

This project has received funding
from the Plan France 2030 through the project ANR-22-PETQ-0006.

This project
has received funding from the European Research Council (ERC) under the
European Union's Horizon 2020 research and innovation program (grant
agreement No. [884762]).

The numerical simulations were performed using the computer cluster of Inria Paris.
The simulations in the full model picture were performed using the QuTiP open-source package.

\bibliographystyle{unsrt}
%\bibliographystyle{plain}
%\bibliography{RouchonJabref}

\clearpage
\appendix

\section{High-order expansion and simulations}

\subsection{Expansion order exceeding 2} \label{ssec:order_n}

Take $n\geq 2$ and assume that we have computed all the terms $F_{d',d}^{(r)}$
and $\bS_{d'}^{(r)}$ of order $r< n$ with $\Tr{\bJ_{d'} \bS_{d}^{(r)}}=0$ for
all $d$ and $d'$. Invariance condition of order $n$ reads
\begin{equation*}
  \begin{split}
 &\forall d  \in\{1,\ldots,\bar d\},\\
 &  \sum_{d''=1}^{\bar d} \sum_{r=1}^{n}
 F_{d'',d}^{(r)} \bS_{d''}^{(n-r)} = \cL_0(\bS_{d}^{(n)}) +
 \cL_1(\bS_{d}^{(n-1)}).
  \end{split}
\end{equation*}
Left multiplication by operator $\bJ_{d'}$ and
taking the trace yields
\begin{equation}\label{eq:Fn}
 F_{d',d}^{(n)} = \Tr{\bJ_{d'} \cL_1( \bS_{d}^{(n-1)})}= \Tr{\cL_1^{*}(\bJ_{d'}) ~\bS_{d}^{(n-1)})}.
\end{equation}

For $\bS_{d}^{(n)}$ we take the solution of $$ \cL_0(\bX) = \sum_{d''=1}^{\bar
  d} \sum_{r=1}^{n} F_{d'',d}^{(r)} \bS_{d''}^{(n-r)} - \cL_1(\bS_{d}^{(n-1)}) $$
such that, for all $d'$, $\Tr{\bJ_{d'} \bX}=0$:
\begin{multline}\label{eq:Sn}
 \bS_{d}^{(n)} = \ocR\left( \cL_1(\bS_{d}^{(n-1)}) - \sum_{d''=1}^{\bar d} \sum_{r=1}^{n} F_{d'',d}^{(r)}
 \bS_{d''}^{(n-r)} \right)
 \\
 = \int_{0}^{+\infty} e^{s\cL_0}\left( \cL_1(\bS_{d}^{(n-1)}) - \sum_{d''=1}^{\bar d} \sum_{r=1}^{n} F_{d'',d}^{(r)}
 \bS_{d''}^{(n-r)} \right) ~ds
.
\end{multline}
Since $\ocK(\bS_{d''}^{n-r})=0$ for any $r\in\{1,n-1\}$ and $\ocK\big( \cL_1(\bS_{d}^{(n-1)})\big) = \sum_{d''=1}^{\bar d} F_{d'',d}^{(n)} \bS_{d''}^{(0)}$, the above integral is absolutely convergent.

With such asymptotic expansion, we get an order $n$ approximation of the
dynamics on the invariant slow-manifold $\cD_\epsilon$, a reduced dynamical
model of~\eqref{eq:dynL0L1} based on the following $\bar d$ dimensional linear
system:
\begin{equation}\label{eq:dynX}
 \dotex x(t) =\left(\sum_{r=1}^{n} \epsilon^r F^{(r)}\right) x(t)
\end{equation}
where $\rho_t = \sum_{d=1}^{\bar d} x_{d}(t) \left(\sum_{r=0}^{n} \epsilon^r \bS_d^{(r)}\right) $ satisfies~\eqref{eq:dynL0L1} up to $\epsilon^{n+1}$ terms. Here $F^{(r)}$ is the matrix of real entries $F^{(r)}_{d,d'}$. Since $x_{d}=\Tr{\bJ_d \rho_t}$, the dynamical system~\eqref{eq:dynX} is an approximation of order $n$ for the reduced model slow dynamics of the nominal invariant operators $\bJ_d$:
Up-to $\epsilon^{n+1}$ corrections we have in the reduced model picture:
\begin{equation*} \begin{split}
  & \forall d\in\{1, \ldots,\bar d\}, \\
 & \dotex \bJ_d \triangleq \cL_{0}^*(\bJ_d) + \epsilon \cL_{1}^*(\bJ_d) = \sum_{d'=1}^{\bar d} \sum_{r=1}^{n} \epsilon^r F^{(r)}_{d,d'} \bJ_{d'} + O(\epsilon^{n+1}).
\end{split}\end{equation*}

\subsection{Z-gate simulations up to order 5}
\label{ssec:Zgate_order5}

\begin{figure*}
 \centering
 \includegraphics[width=\textwidth]{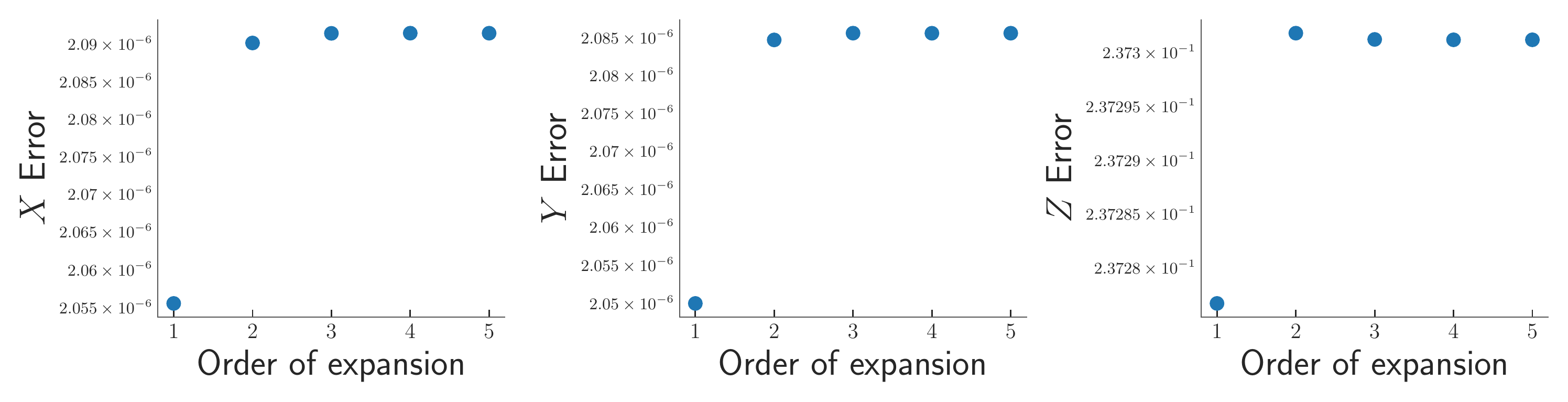}
 \caption{
  Convergence of the $X$, $Y$ and $Z$ error probabilities by increasing the order of the pertubative analysis from 1 to 5.
  The error probabilities are computed for a cat qubit on which we perform a Z gate as in Sec.~\ref{ssec:Zgate} with $\nbar=4$.
  The second-order expansion is already very accurate, and the third-order expansion is almost indistinguishable from higher order expansions.
 }
 \label{fig:order_expansion}
\end{figure*}

\begin{figure}
  \centering
  \begin{subcaptiongroup}
  \subcaptionlistentry{}
  % \label{fig:Z_final_leakage}
  \begin{overpic}[
   width=0.35\textwidth,
   % width=100pt,
   % grid
   ]{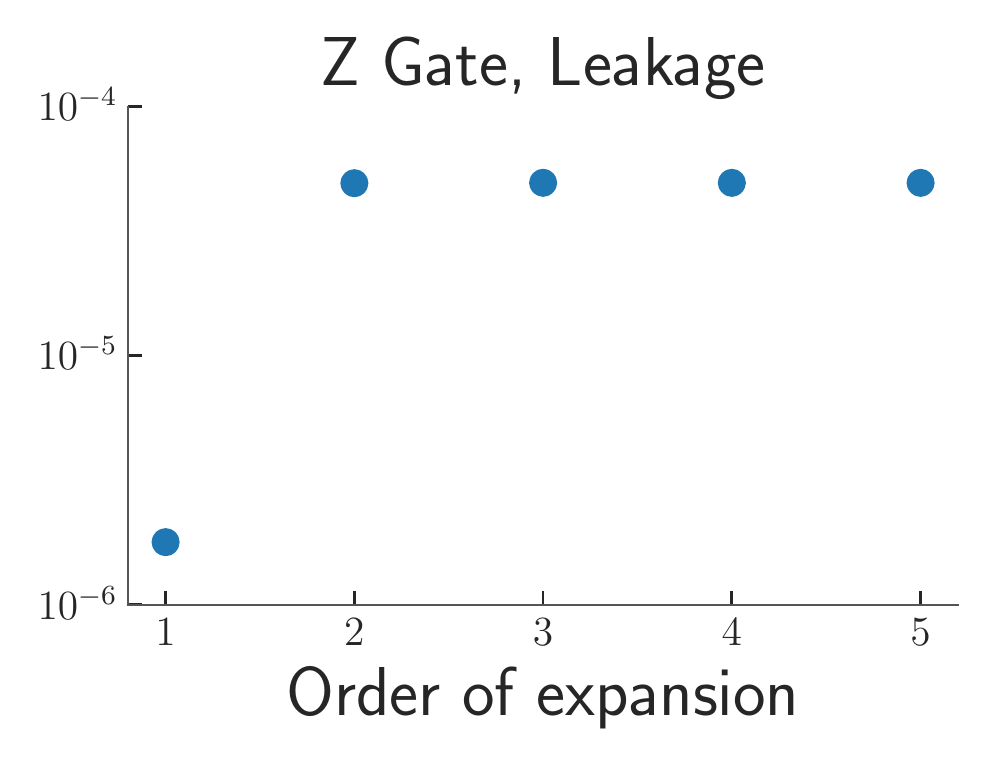}
   % ]{example-image-a}
  % \put(5,70){\captiontext*{}}
  \end{overpic}
  \end{subcaptiongroup}
  \captionsetup{subrefformat=parens}
  \caption{
   Up to $5^{th}$-order leakage of the Z gate starting from the $\Cp$ and ending in $\Cm$ for a cat-qubit with $\nbar=4$ and the same simulation parameters as in Sec.~\ref{ssec:Zgate}.
  All orders $\geq 2$ are superposed.
  % Contributions to cat qubit the code space $\bI_c$ of the $d=1$ to $d=4$ different basis vectors $\bS_{d}^{(r)}$ for orders of expansion $r=1$ to $r=5$: $\Tr{\bI_c \bS_{d}^{(r)}}$ for the Z gate with $\nbar=4$ and the same simulation parameters as in Sec.~\ref{sec:applications-z}.
  }
  \label{fig:leakage}
 \end{figure}

An example of such higher order approximation using Eqs.~\ref{eq:Sn} and \ref{eq:dynX} is given in figure~\ref{fig:order_expansion} and figure~\ref{fig:leakage} for the case of a cat-qubit on which we perform a Z gate as in Sec.~\ref{ssec:Zgate} with $\nbar=4$.
For the error probabilities, we see that the second-order expansion is already very accurate, and the third-order expansion is almost indistinguishable from higher order expansions.
Regarding leakage, we see that first-order leakage is not enough to capture the leakage dynamics, but that second-order leakage is already very accurate and indistinguishable from higher order expansions figure~\ref{fig:leakage}.

\section{Second-order approximation with slow time dependency} \label{sec:TimeVarying}

 Here we only derive the second-order approximation with slow time dependency.
We are looking for solutions of the perturbed system
\begin{equation}
 \dotex \rho_t = \cL_0(\rho_t) + \epsilon \cL_1(\epsilon t, \rho_t)
\end{equation}
based on the following asymptotic expansion:
$
 \rho_t=\sum_{d=1}^{d_0} x_{d, t}\left( \bS_{d}^{(0)} + \bS_{d}^{(1)}\right)
$
where
$$X_{t+1}=\left(F^{(0)}+ F^{(1)}+ F^{(2)} \right) X_t$$
with $F^{(0)}=I$, $X_t=\left(x_{1, t}, \ldots, x_{d_0, t}\right)^T$.
We assume that the GKSL superoperator $\cL_1$ in~\eqref{eq:dynL0L1} depends
slowly on time, i.e., that the operators $\bH_1$ and $\bL_{1,\nu}$ are smooth
functions of $\epsilon t$:
\begin{equation*}
  \begin{split}
 & \cL_{1}(\epsilon t,\rho)= - i [\bH_1(\epsilon
   t),\rho] + \sum_{\nu} \bL_{1,\nu}(\epsilon t) \rho \bL_{1,\nu}^\dag(\epsilon t) \\
 & - \tfrac{1}{2} \Big( \bL_{1,\nu}^\dag(\epsilon t) \bL_{1,\nu}(\epsilon t)\rho+
 \rho \bL_{1,\nu}^\dag(\epsilon t) \bL_{1,\nu}(\epsilon t) \Big).
  \end{split}
\end{equation*}
Then for
each $n$, $F^{(n)}_{d',d'}$ and $\bS^{(n)}_d$ depend also on $\epsilon t$. Thus, the invariance condition~\eqref{eq:InvCond} becomes $$ \sum_{d=1}^{\bar d}\left(
 \frac{dx_d}{dt} ~ \bS_{d}(\epsilon) + x_ d \dotex \bS_{d}(\epsilon)\right) =
 \left(\cL_0+ \epsilon \cL_1\right)\left(\sum_{d=1}^{\bar d} x_d
 \bS_{d}(\epsilon) \right) $$ where $ F_{d,d'}(\epsilon t, \epsilon)
 =\sum_{n\geq 0} \epsilon^n F_{d,d'}^{(n)}(\epsilon t)$ and $ \bS_{d}(\epsilon
 t,\epsilon) = \sum_{n\geq 0} \epsilon^n \bS_{d}^{(n)}(\epsilon t)$. One has to
identify terms with same orders versus $\epsilon$ in
\begin{multline}
 \forall d
 \in\{1,\ldots,\bar d\},\\
 \sum_{n\geq 0}\epsilon^{n} \dotex \bS_{d}^{(n)} +
 \mathlarger{\mathlarger{\sum}}_{d'=1}^{\bar d} \left(\sum_{n\geq 0} \epsilon^n
 F_{d',d}^{(n)}\right) \left(\sum_{n'\geq 0}\epsilon^{n'} \bS_{d'}^{(n')}
 \right) \\
  = \left(\cL_0 +\epsilon\cL_1\right) \left(\sum_{n\geq 0}
 \epsilon^n\bS_{d}^{(n)} \right) ,
\end{multline}
using the fact that, for each $n$, $\dotex
 \bS^{(n)}(\epsilon t)$ is of order $\epsilon$. .

The zero-order condition is satisfied with $F^{(0)}_{d,d'}=0$ and
$\bS_{d}^{(0)} = \bS_d$. First-order condition remains unchanged and yields as
in~\eqref{eq:F1} and~\eqref{eq:S1} to
$$ F_{d',d}^{(1)}(\epsilon t) =
 \Tr{\bJ_{d'} \cL_1(\epsilon t, \bS_{d})}
$$
with

\begin{equation}
 \bS_{d}^{(1)}(\epsilon t)=
 \int_{0}^{+\infty} e^{s\cL_0}\left(\cL_1(\epsilon t, \bS_d) -
 \ocK\big(\cL_1(\epsilon t,\bS_d) \big) \right) ~ds
 \label{eq:S1_t}
\end{equation}
where
$\Tr{\bJ_{d'}\bS_{d}^{(1)}(\epsilon t)} = 0$ and thus $\Tr{\bJ_{d'}~\dotex
  \bS_{d}^{(1)}(\epsilon t)} = 0$, for all $d'$ and $t$.
The second-order
condition is:
\begin{multline}
 \forall d \in\{1,\ldots,\bar d\},\\
  \frac{d }{ d(\epsilon
  t)} \bS^{(1)}_d(\epsilon t)+ \sum_{d''=1}^{\bar d} \left(
 F_{d'',d}^{(1)}(\epsilon t)\bS_{d''}^{(1)}(\epsilon t)+
 F_{d'',d}^{(2)}\bS_{d''} \right)
 \\
 = \cL_0(\bS_{d}^{(2)}) + \cL_1(\epsilon
 t,\bS_{d}^{(1)}(\epsilon t)).
\end{multline}
Multiplying by $\bJ_{d'}$ and tacking the
trace show that the second-order correction formula is identical to the one for
time-invariant $\cL_1$. To summarize, we have either for time-invariant or
slowly time-varying $\cL_1$, the following second-order approximation formula
for the dynamics of $x$:
\begin{equation}\label{eq:dyn2X}
 \forall d'\in\{1,\ldots,\bar d\}, \quad \dotex x_{d'} = \sum_{d=1}^{\bar d} \left(\epsilon F_{d',d}^{(1)}(\epsilon t) + \epsilon^2 F_{d',d}^{(2)}(\epsilon t)\right) x_d
\end{equation}
with
\begin{multline}\label{eq:F12_time}
 F_{d',d}^{(1)}(\epsilon t) = \Tr{\bJ_{d'} \cL_1(\epsilon t, \bS_{d})}, \\
  F^{(2)}_{d',d}(\epsilon t)
 = \Tr{ \cL_1^{*} (\epsilon t, \bJ_{d'}) ~ \ocR\Big( \cL_1(\epsilon t, \bS_{d})\Big) }
\end{multline}
where $\ocR$ is defined in~\eqref{eq:R}.

\section{Time discretization of  continuous-time quantum master equation} \label{sec:kraus_map}

We propose here an adapted numerical scheme to convert the continuous-time
dynamics~\eqref{eq:dynL0L1} into a discrete-time dynamic~\eqref{eq:dynK0K1}.

Take a time-step $\dt>0$ very small compared to evolution time-constant
of~\eqref{eq:dynL0L1}. An exact quantum channel approximation of $e^{\dt\cL_0}$
identical, up to $\dt^2$ terms to the explicit Euler scheme, is the following
(see~\cite[appendix B]{JordanChantasriRouchonEtAl2016}):
\begin{multline}\label{eq:DynDiscrete}
 \rho_{t+\dt}=\cK_0(\rho_t) \\
  \Big( = e^{\dt\, \cL_0}(\rho_t)+ O(\dt^2) = \rho_t + \dt ~\cL_0(\rho_t) + O(\dt^2) \Big)
\end{multline}
where $\cK_0$ admits the following Kraus structure:
\begin{equation} \begin{split}
 &\cK_0(\rho)= \bU_0 \left( \mathbf{\bM}_0 ~ \bU_0\rho \bU_0^\dag~ \mathbf{\bM}_0^\dag \right.\\
  &\left. + \dt \left(\sum_{\nu} \mathbf{\bL}_{0,\nu} ~\bU_0\rho \bU_0^\dag ~ \mathbf{\bL}^\dag_{0,\nu}\right) \right)\bU_0^\dag
\end{split}\end{equation}
with
\begin{equation} \begin{split}
  \bU_0= e^{-i\dt \bH_0/2}, \quad \mathbf{\bM}_0=\bM_0\bW_0^{-1/2}, \quad \mathbf{\bL}_{0,\nu}= \bL_{0,\nu}~ \bW_0^{-1/2}
\end{split}\end{equation}
where
\begin{equation*} \begin{split}
  &\bM_0= I- \sum_\nu \tfrac{\dt}{2} \bL_{0,\nu}^\dag \bL_{0,\nu}, \\
  & \bW_0 = \bM_0^\dag \bM_0 + \dt \sum_{\nu} \bL_{0,\nu}^\dag \bL_{0,\nu}
.
\end{split}\end{equation*}
Take as perturbation $\cK_1$ the simplest approximation:
\begin{equation} \begin{split}
  \cK_1(\rho) = \dt \cL_1(\rho)
.
\end{split}\end{equation}

\section{Adiabatic elimination in  discrete-time }\label{sec:discrete}

\subsection{Single system}
\label{sec:discrete_single}

When $t$ is an integer,~\eqref{eq:dynL0L1} is replaced by
\begin{equation}\label{eq:dynK0K1}
 \rho_{t+1}= \cK_0(\rho_t) + \epsilon \cK_1(\rho_t)
\end{equation}
where $\cK_0$ is a quantum channel stabilizing the subspace $\cD_0$ spanned by the orthonormal basis $\bS_d$ and with invariant operator $\bJ_d=\lim_{t\mapsto +\infty} (\cK_0^*)^t(\bS_d)$ where $(\cK_0^*)^t$ corresponds to $t$ iterates of the adjoint map $\cK_0^*$.
%Notice that for $\epsilon \geq 0$, $\cK_0+\epsilon \cK_1$ is a quantum channel. In particular $\Tr{\cK_1(\rho)}=0$ for any $\rho$.
For any $\rho_0$ we have
\begin{equation} \begin{split}
  \lim_{t\mapsto +\infty} (\cK_0)^t(\rho_0)= \ocK(\rho_0) =\sum_{d} \Tr{\bJ_d \rho_0} \bS_d
. \end{split}\end{equation}
Invariance condition~\eqref{eq:InvCond} reads then
\begin{equation} \begin{split}
  \sum_{d=1}^{\bar d} x_d(t+1) ~ \bS_{d}(\epsilon)
 = \left(\cK_0+ \epsilon \cK_1\right)\left(\sum_{d=1}^{\bar d} x_d(t) \bS_{d}(\epsilon)
 \right)
\end{split}\end{equation}
with $x_d(t+1)= \sum_{d'}F_{d,d'}(\epsilon) x_{d'}(t)$. Combined with the series expansion of $ \bS_{d}(\epsilon)$ and $F_{d,d'}(\epsilon)$
it yields:
\begin{equation*} \begin{split}
  & \forall d \in\{1,\ldots,\bar d\},\\
  & \sum_{d'=1}^{\bar d} \left(\sum_{n\geq 0} \epsilon^n F_{d',d}^{(n)}\right) \left(\sum_{n'\geq 0}\epsilon^{n'} \bS_{d'}^{(n')} \right) \\
  & = \left(\cK_0 +\epsilon\cK_1\right)
 \left(\sum_{n\geq 0} \epsilon^n\bS_{d}^{(n)} \right).
\end{split}\end{equation*}
The zero-order term is satisfied with $F^{(0)}_{d,d'}=\delta_{d,d'}$ and $\bS_{d}^{(0)} = \bS_d$. First-order conditions read
\begin{equation*} \begin{split}
  &\forall d \in\{1,\ldots,\bar d\},\\
  & \bS_d^{(1)}+ \sum_{d''=1}^{\bar d} F_{d'',d}^{(1)} \bS_{d''}^{(0)} = \cK_0(\bS_{d}^{(1)} ) + \cK_1( \bS_{d}^{(0)})
.
\end{split}\end{equation*}
Left multiplication by operator $\bJ_{d'}$ and taking the trace yields
\begin{equation}\label{eq:F1discrete}
 F_{d',d}^{(1)} = \Tr{\bJ_{d'} \cK_1( \bS_{d})}
\end{equation}
since $\Tr{\bJ_{d'} \bS_{d''}^{(0)}}= \delta_{d',d''}$ and $\Tr{\bJ_{d'}\cK_0(\bW)}=\Tr{\bJ_{d'} \bW}$ for any operator $\bW$ because $\cK_0^*(\bJ_{d'})=\bJ_{d'}$.
Thus, $\bS_{d}^{(1)}$ is a solution $\bX$ of the following equation:
$$
  \bX = \cK_0(\bX ) + \cK_1( \bS_{d}^{(0)}) - \sum_{d''=1}^{\bar d} F_{d'',d}^{(1)} \bS_{d''}^{(0)}
$$
Since the quantum channel $\cK_0$ is a  contraction with a  rate assumed to be strictly less than 1,  the following solution is chosen,
$$
  \bS_{d}^{(1)} = \sum_{s\geq 0} (\cK_0)^s \left(\cK_1(\bS_d^{(0)}) -\ocK\big(\cK_1(\bS_d^{(0)}) \big) \right)
  ,
$$
based on this absolutely converging series and satisfying
$\Tr{\bJ_{d'} \bS_{d}^{(1)}}=0$ for all $d'$.
This defines the superoperator
$$
  \ocR(\bW)= \sum_{s=0}^{+\infty} (\cK_0)^s \big(\bW - \ocK(\bW)\big)
$$
where $ (\cK_0)^0$ stands for identity.

Take $n\geq 2$ and assume that we have computed all the terms $F_{d',d}^{(r)}$
and $\bS_{d'}^{(r)}$ of order $r< n$ with $\Tr{\bJ_{d'} \bS_{d}^{(r)}}=0$ for
all $d$ and $d'$. Invariance condition of order $n$ reads
\begin{multline*}
\forall d \in\{1,\ldots,\bar d\},\\
 \bS_d^{(n)}+ \sum_{d''=1}^{\bar d} \sum_{r=1}^{n}
 F_{d'',d}^{(r)} \bS_{d''}^{(n-r)} = \cK_0(\bS_{d}^{(n)}) +
 \cK_1(\bS_{d}^{(n-1)}).
\end{multline*}

Left multiplication by operator $\bJ_{d'}$ and
taking the trace yields $$ F_{d',d}^{(n)} = \Tr{\bJ_{d'} \cK_1(
  \bS_{d}^{(n-1)})}= \Tr{\cK_1^{*}(\bJ_{d'}) ~\bS_{d}^{(n-1)}}. $$ For
$\bS_{d}^{(n)}$ we take the solution such that, for all $d'$, $\Tr{\bJ_{d'}
  \bS_{d}^{(n)}}=0$:
\begin{multline*}
 \bS_{d}^{(n)} = \ocR\left( \cK_1(\bS_{d}^{(n-1)}) -
 \sum_{d''=1}^{\bar d} \sum_{r=1}^{n} F_{d'',d}^{(r)} \bS_{d''}^{(n-r)} \right)
 \ldots \\ \ldots
 =
 \sum_{s\geq 0}(\cK_0)^s\left( \cK_1(\bS_{d}^{(n-1)}) - \sum_{d''=1}^{\bar d}
 \sum_{r=1}^{n} F_{d'',d}^{(r)} \bS_{d''}^{(n-r)} \right).
\end{multline*}

The discrete-time reduced model is then
\begin{equation}\label{eq:dynXdiscrete}
 x(t+1) = x(t) + \left(\sum_{r=1}^{n} \epsilon^rF^{(r)}\right) x(t)
\end{equation}
with
$
 \rho_t = \sum_{d=1}^{\bar d} x_{d}(t) \left(\sum_{r=0}^{n} \epsilon^r\bS_d^{(r)}\right)
$
satisfying~\eqref{eq:dynK0K1} up-to $\epsilon^{n+1}$ correction and for any $d$, $x_{d}(t)=\Tr{\bJ_d \rho_t}$.
Up-to $\epsilon^{n+1}$ corrections, we have the following reduced model dynamics for the invariant operators
\begin{multline*}
  \forall d\in\{1, \ldots,\bar d\}, \\
 \bJ_d(t+1) \triangleq \cK_{0}^*(\bJ_d(t)) + \epsilon \cK_{1}^*(\bJ_d(t)) \\ = \bJ_d(t) + \sum_{d'=1}^{\bar d} \sum_{r=1}^{n} \epsilon^r F^{(r)}_{d,d'} \bJ_{d'}(t) + O(\epsilon^{n+1}).
\end{multline*}

The discrete-time version of equation~\eqref{eq:F2} providing the second-order
approximation reads
\begin{multline}\label{eq:F12discrete}
 F_{d',d}^{(1)} = \Tr{\bJ_{d'} \cK_1(\bS_{d})}, \\ F^{(2)}_{d',d}
 = \Tr{ \cK_1^{*} (\bJ_{d'}) ~ \ocR \Big( \cK_1(\bS_{d})\Big) }
\end{multline}
and remains valid for slowly time-varying perturbation, i.e., for $\cK_1(\epsilon t, \rho)$ where the dependence versus $\epsilon t$ of $\cK_1$ is smooth:
\begin{multline}\label{eq:F12discrete_time_varying}
 F_{d',d}^{(1)}(\epsilon t) = \Tr{\bJ_{d'} \cK_1(\epsilon t, \bS_{d})}, \\ F^{(2)}_{d',d}(\epsilon t)
 = \Tr{ \cK_1^{*} (\epsilon t,\bJ_{d'}) ~ \ocR \Big( \cK_1(\epsilon t,\bS_{d})\Big) }
\end{multline}

\subsection{Composite systems}
\label{sec:discrete_composite}

Discrete-time bipartite structure is based on
\begin{equation} \begin{split}
  \cK_0=\cK_{A,0}\otimes\cK_{B,0}
\end{split}\end{equation}
 where $\cK_{A,0}$ and $\cK_{B,0}$ are local
quantum maps on $\cH_A$ and $\cH_B$ stabilizing the local subspaces $\cD_{A,0}$
and $\cD_{B,0}$. Their dimensions are $\bar d_A$ and $\bar d_B$ with
$(\bS_{A,d_A})_{1\leq d_A\leq \bar d_A}$ and $(\bS_{B,d_B})_{1\leq d_B\leq \bar
   d_B}$ as orthonormal basis of Hermitian operators. We assume that $\cK_{A,0}$
and $\cK_{B,0}$ ensure exponential convergence towards $\cD_{A,0}$ and
$\cD_{B,0}$: for any operators on $\cH$, $$ \lim_{t\mapsto
  +\infty}(\cK_{A,0})^t\otimes (\cK_{B,0})^t (\rho_0) = \ocK(\rho_0) $$ where
$\ocK$ remains given by~\eqref{eq:cKAB0} with $\bJ_{A,d_A}$ and $\bJ_{B,d_B}$
as follows:
\begin{multline*}
\bJ_{A,d_A} = \lim_{t\mapsto +\infty}
 (\cK^*_{A,0})^t(\bS_{A,d_A}), \\ \bJ_{B,d_B} = \lim_{t\mapsto +\infty}
 (\cK^*_{B,0})^t(\bS_{B,d_B}).
\end{multline*}
Assume the super operator $\cK_1$ only involve
finite sums of tensor products of operators on $\cH_A$ and $\cH_B$. This means
that for any $\bX_A$ and $\bX_B$ local operators on $\cH_A$ and $\cH_B$,
\begin{equation}\label{eq:K1AB}
 \cK_{1} (\bX_A\otimes \bX_B) = \sum_{\nu=1}^{\bar \nu} \bL_{A,\nu}\bX_A \bR_{A,\nu} \otimes \bL_{B,\nu}\bX_B \bR_{B,\nu}
\end{equation}
where $\bar \nu$ is a positive integer, where $\bL_{A,\nu}$, $\bR_{A,\nu}$ are operators on $\cH_A$ and where $\bL_{B,\nu}$, $\bR_{B,\nu}$ are operators on $\cH_B $.

The discrete-time analogue of~\eqref{eq:F1AB} reads:
\begin{multline}\label{eq:F1ABdiscrete}
 F^{(1)}_{(d'_A,d'_B),(d_A,d_B)} \\
 = \sum_{\nu=1}^{\bar\nu} \Tr{\bJ_{A,d'_A} \bS_{A,d_A,\nu}} \Tr{\bJ_{B,d'_B} \bS_{B,d_B,\nu}}
 \\ = \sum_{\nu=1}^{\bar\nu} \Tr{\bS_{A,d_A} \bJ_{A,d'_A,\nu}} \Tr{\bS_{B,d_B} \bJ_{B,d'_B,\nu}},
\end{multline}
where for $X=A,B$
\begin{equation} \begin{split}
  \bJ_{X,d'_X,\nu}= \bR_{X,\nu}\bJ_{X,d'_X} \bL_{X,\nu}, \quad \bS_{X,d_X,\nu}= \bL_{X,\nu}\bS_{X,d_X} \bR_{X,\nu}
.
\end{split}\end{equation}
Similarly, we derive from~\eqref{eq:F2AB} the second-order discrete-time matrix $F^{(2)}$:
\begin{multline} \label{eq:F2ABdiscrete}
 F^{(2)}_{(d'_A,d'_B),(d_A,d_B)} \\ =
 \mathlarger{\mathlarger{\sum}}_{\nu,\nu'=1}^{\bar\nu}
 \mathlarger{\sum}_{s=0}^{+\infty}
 \Bigg(
 \Tr{\bJ_{A,d'_A,\nu'}~(\cK_{A,0})^s\big(\bS_{A,d_A,\nu} \big)} \\
 \times \Tr{ \bJ_{B,d'_B,\nu'} ~ (\cK_{B,0})^s \big( \bS_{B,d_B,\nu} \big)} \ldots
 \\
 \ldots - G_{A,d'_A,d_A,\nu,\nu'}G_{B,d'_B,d_B,\nu,\nu'} \Bigg),
\end{multline}
where $$
 G_{X,d'_X,d_X,\nu,\nu'} = \sum_{d''_X} \Tr{\bJ_{X,d'_X}~ \bS_{X,d''_X,\nu'}} \Tr{\bJ_{X,d''_X} \bS_{X,d_X,\nu}}
$$ for $X=A,B$.

\section{Propagator simulation results} \label{sec:error_models}

\begin{figure}[!h]
 \centering
 \begin{subcaptiongroup}
  \subcaptionlistentry{}
  \label{fig:ScrodZZchi}
  \begin{overpic}[
    width=0.49\textwidth,
    % grid
    % ]{fig/comp_CNOT_bit-flip_nbsteps20000.pdf}
   ]{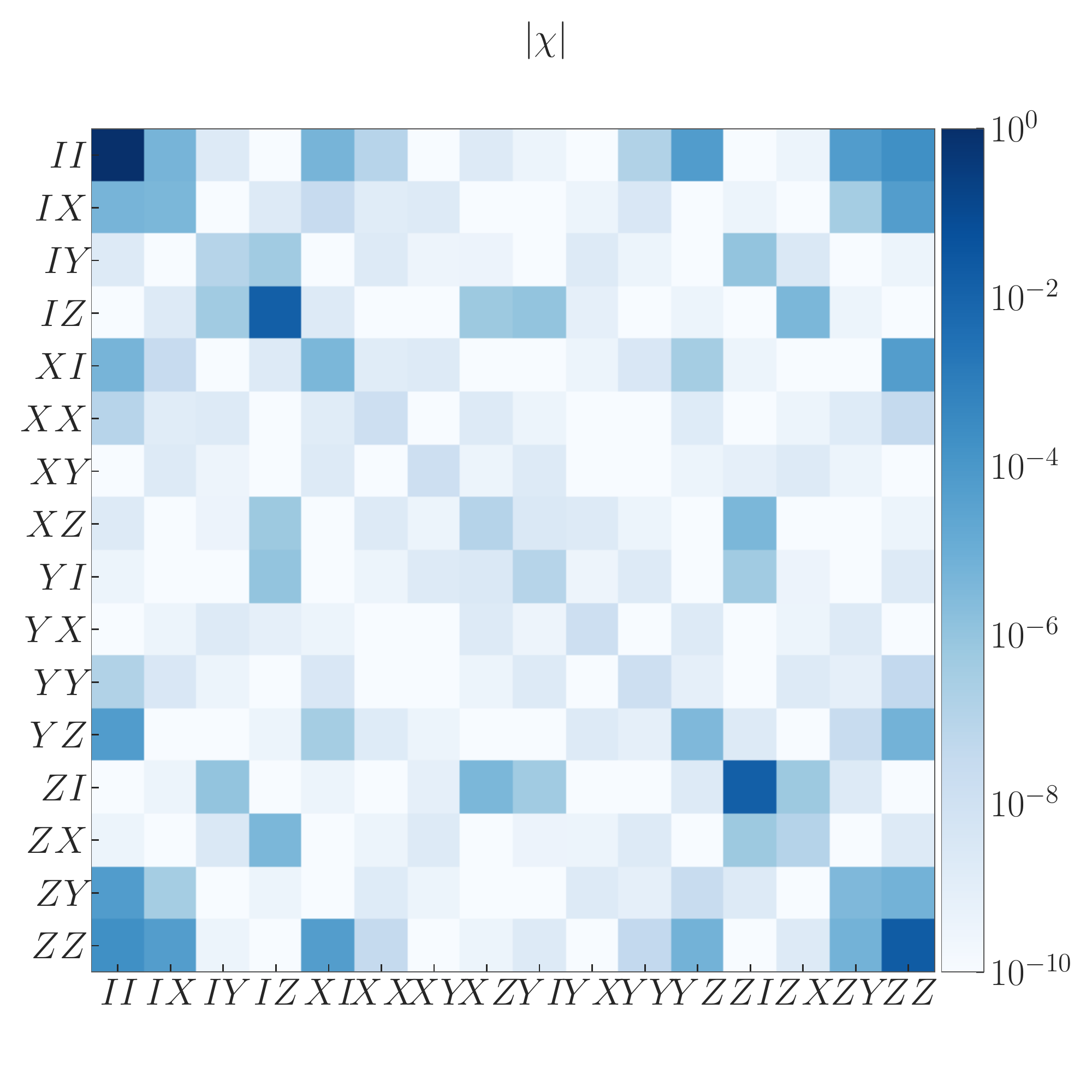}
   % ]{example-image-a}
   \put(0, 90){\captiontext*{}}
  \end{overpic}
  \subcaptionlistentry{}
  \label{fig:HeisZZchi}
  \begin{overpic}[
    width=0.49\textwidth,
    % grid
    % ]{fig/comp_CNOT_bit-flip_nbsteps20000.pdf}
   ]{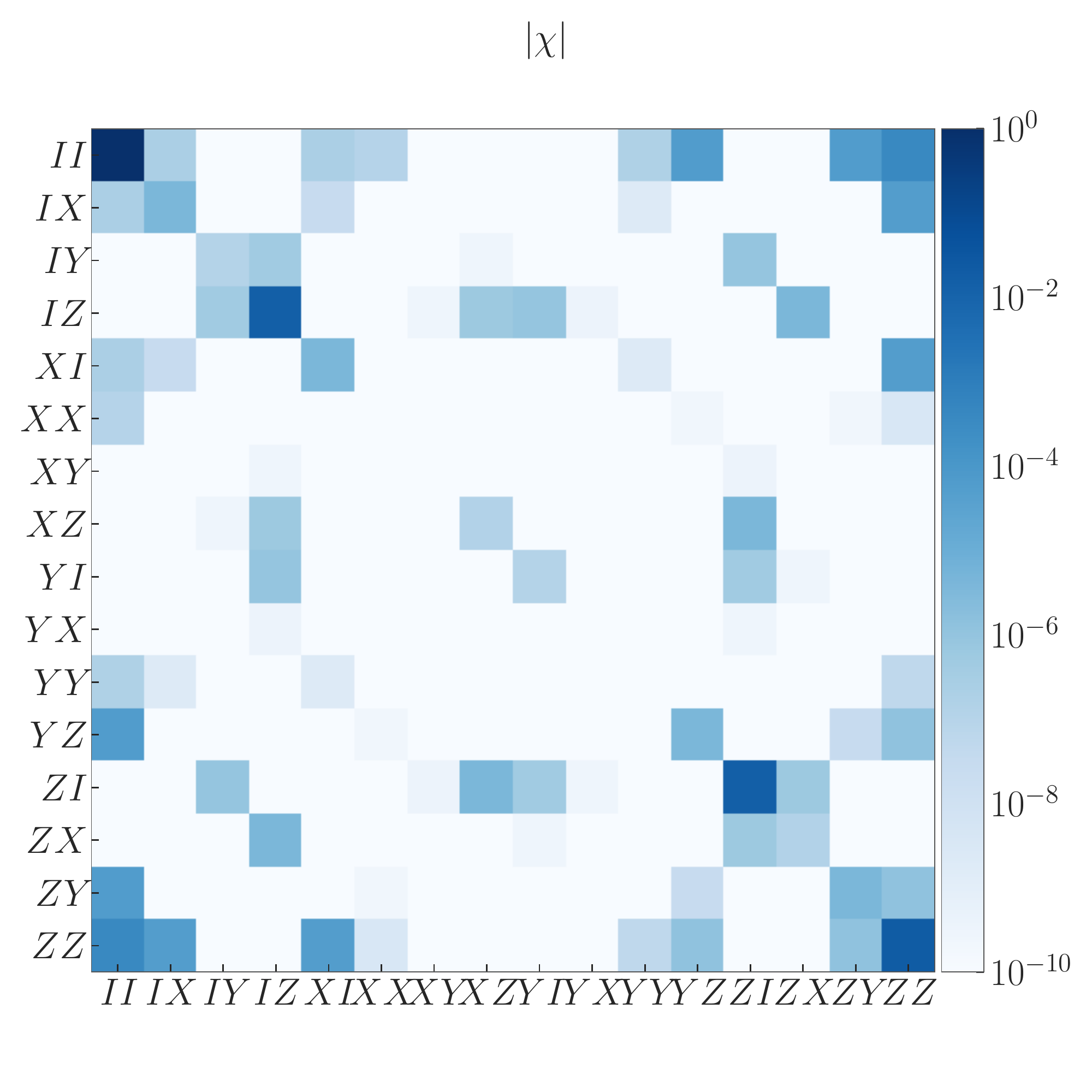}
   % ]{example-image-a}
   \put(0, 90){\captiontext*{}}
  \end{overpic}
 \end{subcaptiongroup}
 \captionsetup{subrefformat=parens}
 \caption{$\chi$ error-matrix  of the ZZ gate with \subref{fig:ScrodZZchi} full model, Galerkin truncation to $41$ photons and \subref{fig:HeisZZchi} second-order reduced model where $\alpha=2, \kappa_1=\frac{\kappa_2}{100}, \epsilon_Z=\frac{\kappa_2}{20}$.}
 \label{fig:ZZchi}
\end{figure}

\begin{figure*}
  \centering
  \begin{subcaptiongroup}
  \begin{overpic}[
   width=0.8\textwidth,
   % grid
   % ]{fig/comp_CNOT_bit-flip_nbsteps20000.pdf}
   ]{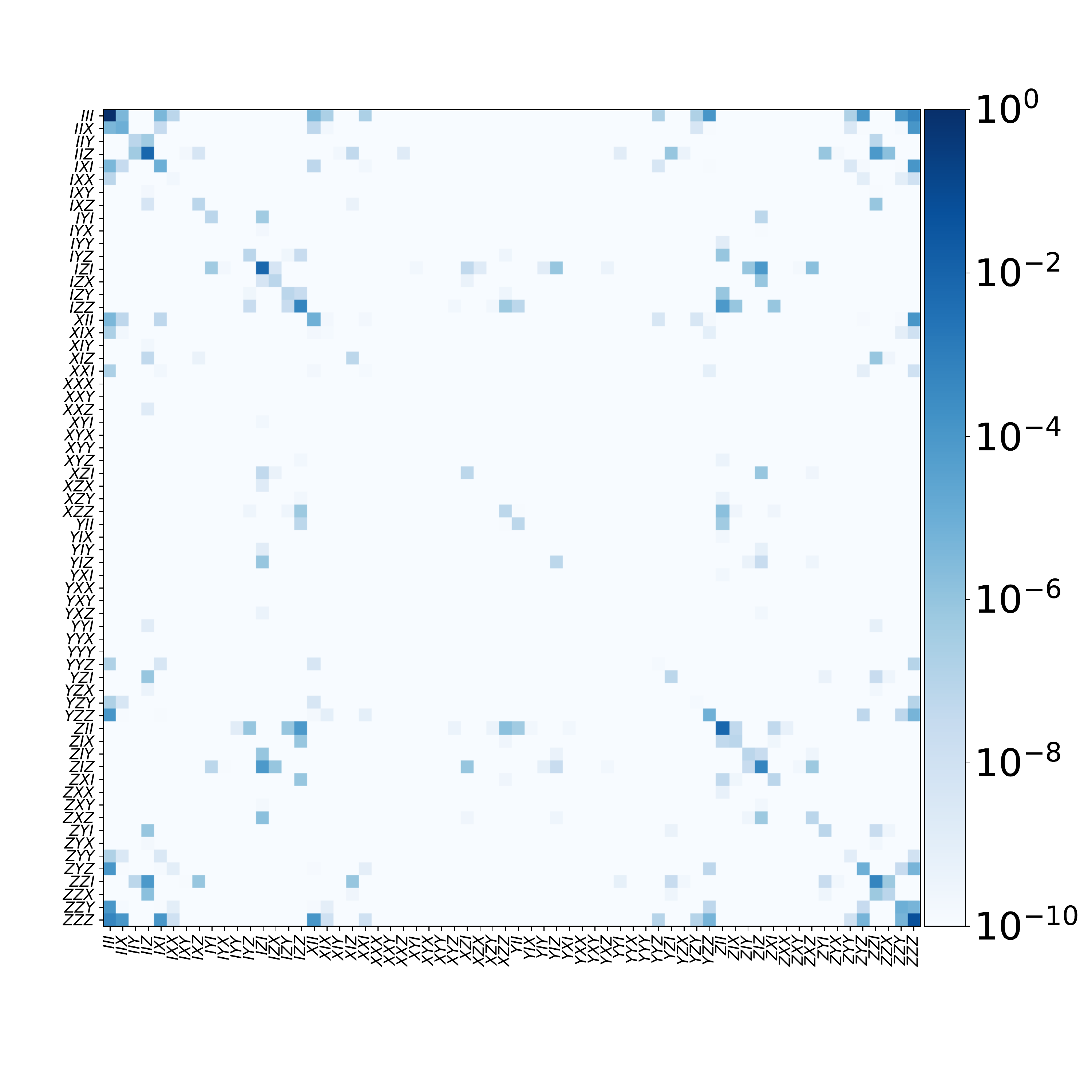}
   % ]{example-image-a}
  % \put(0, 90){\captiontext*{}}
  \put(48.5, 92){$|\chi|$}
  \end{overpic}
  \end{subcaptiongroup}
  \captionsetup{subrefformat=parens}
  \caption{
  $\chi$ error-matrix  of the ZZZ gate obtained with simulations based on the second-order reduced model where $\alpha=2, \kappa_1=\frac{\kappa_2}{100}, \epsilon_{ZZZ}=\frac{\kappa_2}{20}$.}
  \label{fig:ZZZchi}
 \end{figure*}

\begin{figure}[!h]
 \centering
 \begin{subcaptiongroup}
  \subcaptionlistentry{}
  \label{fig:ScrodCXchi}
  \begin{overpic}[
    width=0.49\textwidth,
    % grid
    % ]{fig/comp_CNOT_bit-flip_nbsteps20000.pdf}
   ]{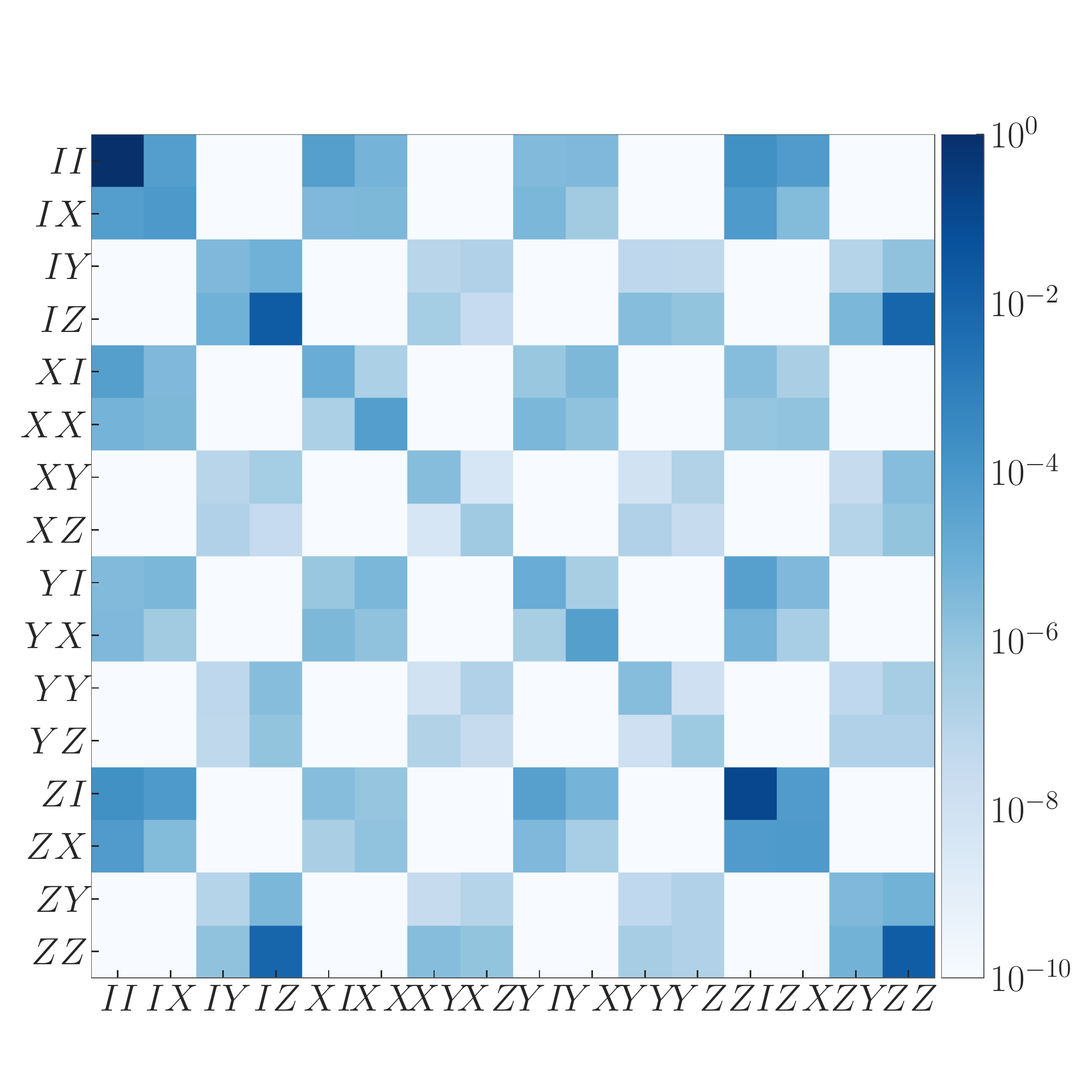}
   % ]{example-image-a}
   \put(0, 90){\captiontext*{}}
  \end{overpic}
  \subcaptionlistentry{}
  \label{fig:HeisCXchi}
  \begin{overpic}[
    width=0.49\textwidth,
    % grid
    % ]{fig/comp_CNOT_bit-flip_nbsteps20000.pdf}
   ]{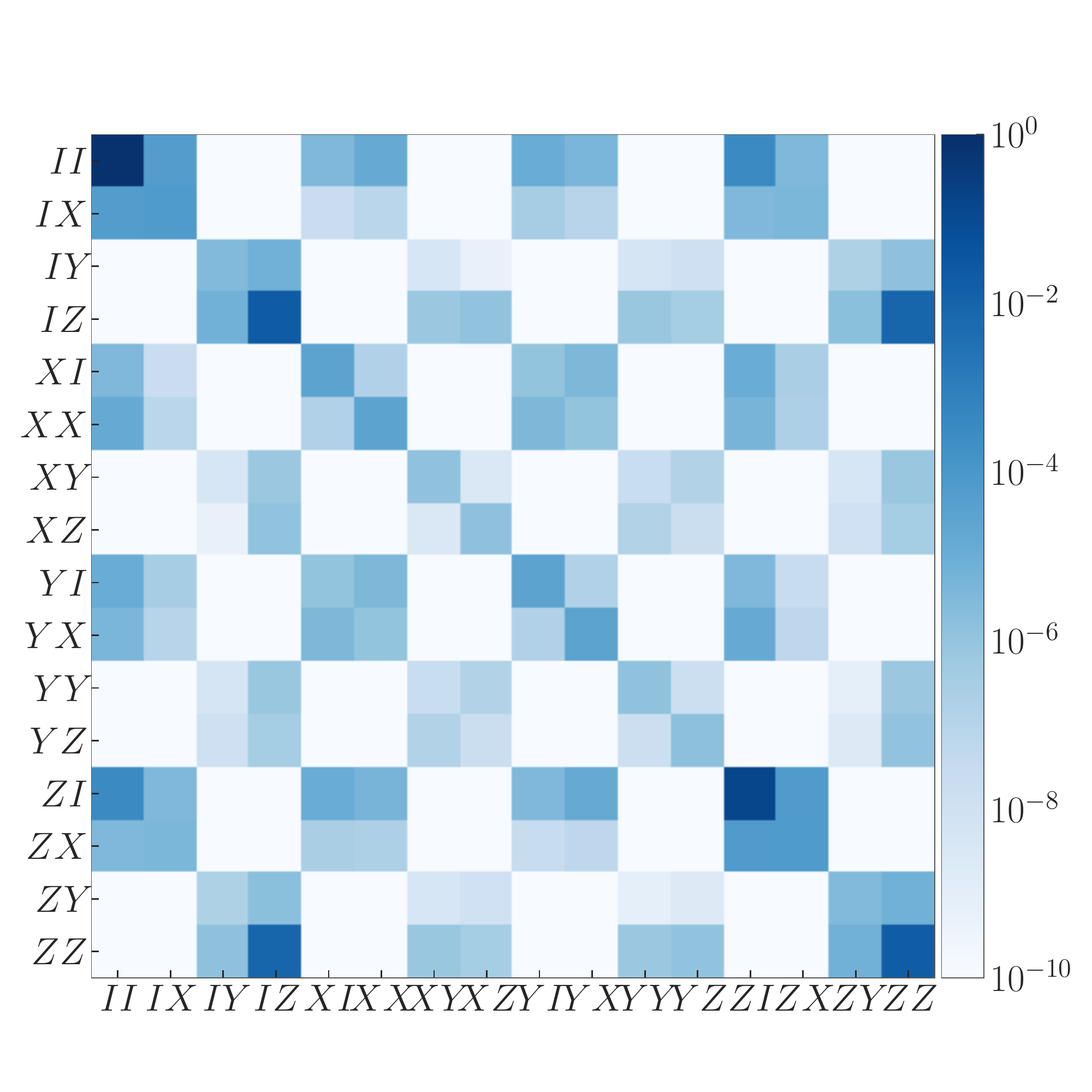}
   % ]{example-image-a}
   \put(0, 90){\captiontext*{}}
  \end{overpic}
 \end{subcaptiongroup}
 \captionsetup{subrefformat=parens}
 \caption{$\chi$ error-matrix of the CNOT gate with \subref{fig:ScrodCXchi} the full model and \subref{fig:HeisCXchi} the second-order reduced model where $\alpha=2$, and $ \kappa_1=\kappa_2/100 $}
 \label{fig:CNOTchi}
\end{figure}

\begin{figure*}
 \centering
 \begin{subfigure}{0.8\textwidth}
  \includegraphics[width=\textwidth]{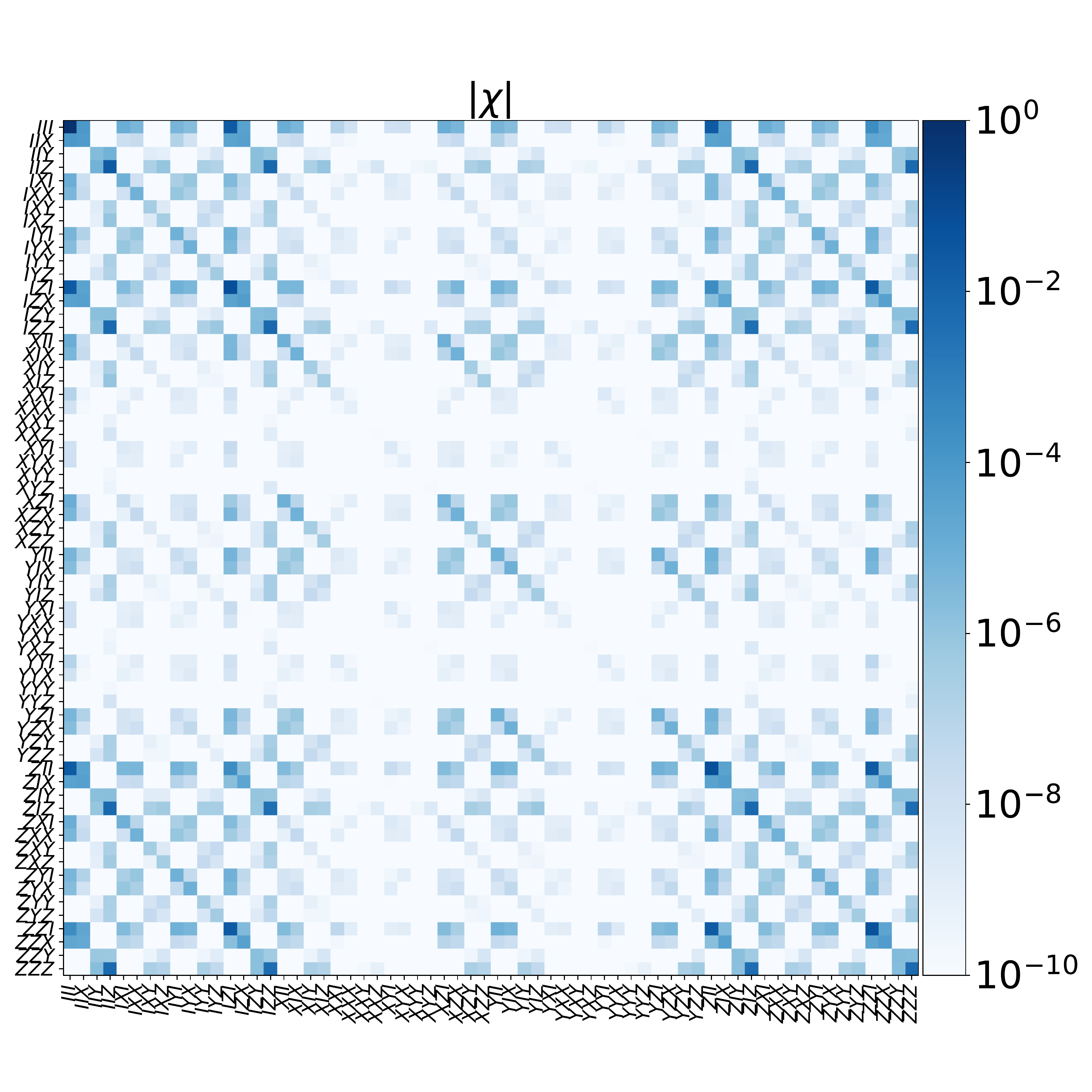}
 \end{subfigure}
 \caption{
  $\chi$ error-matrix of the CCNOT gate obtained with simulations based on the second-order reduced model where $\alpha=4, \kappa_1=\frac{\kappa_2}{100}$.
  The diagonal elements show the Pauli errors.
  The top left coefficient displays the fidelity of the gate at $82\%$.
  The bottom right coefficient displays the $ZZZ$ error of the gate at $0.5\%$.
 }
 \label{fig:CCNOTchi}
\end{figure*}

In this appendix, we give examples of $\chi$ error-matrices defined in~\eqref{eq:chiAB}  for the ZZ-gate  in figure~\ref{fig:ZZchi}, the ZZZ gate in figure~\ref{fig:ZZZchi}, the CNOT gate in figure~\ref{fig:CNOTchi} and the Toffoli gate in figure~\ref{fig:CCNOTchi} from which we extracted the Pauli error models shown in the main text, i.e., the diagonal of the $\chi$ error-matrix used  in  quantum process tomography.

% \subsection{ZZZ gate}

% \subsection{CNOT gate}

\section{Leakage computation}
\label{sec:leakage}

\subsection{Single-mode leakage} \label{sec:leakage_one_mode}

The equation~\eqref{eq:S1} allows to perform a first-order computation of the leakage, defined as the population outside the code space.
If we define $\bI_c$ to be the projector on the code space of our system, then the population of the state $\rho_t$ at a given time $t$ inside the code space is $\Tr{\bI_c \rho_t} $.
In the context of cat-qubit, the code space projector is defined by
$$
\bI_c =(\CpCp +\Cm \Cmd ) \sim \sqrt{2} \bS_1.
$$
So for any state written at first-order
$
\rho_t=\sum_{d=1}^{d_0} x_{d, t}\left(\bS_{d}^{(0)}+\bS_{d}^{(1)}\right)
$,
the leakage $l$ is given by:
\begin{equation}
l(t)= 1- \Tr{\bI_c \rho_t} = 1 - \sum_{d=1}^{d_0} x_{d, t} c_{d}
\label{eq:leakage_1}
\end{equation}
where
\begin{equation} \label{eq:leakage_one_mode}
c_{d} = \Tr{\bI_c (\bS_{d}^{(0)}+\bS_{d}^{(1)})} = \sqrt{2} \delta_{1, d}+ \Tr{\bI_c ~\ocR\big(\cL_1(\bS_d) \big)}.
\end{equation}

\subsection{Composite-system leakage} \label{sec:leakage_composite}

In the case of a composite system, we can still compute the leakage at first-order, using the generalization of equation~\eqref{eq:S1}.
For $(d_A, d_B)$, we define $c_{d_A, d_B}$ as:
$c_{d_A, d_B} = \Tr{\bI_{c, A} \bI_{c, B} \bS_{d_A, d_B}(\epsilon)} $
where $\bS_{d_A, d_B}(\epsilon) = \bS_{d_A, d_B}^{(0)} + \bS_{d_A, d_B}^{(1)}$ with $\bS_{d_A, d_B}^{(0)} = \bS_{A,d_A}^{(0)} \bS_{B,d_B}^{(0)}$ and
$$
\bS_{d_A, d_B}^{(1)} = \ocR\big(\cL_1(\bS_{d_A, d_B}^{(0)}) \big) = \sum_{\nu=1}^{\bar\nu}\ocR\Big(\bS_{A,d_A,\nu}\otimes\bS_{B,d_B,\nu}\Big).
$$

So we find that $c_{d_A, d_B} = 2 \delta_{1, d_A} \delta_{1, d_B} + \sum_{\nu=1}^{\bar\nu}\Tr{\bI_{c, A} \bI_{c, B} \ocR\Big(\bS_{A,d_A,\nu}\otimes\bS_{B,d_B,\nu}\Big)}$.

And finally, the leakage $l$ of a state $\rho_t = \sum_{d_A,d_B} x_{d_A, d_B, t} \bS_{d_A, d_B} $ is given by:
\begin{equation}
l(t)= 1- \Tr{\bI_c \rho_t} = 1 - \sum_{d_A, d_B=1}^{d_0} x_{d_A, d_B, t} c_{d_A, d_B}
\label{eq:leakage_2}
\end{equation}
However, second-order leakage can be obtained numerically  via the following relation
$$
\Tr{\bI_c \bS_{d}^{(2)}} =\Tr{\ocR^\star\left(\bI_c\right) \left(\cL_1(\bS_d^{(1)})-\sum_{d''=1}^{\bar d} F_{d'',d}^{(1)} \bS_{d''}^{(1)} \right)}
$$
 with only  local computations.
\begin{figure}
 \centering
 \begin{subcaptiongroup}
 \subcaptionlistentry{}
 \label{fig:Zleakage}
 \begin{overpic}[
  width=0.32\textwidth,
  % grid
  % ]{fig/Z_gate_leakage_first_order.pdf}
  ]{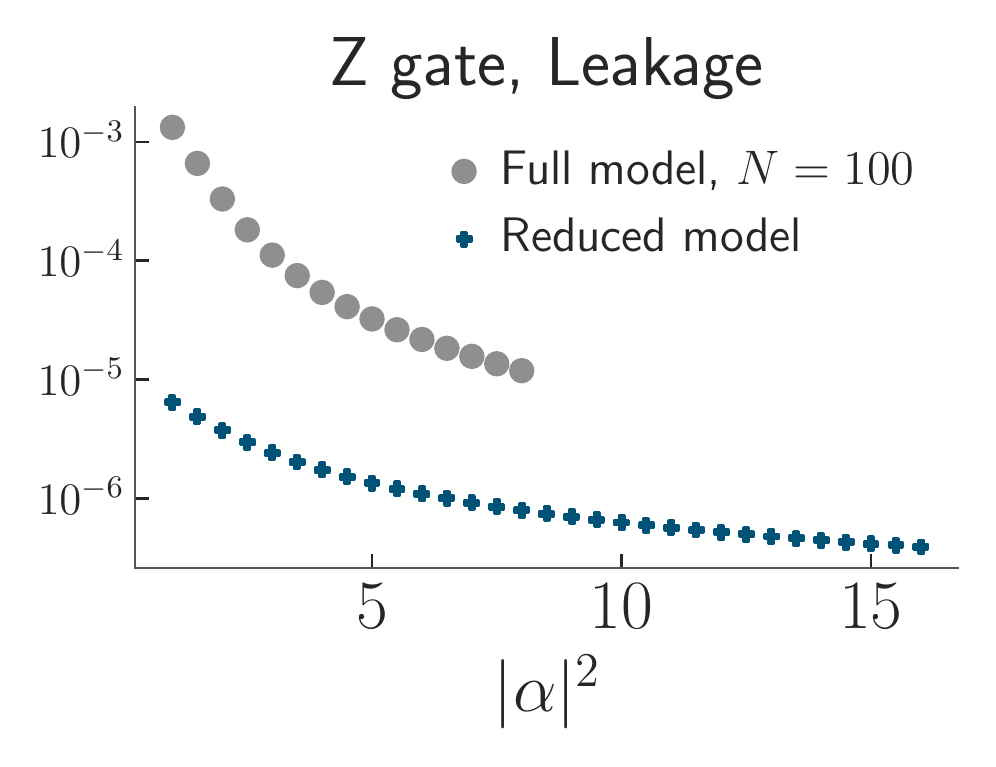}
  % ]{example-image-a}
 \put(5, 70){\captiontext*{}}
 \end{overpic}
 \subcaptionlistentry{}
 \label{fig:ZZleakage}
 \begin{overpic}[
  width=0.32\textwidth,
  % grid
  % ]{fig/Z_gate_leakage_first_order.pdf}
  ]{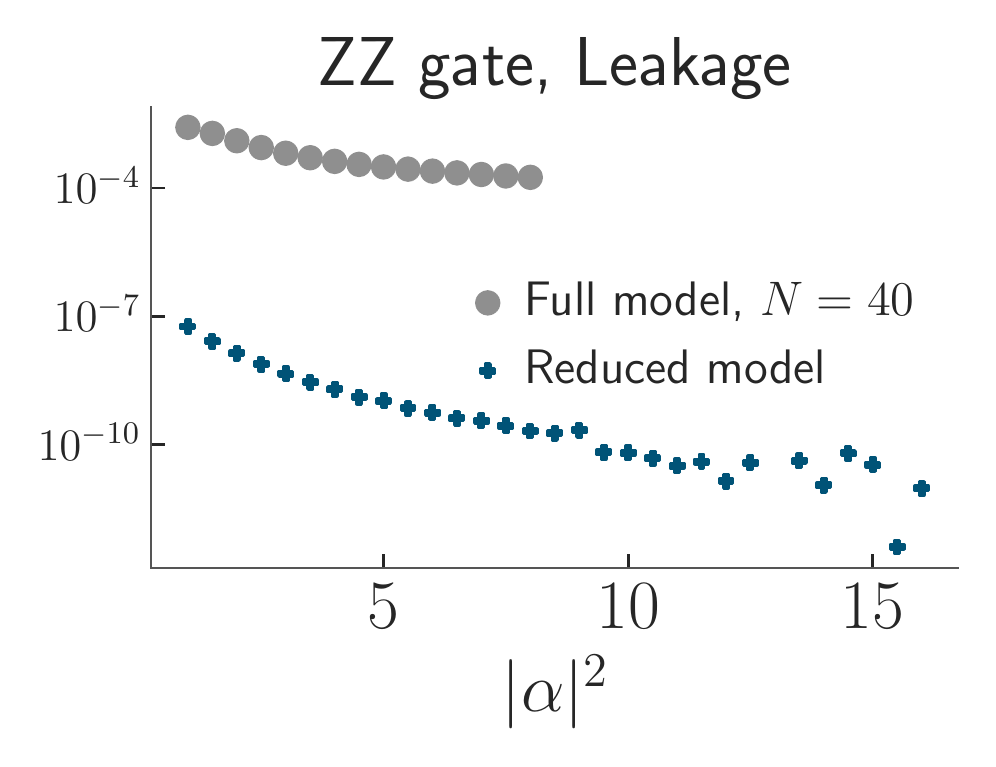}
  % ]{example-image-a}
 \put(5, 70){\captiontext*{}}
 \end{overpic}
 \subcaptionlistentry{}
 \label{fig:ZZZleakage}
 \begin{overpic}[
  width=0.32\textwidth,
  % grid
  % ]{fig/Z_gate_leakage_first_order.pdf}
  ]{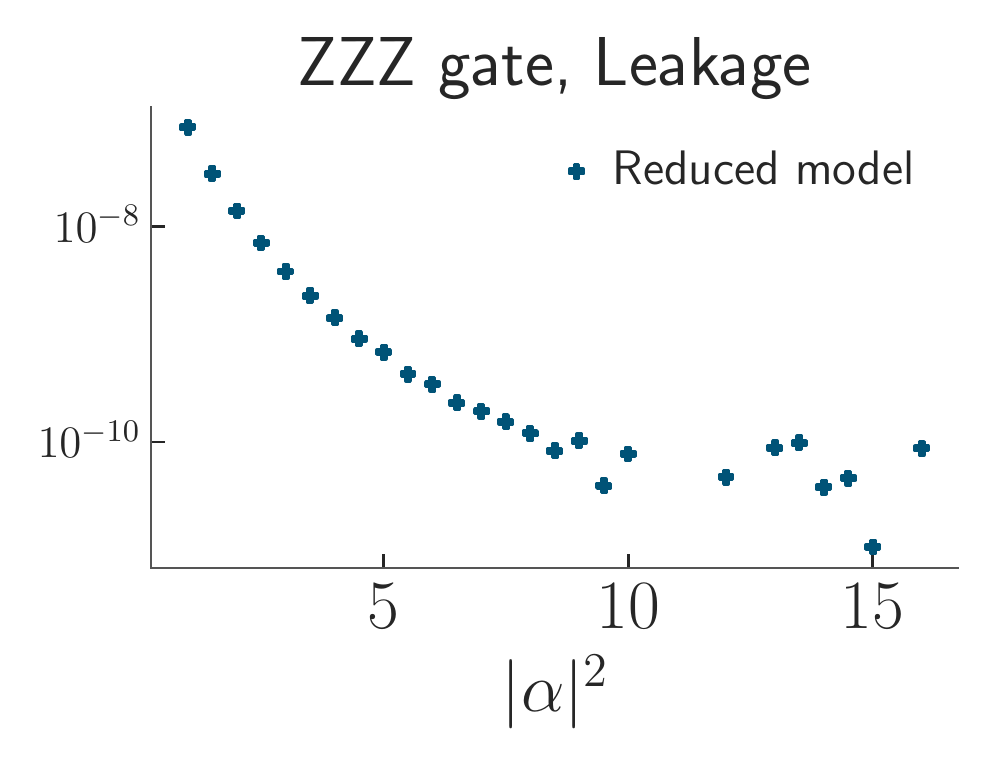}
  % ]{example-image-a}
 \put(5, 70){\captiontext*{}}
 \end{overpic}
 \end{subcaptiongroup}
 \captionsetup{subrefformat=parens}
 \caption{
Leakage of a \subref{fig:Zleakage} Z gate, \subref{fig:ZZleakage} ZZ gate, and \subref{fig:ZZZleakage} ZZZ gate obtained via full model simulations (shown as gray circles) and the reduced model simulations (colored plus) with $\kappa_1=\kappa_2/100, \epsilon_Z=\kappa_2 / 20$ for different mean photon number $\nbar$.
 }
 \label{fig:leakage_nZ}
\end{figure}

\subsection{Hybrid system leakage}
\label{sec:hybrid_leakage_composite}

For a composite state $\rho_{t}=\sum_{d_A}\bS_{A,d_A} \rho_{B, d_A}$ where one subsystem is not actively stabilized, one cannot in general define the leakage on the full system, but only on the stabilized subsystems, or use its full Hilbert space as the code space.
But if there is an explicit code space for all the subsystems, then we can apply the definition of the leakage for a composite system introduced in section~\ref{sec:leakage_composite} to this hybrid case.
For a bipartite system, we still write the code space as $I_{c, A}\otimes I_{c, B}$.
At first-order, we have $\rho_{t}=\sum_{d_A}\bS_{A,d_A} \rho_{B, d_A} = \sum_{d_A}(\bS_{A,d_A}^{(0)} +\ocR\big(\cL_1(\bS_{A,d_A}) \big) ) \rho_{B, d_A} $
And so the leakage $l$ can be expressed as
$$
l = \sum_{d_A} c_{d_A} \Tr{I_{c, B} \rho_{B, d_A}}
$$
where $c_{d_A}$ has been defined in Eq.~\eqref{eq:leakage_one_mode}:
\begin{multline*}
c_{d_A} = \Tr{\bI_{c, A} (\bS_{A, d_A}^{(0)}+\bS_{A, d_A}^{(1)})} \\
 = \sqrt{2} \delta_{1, d_A}+ \Tr{\bI_{c, A} ~\ocR\big(\cL_1(\bS_{A, d_A}) \big)}.
\end{multline*}

\begin{figure*}
 \centering
 \begin{subcaptiongroup}
 \subcaptionlistentry{CNOT leakage}
 \label{fig:CNOTleakage}
 \begin{overpic}[
  width=0.35\textwidth,
  % grid
  % ]{fig/comp_CNOT_bit-flip_nbsteps20000.pdf}
  % ]{fig/CNOT_leakage.pdf}
  ]{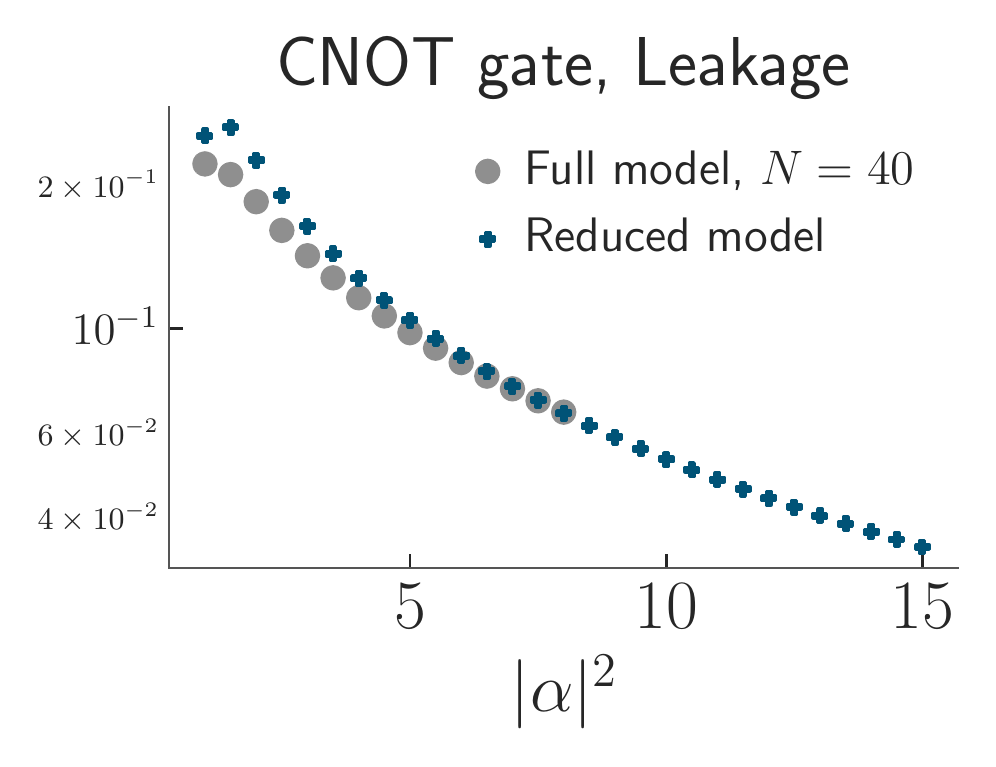}
  % ]{example-image-a}
 \put(5, 70){\captiontext*{}}
 \end{overpic}
 \subcaptionlistentry{CCNOT leakage}
 \label{fig:CCNOTleakage}
 \begin{overpic}[
  width=0.35\textwidth,
  % grid
  % ]{fig/comp_CNOT_bit-flip_nbsteps20000.pdf}
  % ]{fig/CNOT_leakage.pdf}
  ]{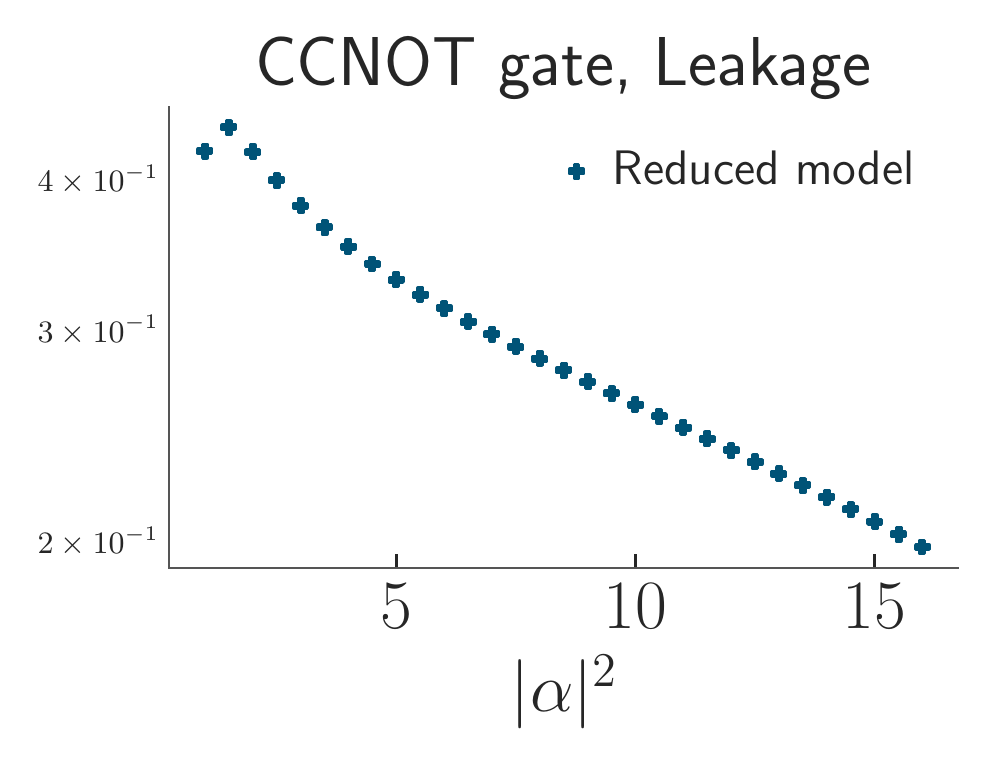}
 \put(5, 70){\captiontext*{}}
 \end{overpic}
 \end{subcaptiongroup}
 \captionsetup{subrefformat=parens}
 \caption{
Leakage of a \subref{fig:CNOTleakage} CNOT gate, and \subref{fig:CCNOTleakage} CCNOT gate obtained via full model simulations (shown as gray circles) and the reduced model simulations (colored plus) with $\kappa_1=\kappa_2/100$ for different mean photon number $\nbar$.}
 \label{fig:CNOT_CCNOT_leakage}
\end{figure*}

\section{Analytic error models} \label{sec:SFB}

In this section, we briefly recall the formalism of the Shifted Fock Basis (SFB) introduced in~\cite{AmazonPRXQ2022} in the context of cat-qubits.
We use the decomposition of the cat-qubit into a two-Level System (TLS)  and a gauge to derive analytical formulas of phase-flip errors for the ZZZ gate.
% and the CNOT gates.

The basis is defined as the displacement along the $+\alpha$ and $-\alpha$ directions of the Fock states $\ket{n}$:
$$
\ket{\pm}_L \otimes \ket{n}_g:= \mathcal{N}_{ \pm}\left[\hat{D}(\alpha) \pm(-1)^n \hat{D}(-\alpha)\right] \ket{\hat{n}=n}.
$$
We can equivalently think about it as a separation of the full Hilbert space as a direct sum between the even an odd parity spaces or, after relabelling, as a tensor product structure of a logical two level system, a qubit encoding the logical state of the cat mode, and a gauge mode $\bg$ of another oscillator:
$$
\mathcal{H}=\mathbb{C}_L^2 \otimes \mathcal{H}_g.
$$
For example, using this basis, the Schr\"{o}dinger cat states $\ket{\mathcal{C}_{\alpha}^{\pm}}$ are given by:
$$
\ket{\mathcal{C}_{\alpha}^{\pm}}= \ket{\pm}_L \otimes \ket{0}_g.
$$

We will use the following approximation of the annihilation operator, valid for large cat-qubits:
\begin{equation}
 \hat{a} \stackrel{|\alpha|^2 \gg d}{\longrightarrow} \hat{Z} \otimes(\bg_a+\alpha).
\end{equation}

This decomposition of the annihilation operator of the full mode as a $\hat{Z}$ operator acting on a qubit tensored with a gauge mode $\bg_a$ is well suited in the pertubative regime where the cat-qubit can be excited to its first excited state, but will quickly decay back to its ground state because of the engineered two-photon dissipation.

Indeed, the operator of the dissipation mechanism $\ba^2 - \nbar $  is more intuitive than the annihilation operator because it corresponds to   $ 2 \alpha \bI \otimes \bg$, i.e.  just to  cool down the gauge mode $\bg$ to vacuum.

In the following, we will use these correspondence  in order to compute   analytical expressions of the $Z$ errors of the Z and ZZ gates used in Sec.~\ref{ssec:Zgate} and Sec.~\ref{ssec:ZZgate}, and explicitly derive the analytical expressions of the $Z$ errors of the ZZZ gate involving three cat-qubits used in Sec.~\ref{ssec:ZZZgate}.

\subsection{Z and ZZ gates}\label{ssec:sfb_Z_ZZ}

The analytical expressions of the $Z$ errors of the Z gate were derived using the SFB in \cite{AmazonPRXQ2022} by adiabatically eliminating the gauge which decays to the ground state manifold with the two-photon dissipation and induces phase flips via the coupling Hamiltonian: $p_Z=|\alpha|^2 \kappa_1 T + \frac{\epsilon_Z^2
T}{|\alpha|^2 \kappa_2}$.
For a ZZ gate, the gauges of both modes are adiabatically eliminated independently.
The two-photon dissipators with a decay rate $\kappa = 4 \nbar \kappa_2$ and the coupling Hamiltonian of rate $g=\epsilon_{ZZ}$ simplify into a single dissipator with a rate $2 \cdot 4 g^2 / \kappa = 2 \epsilon_{ZZ}^2 /  \kappa_2$ causing a $ZZ$ errors with probability $p_{Z_aZ_b}=\frac{\pi^2}{8 \alpha^4 \kappa_2 T} = \frac{\pi\epsilon_{ZZ}}{2\alpha^2\kappa_2}$ while the $Z_a$ and $Z_b$ errors are pure photon loss errors $p_{Z_a}=p_{Z_b}=\alpha^2 \kappa_1 T= \frac{\pi \kappa_1}{4\epsilon_{ZZ}}$.
To obtain the total value of $p_{Z_a Z_b}$, one has to add the errors due to single photon losses on the two qubits $p_{Z_a} p_{Z_b} $.

\subsection{ZZZ gate}\label{ssec:sfb_ZZZ}
We first recall the full master equation then write its expression in the SFB
before performing an adiabatic elimination of the three gauges to derive the $ZZZ$ error rate.
As detailed in~\ref{ssec:ZZZgate}, the master equation of this tripartite systems made of three cat-qubits with
annihilation operators $\ba$, $\bb$ and $\bc$ and gauges $\bg_a$, $\bg_a$ and $\bg_c$ is composed of the stabilization $\cL_0$ and perturbations $\epsilon\cL_1$ that can be split between errors and gate dynamics:
\begin{align*}
 \cL_0(\rho) = \cD_{\ba^2-\alpha^2}(\rho)+\kappa_2 \cD_{\bb^2-\alpha^2}(\rho)+\kappa_2
 \cD_{\bc^2-\alpha^2}(\rho) ,\\
 \epsilon\cL_1(\rho)=\kappa_1
 \cD_{\ba}(\rho)+\kappa_1 \cD_{\bb}(\rho)+\kappa_1
 \cD_{\bc}(\rho)-i\left[\bH_{1}, \rho\right]
\end{align*}
 where $\bH_{1}=\epsilon_{Z Z
 Z}\left(\ba \bb \bc^{\dagger}+\ba^{\dagger} \bb^{\dagger} \bc\right) $ is
applied for a gate-time $ T=\frac{\pi}{4|\alpha|^3 \epsilon_{Z Z Z}}$.

In the SFB, the dissipation writes:
$$
\ba^2-\alpha^2 = \bg_a^2 + 2 \alpha \bg_a \sim 2 \alpha \bg_a
$$
and so the stabilization $\cL_0$ becomes $4 \nbar \kappa_2 (\cD_{\bg_a} +\cD_{\bg_b} +\cD_{\bg_c}) (\rho)$.
The one photon loss becomes:
$\kappa_1 \nbar \cD_{\bZ_a} (\rho) $.

The gate dynamics $\bH_{1}$ becomes
\begin{multline*}
 2|\alpha|^{3} \epsilon_{Z Z
Z}\bZ_a\bZ_b\bZ_c \\
+ \epsilon_{Z Z
Z} \nbar \bZ_a\bZ_b\bZ_c \otimes \left( \bg_a +\bg_a^{\dagger}+\bg_b +\bg_b^{\dagger}+\bg_c +\bg_c^{\dagger} \right).
\end{multline*}
The first term of the gate dynamics produces the desired rotation. It comes with excitations on the gauges, each with a coupling strength $g = \epsilon_{Z Z
Z} \nbar$, inflicting a $ZZZ$ error on the cat-qubits. This excitation decays back to the code space (i.e. ground state of the gauges) with a decay rate $\kappa = 4 \nbar \kappa_2 $ due to $\cL_0$.
In the regime $\kappa \gg g$, the gauges remain mainly on their ground states and thus can be adiabatically eliminated, by adding an effective $ZZZ$ error rate on the qubits with a rate $3 \times 4 g^2 / \kappa$, the factor 3 coming from the three gauges indistinctively.
The effective master equation of the effective system $\rho$ therefore becomes:
\begin{multline*}
\dotex \rho = \kappa_1 \nbar (\cD_{\bZ_a} + \cD_{\bZ_b} + \cD_{\bZ_c} ) (\rho(t)) \\
 + \frac{3 \epsilon_{Z Z
Z} \nbar}{\kappa_2} \cD_{\bZ_a \bZ_b \bZ_c} (\rho(t))
\\ - i \left[ 2|\alpha|^{3} \epsilon_{Z Z
Z}\bZ_a\bZ_b\bZ_c , \rho(t)\right]
\end{multline*}
The effective Hamiltonian term describes the gate dynamics. We perform a rotation around the ZZZ axis of the qubits with an angle $\theta =4|\alpha|^3 \epsilon_{Z Z Z} T$.
The first terms due to one photon losses induces $Z$ errors on the three cat-qubits: $p_{Z_a} =p_{Z_b} =p_{Z_c} = \nbar \kappa_1 T$.
The $Z_aZ_bZ_c$ errors due to the middle term is given by: $3 \frac{\epsilon_{Z Z
Z}^2 \nbar}{\kappa_2} T = \frac{3 \pi \epsilon_{ZZZ}}{4 \alpha \kappa_2}$ for a $\pi$ rotation, to which one has to add the errors due to single photon losses on the three qubits $p_{Z_a} p_{Z_b} p_{Z_c}$ to obtain the total value of $p_{Z_a Z_b Z_c}$.

\end{document}

%% file: fig/chi_matrix_1_qubit.tikz
\begin{tikzpicture} % Adjust the scale if you want a larger or smaller figure

 % Define the labels for columns and rows
 \def\columnlabels{{"$I$","$X$","$Y$","$Z$"}}
 \def\rowlabels{{"$I$","$X$","$Y$","$Z$"}}
 % Define the labels for the diagonal squares
 \def\diagonallabels{{"$\mathcal{F}$","$p_X$","$p_Y$","$p_Z$"}}
  \newcommand{\opacity}{80}
 
 % Loop through rows and columns to draw the squares and labels
%  \foreach \row in {0,1,2,3} {
 \foreach \row in {0,1,2,3} {
  \foreach \col in {0,1,2,3} {
   % Define the coordinates of the current square
   \pgfmathsetmacro\x{0.5*\col}
   \pgfmathsetmacro\y{0.5*\row}
   \pgfmathtruncatemacro\rowupdated{3-\row}
   
   % Fill the square with the appropriate color
   \ifnum\rowupdated=\col
  %  \ifnum\row=\col
    \ifnum\row=3 % White for the first square
     \fill[white] (\x,\y) rectangle (\x+0.5,\y+0.5);
    \fi
    \ifnum\row=2 % Red for the second square
     \fill[rougeX!\opacity!white] (\x,\y) rectangle (\x+0.5,\y+0.5);
    \fi
    \ifnum\row=1 % Green for the third square
     \fill[vertY!\opacity!white] (\x,\y) rectangle (\x+0.5,\y+0.5);
    \fi
    \ifnum\row=0 % Blue for the fourth square
     \fill[bleuZ!\opacity!white] (\x,\y) rectangle (\x+0.5,\y+0.5);
    \fi
    
    % Add the label on the diagonal square
    \ifnum\row=3 % White for the first square
      \node at (\x+0.25, \y+0.25) {\pgfmathparse{\diagonallabels[\rowupdated]}\pgfmathresult};
    \else
      % \node at (\x+0.25, \y+0.25) {\textcolor{white}{\pgfmathparse{\diagonallabels[\rowupdated]}\pgfmathresult}};
      \node at (\x+0.25, \y+0.25) {\pgfmathparse{\diagonallabels[\rowupdated]}\pgfmathresult};
    \fi

   \else
    \fill[lightgray!\opacity!white] (\x,\y) rectangle (\x+0.5,\y+0.5);
   \fi
   
   % \draw (\x,\y) rectangle (\x+0.5,\y+0.5); % Draw the square border
   
   % Add column labels on top
   \ifnum\row=3
    \node at (\x+0.25, \y+0.7) {\pgfmathparse{\columnlabels[\col]}\pgfmathresult};
   \fi
   
   % Add row labels on the left
   \ifnum\col=0
    \node at (\x-0.2, \y+0.25) {\pgfmathparse{\rowlabels[\rowupdated]}\pgfmathresult};
   \fi
  }
 }

\end{tikzpicture}

%% file: fig/chi_matrix_2_qubits.tikz
\begin{tikzpicture}[scale=1.5] % Adjust the scale if you want a larger or smaller figure

  % Define the labels for columns and rows
  \def\columnlabels{{"$I_b$","$X_b$","$Y_b$","$Z_b$","$I_b$","$X_b$","$Y_b$","$Z_b$","$I_b$","$X_b$","$Y_b$","$Z_b$","$I_b$","$X_b$","$Y_b$","$Z_b$"}}
  \def\rowlabels{{"$I_b$","$X_b$","$Y_b$","$Z_b$","$I_b$","$X_b$","$Y_b$","$Z_b$","$I_b$","$X_b$","$Y_b$","$Z_b$","$I_b$","$X_b$","$Y_b$","$Z_b$"}}
  % Define the labels for the diagonal squares
  \def\diagonallabels{{"$\mathcal{F}$","$p_{X_b}$","$p_{Y_b}$","$p_{Z_b}$",
                      "$p_{X_a}$","$p_{X_aX_b}$","$p_{X_aY_b}$","$p_{X_aZ_b}$",
                      "$p_{Y_a}$","$p_{Y_aX_b}$","$p_{Y_aY_b}$","$p_{Y_aZ_b}$",
                      "$p_{Z_a}$","$p_{Z_aX_b}$","$p_{Z_aY_b}$","$p_{Z_aZ_b}$"}}
  
  \def\diagonalcolors{{
    "white","rougeX","rougeX","bleuZ",
    "rougeX",
    "rougeX",
    "rougeX",
    "rougeX",
    "rougeX",
    "rougeX",
    "rougeX",
    "rougeX",
    "bleuZ",
    "rougeX",
    "rougeX",
    "roseZZ",
    }}

    \newcommand{\opacity}{80}

  \foreach \row [evaluate=\row as \usecolor using {\diagonalcolors[15-\row]}]in {0,...,15} {
    \foreach \col in {0,...,15} {
      \pgfmathsetmacro\x{0.5*\col}
      \pgfmathsetmacro\y{0.5*\row}
      \pgfmathtruncatemacro\rowupdated{15-\row}
      
      % Fill the square with the appropriate color
      \ifnum\rowupdated=\col
      \else
        \fill[lightgray!\opacity!white] (\x,\y) rectangle (\x+0.5,\y+0.5);
      \fi
    }
  }
  
  % Loop through rows and columns to draw the squares and labels
  \foreach \row [evaluate=\row as \usecolor using {\diagonalcolors[15-\row]}]in {0,...,15} {
    \foreach \col in {0,...,15} {
      % Define the coordinates of the current square
      \pgfmathsetmacro\x{0.5*\col}
      \pgfmathsetmacro\y{0.5*\row}

      \pgfmathtruncatemacro\rowupdated{15-\row}
      
      % Fill the square with the appropriate color
      \ifnum\rowupdated=\col
    %   \draw (\x,\y) rectangle (\x+0.5,\y+0.5); % Draw the square border
        % \pgfmathtruncatemacro\colorindex{\row % Index to loop through the colors
        %   - 4*floor(\row/4)} % Make sure it stays within 0 to 3
        \fill[\usecolor!\opacity!white] (\x,\y) rectangle (\x+0.5,\y+0.5);
        
        % \ifnum\row=0 % White for the first square
        %   \fill[white] (\x,\y) rectangle (\x+0.5,\y+0.5);
        % \fi
        % \ifnum\row=1 % Red for the second square
        %   \fill[red] (\x,\y) rectangle (\x+0.5,\y+0.5);
        % \fi
        % \ifnum\row=2 % Green for the third square
        %   \fill[green] (\x,\y) rectangle (\x+0.5,\y+0.5);
        % \fi
        % \ifnum\row=3 % Blue for the fourth square
        %   \fill[blue] (\x,\y) rectangle (\x+0.5,\y+0.5);
        % \fi
        
        % Add the label on the diagonal square
        \ifnum\row=15 % White for the first square
            \node at (\x+0.25, \y+0.25) {\pgfmathparse{\diagonallabels[\rowupdated]}\pgfmathresult};
        \else
            \node at (\x+0.25, \y+0.25) {\pgfmathparse{\diagonallabels[\rowupdated]}\pgfmathresult};
        \fi
        % \node at (\x+0.25, \y+0.25) {\pgfmathparse{\diagonallabels[\rowupdated]}\pgfmathresult};
        
    %   \else
    %     \fill[lightgray] (\x,\y) rectangle (\x+0.5,\y+0.5);
      \fi
      
    %   \draw (\x,\y) rectangle (\x+0.5,\y+0.5); % Draw the square border
      
      % Add column labels on top
      \ifnum\row=15
        \node at (\x+0.25, \y+0.7) {\pgfmathparse{\columnlabels[\col]}\pgfmathresult};
      \fi
      
      % Add row labels on the left
      \ifnum\col=0
        \node at (\x-0.2, \y+0.25) {\pgfmathparse{\rowlabels[\rowupdated]}\pgfmathresult};
      \fi
    }
  }

  \node at (1, 8.5) {$I_a$};
  \node at (3, 8.5) {$X_a$};
  \node at (5, 8.5) {$Y_a$};
  \node at (7, 8.5) {$Z_a$};

  \draw (2, 8.8) -- (2, 8.1);
  \draw (4, 8.8) -- (4, 8.1);
  \draw (6, 8.8) -- (6, 8.1);

  \node at (-0.6, 7) {$I_a$};
  \node at (-0.6, 5) {$X_a$};
  \node at (-0.6, 3) {$Y_a$};
  \node at (-0.6, 1) {$Z_a$};

  \draw (-1, 6) -- (-0.1, 6);
  \draw (-1, 4) -- (-0.1, 4);
  \draw (-1, 2) -- (-0.1, 2);

  \path
    (1, 7.25) edge[bend left] node [left] {} (6, 5);
  \path
    (1.5, 6.75) edge[bend left] node [left] {} (6, 5);

  \foreach \offset in {0.5,1,...,2.} {
    \pgfmathsetmacro\x{2.+\offset}
    \pgfmathsetmacro\y{6.25-\offset}
  \path
    (\x, \y) edge[bend left] node [left] {} (6, 5);
    \pgfmathsetmacro\x{6.5-\offset}
    \pgfmathsetmacro\y{1.75+\offset}
  \path
    (\x, \y) edge[bend right] node [left] {} (6, 5);
    }

  \path
    (7, 1.25) edge[bend right] node [left] {} (6, 5);
  \path
    (7.5, 0.75) edge[bend right] node [left] {} (6, 5);

    \node[fill,circle, color=rougeX!\opacity!white] at (6, 5) {\textcolor{black}{$p_X$}};

\end{tikzpicture}